\documentclass[a4paper,11pt]{article}
\pdfoutput=1 

\usepackage{jheppub} 
\usepackage[utf8]{inputenc}
\usepackage[T1]{fontenc} 

\usepackage{placeins} 

\usepackage{chngpage}

\usepackage{amsmath,amsfonts}
\usepackage{slashed}
\usepackage{feynmp}
\usepackage{float}
\usepackage{graphicx}
\usepackage{array}
\usepackage{colortbl}
\usepackage{booktabs}
\usepackage{tcolorbox}
\newcolumntype{x}[1]{>{\centering\let\newline\\\arraybackslash\hspace{0pt}}p{#1}}

\usepackage{multicol}

\usepackage{fontawesome}
\usepackage{hyperref} 
\hypersetup{
     colorlinks = true,
     linkcolor = blue,
     anchorcolor = blue,
     citecolor = blue,
     filecolor = blue,
     urlcolor = blue,
     bookmarks=true,
     linktocpage=true
     }

\newcommand{\eps}{\varepsilon}
\newcommand{\vp}{H}
\newcommand{\tvp}{\widetilde{H}}

\newcommand{\vpj }{\mbox{${\vp^\dag i\,\raisebox{2mm}{\boldmath ${}^\leftrightarrow$}\hspace{-4mm} D_\mu\,\vp}$}}
\newcommand{\vpjt}{\mbox{${\vp^\dag i\,\raisebox{2mm}{\boldmath ${}^\leftrightarrow$}\hspace{-4mm} D_\mu^{\,I}\,\vp}$}}

\newcommand{\OO}{\ensuremath{\mathcal{O}}}

\renewcommand{\phi}{\ensuremath{\varphi}}

\newcommand{\sss}{\scriptscriptstyle}
\newcommand{\bpm}{\begin{pmatrix}}      
\newcommand{\epm}{\end{pmatrix}}

\newcommand\mz{m_{\sss Z}}

\newcommand\mh{m_{\sss H}}

\newcommand\GF{G_{\sss F}}

\newcommand{\Op}[1]{\OO_{\sss #1}}
\newcommand{\Opp}[2]{\OO_{\sss #1}^{\sss #2}}

\newcommand{\Cp}[1]{C_{ #1}}
\newcommand{\Cpp}[2]{C_{ #1}^{ #2}}

\newcommand{\aEW}{\alpha_{\sss EW}}
\newcommand{\red}[1]{ {\textcolor{red}{#1}} }
\definecolor{lightgray}{rgb}{0.83, 0.83, 0.83}
\definecolor{lightpurp}{rgb}{0.901,0.796,0.882}

\newcommand{\lgc}{\cellcolor{lightgray} }

\newcommand{\lgr}{\rowcolor{lightgray} }

\newcommand{\lgbox}[1]{ \fcolorbox{white}{lightgray}{#1}}

\def\wt{\widetilde}

\usepackage[framemethod=TikZ]{mdframed}
\mdfdefinestyle{boxed}{%
    linecolor=black,
    outerlinewidth=0.5pt,
    roundcorner=5pt,
    innertopmargin=5pt,
    innerbottommargin=10pt,
    innerrightmargin=10pt,
    innerleftmargin=10pt,
    backgroundcolor=gray!05!white
    }

\title{\boldmath Top, Higgs, Diboson and Electroweak Fit to the Standard Model Effective Field Theory}

\author[a,b,c]{John Ellis,}
\author[d]{Maeve Madigan,}
\author[a]{Ken Mimasu,}
\author[e,f]{Veronica Sanz}
\author[b,d,g]{and Tevong You}

\affiliation[a]{Theoretical Particle Physics and Cosmology Group, Department of Physics, \\
King's~College~London, London WC2R 2LS, UK}
\affiliation[b]{Theoretical Physics Department, CERN, CH-1211 Geneva 23, Switzerland}
\affiliation[c]{National Institute of Chemical Physics \& Biophysics, R{\" a}vala 10, 10143 Tallinn, Estonia}
\affiliation[d]{DAMTP, University of Cambridge, Wilberforce Road, Cambridge CB3 0WA, UK}
\affiliation[e]{Instituto de F{\' i}sica Corpuscular (IFIC), Universidad de Valencia-CSIC, E-46980 Valencia, Spain}
\affiliation[f]{Department of Physics and Astronomy, University of Sussex, Brighton BN1 9QH, UK}
\affiliation[g]{Cavendish Laboratory, University of Cambridge, J.J. Thomson Avenue, Cambridge CB3 0HE, UK}
\emailAdd{john.ellis@cern.ch}
\emailAdd{mum20@cam.ac.uk}
\emailAdd{ken.mimasu@kcl.ac.uk}
\emailAdd{veronica.sanz@uv.es}
\emailAdd{tevong.you@cern.ch}

\abstract{The search for effective field theory deformations of the Standard Model (SM) is a major goal of particle physics that can benefit from a global approach in the framework of the Standard Model Effective Field Theory (SMEFT). For the first time, we include LHC data on top production and differential distributions together with Higgs production and decay rates and Simplified Template Cross-Section (STXS) measurements in a global fit, as well as precision electroweak and diboson measurements from LEP and the LHC, in a global analysis with SMEFT operators of dimension 6 included linearly. 
We present the constraints on the coefficients of these operators, both individually and when marginalised, in flavour-universal and top-specific scenarios, studying the interplay of these datasets and the correlations they induce in the SMEFT. We then explore the constraints that our linear SMEFT analysis imposes on specific ultra-violet completions of the Standard Model, including those with single additional fields and low-mass stop squarks. We also present a model-independent search for deformations of the SM that contribute to between two and five SMEFT operator coefficients. In no case do we find any significant evidence for physics beyond the SM. Our underlying $\tt{Fitmaker}$ public code provides a framework for future generalisations of our analysis, including a quadratic treatment of dimension-6 operators. 
}

\begin{document} 

\begin{flushright}
{\small KCL-PH-TH/2020-73, CERN-TH-2020-202}
\end{flushright}

\maketitle
\flushbottom

\section{Introduction}
\label{sec:intro}

Experimental tests of the Standard Model (SM) and probes of possible new physics
beyond it have been taken to a new level by LHC measurements during its Runs~1 and 2.
Previous to the LHC, the most precise tests of the SM were those provided by
measurements at LEP, notably at the $Z$ peak~\cite{ALEPH:2005ab} and in $W^+ W^-$ diboson production~\cite{Heister:2004wr, Achard:2004zw, Abbiendi:2007rs, Schael:2013ita},
and at the Tevatron, notably in measurements of the $W$~\cite{Aaltonen:2013iut} and top quark~\cite{Aaltonen:2017efp}.
The LHC has added new classes of precise measurements, notably those of
Higgs production and decays~\cite{Aad:2015gba,Khachatryan:2016vau,CMS:2019chr,CMS:2019kqw,Aad:2019mbh,CMS:2019pyn,CMS:1900lgv,CMS:2020gsy,Aad:2020plj,Aad:2020xfq} 
and the production of the top quark in various modes at Run 1~\cite{Chatrchyan:2013faa,CMS:2014ika,Khachatryan:2014iya,Chatrchyan:2014tua,Aad:2015eto,Aad:2015eua,Aad:2015mbv,Khachatryan:2015oqa,Khachatryan:2015sha,Aad:2015upn,Khachatryan:2016ewo,Khachatryan:2016fky,Aaboud:2016hsq,Aaboud:2016iot,Aad:2016ove,Khachatryan:2016ysn,Khachatryan:2016yzq,Sirunyan:2017azo,Aaboud:2017era,Sirunyan:2017iyh,Sirunyan:2017lvd,Aaboud:2017pdi,Aad:2020jvx,Aad:2020zhd} and Run 2~\cite{Sirunyan:2016cdg,Aaboud:2016lpj,CMS:2016xnv,Aaboud:2016ymp,Sirunyan:2017mzl,Sirunyan:2017nbr,Sirunyan:2017uzs,Aaboud:2017ylb,Sirunyan:2018goh,Aaboud:2018jsj,Sirunyan:2018lcp,Sirunyan:2018wem,Sirunyan:2018zgs,ATLAS:2019czt,Sirunyan:2019hqb,Sirunyan:2019jud,Aaboud:2019njj,Sirunyan:2019nxl,CMS:2019too,Sirunyan:2019wxt,Aad:2020axn,ATLAS:2020hrf,Sirunyan:2020kga},
the $W$ mass~\cite{Aaboud:2017svj}, and also triple-gauge coupling measurements in diboson production~\cite{Aaboud:2017qkn,Sirunyan:2019bez, Sirunyan:2019gkh, Aaboud:2019gxl,Aaboud:2019nkz} and $Zjj$ production~\cite{Aad:2020sle}.
The SM is a tight theoretical framework that connects all these measurements.
For this reason, it is necessary to take a global approach to the interpretation of its
tests and to searches for deviations from its predictions that may be
signatures of possible new physics beyond the SM (BSM). 

The interactions within the SM are largely constrained by its SU(3)$\times$SU(2)$\times$U(1)
gauge invariance, the exceptions being the magnitudes of the Yukawa couplings of
fermions to the Higgs boson, and its self-interactions. Currently, all experimental data are consistent with the SU(3)$\times$SU(2)$\times$U(1)
assignments of particles that are specified in the SM. It is therefore natural to assume that
any BSM interactions must be invariant under SU(3)$\times$SU(2)$\times$U(1) and respect
these established representation assignments. The absence of any evidence for additional
particles at the LHC or elsewhere suggests that any BSM particles may be significantly
heavier than the electroweak scale, as represented by the masses of the $W, Z$, top
quark and Higgs boson. In this case, the low-energy effects of the more massive BSM particles may
be approximated by integrating them out to obtain higher-dimensional interactions
between the SM fields~\cite{Appelquist:1974tg}. In this approach the SM is regarded as an effective
theory whose known renormalizable interactions are supplemented by higher-order terms
scaled by inverse powers of the BSM mass scale(s)~\cite{Weinberg:1979sa, Buchmuller:1985jz}.
This SM Effective Field Theory (SMEFT) is a powerful tool for analyzing the consistency
of the SM and searching indirectly for possible BSM physics~\footnote{A more general EFT parametrisation in which the electroweak SU(2)$ \times$U(1) symmetry is non-linearly realised and the Higgs is added as a singlet scalar, known as the Higgs EFT (HEFT), is described, e.g., in Ref.~\cite{Cohen:2020xca} with fits given in Refs.~\cite{Buchalla:2015qju, Brivio:2016fzo, deBlas:2018tjm}. Its geometric interpretation is developed in Refs.~\cite{Alonso:2015fsp, Helset:2020yio}.}.

There have been intensive theoretical efforts in recent years to formulate consistently
the SMEFT and to classify the sets of independent operators appearing at each dimension~\cite{Weinberg:1979sa, Buchmuller:1985jz, Grzadkowski:2010es, Lehman:2014jma, Lehman:2015coa, Henning:2015alf, Gripaios:2018zrz, Criado:2019ugp, Murphy:2020rsh, Li:2020gnx, Li:2020xlh, Liao:2020jmn}. The first non-redundant basis of dimension-6 operators, known as the Warsaw basis, was laid out in Ref.~\cite{Grzadkowski:2010es}, and the matrix of
one-loop anomalous dimensions of these operators has been computed in Refs.~\cite{Jenkins:2013zja, Jenkins:2013wua, Alonso:2013hga}.  
These and other theoretical developments have laid the groundwork for phenomenological SMEFT analyses of available data to go beyond fitting subsets of specific operators towards global fits in a complete basis of operators, beginning with early global SMEFT fits to electroweak precision data~\cite{Han:2004az} that started to include triple-gauge coupling measurements and the Higgs boson~\cite{Pomarol:2013zra, Corbett:2012ja, Ellis:2014jta} soon after its discovery~\cite{Aad:2012tfa, Chatrchyan:2012ufa}. Other LHC fits focused on triple-gauge couplings and/or Higgs measurements for subsets of operators~\cite{Dumont:2013wma, Corbett:2013pja, Chang:2013cia, Elias-Miro:2013mua, Boos:2013mqa, Ellis:2014dva}. Since then here have been a variety of studies of the SMEFT, for example, in electroweak processes~\cite{Falkowski:2014tna, Berthier:2016tkq, Banerjee:2019twi, Biekotter:2020flu, Banerjee:2018bio, Falkowski:2020znk}, flavour physics~\cite{Efrati:2015eaa, Silvestrini:2018dos, Descotes-Genon:2018foz, Aebischer:2018iyb, Hurth:2019ula, Aebischer:2020dsw, Aoude:2020dwv, Faroughy:2020ina, Aebischer:2020lsx}, low-energy precision data~\cite{Falkowski:2015krw, Falkowski:2017pss, Falkowski:2020pma}, diboson measurements~\cite{Panico:2017frx, Baglio:2018bkm,Grojean:2018dqj, Gomez-Ambrosio:2018pnl, Dawson:2018pyl, Kozow:2019txg, Baglio:2019uty, Azatov:2019xxn}, at dimension 8~\cite{Ellis:2017edi, Ellis:2019zex, Alioli:2020kez, Li:2020gnx, Murphy:2020rsh, Ellis:2020ljj, Bellazzini:2018paj} (where collider positivity constraints are particularly relevant~\cite{Bellazzini:2017bkb,Bellazzini:2018paj, Remmen:2019cyz, Fuks:2020ujk, Bonnefoy:2020yee}), and its connection with UV-complete models, both at tree-level~\cite{Giudice:2007fh, Dawson:2017vgm, deBlas:2017xtg, Anisha:2020ggj, Marzocca:2020jze, Dawson:2020oco, Mecaj:2020opd} and one-loop~\cite{Henning:2014gca, Huo:2015exa,Drozd:2015kva, Huo:2015nka, Drozd:2015rsp, Chiang:2015ura, Han:2017cfr, Bakshi:2018ics, Haisch:2020ahr, Gherardi:2020det, Ellis:2020ivx, Chala:2020vqp}. In particular, the most recent global analyses have set constraints on dimension-6 SMEFT operator coefficients imposed by 
precision electroweak data from LEP and the Tevatron, together with Higgs and diboson
data from the LHC including some from Run~2~\cite{Ellis:2018gqa, Almeida:2018cld, Biekotter:2018rhp, Falkowski:2019hvp}~\footnote{See Refs.~\cite{ATLAS:2020naq, CMS:2020gsy} for recent SMEFT interpretations of the Higgs by ATLAS and CMS.}, while separate SMEFT fits of data on the top
quark have also been performed~\cite{Buckley:2015lku, Brivio:2019ius, Bissmann:2019gfc, Hartland:2019bjb,Durieux:2019rbz, vanBeek:2019evb,CMS:2020pnn}. 
\begin{figure}[t] 
\centering
\includegraphics[width=0.65\textwidth]{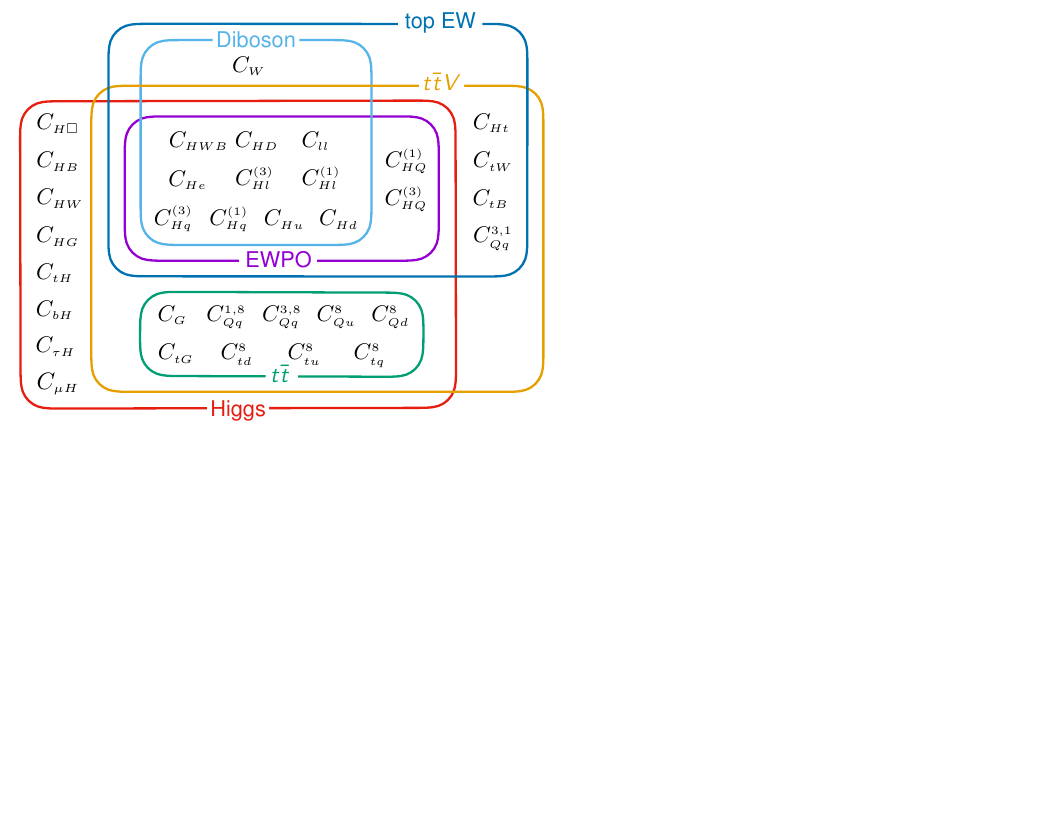}
\caption{
\label{fig:venn}
\it Schematic representation of the datasets and their overlapping dependences on the 34 Wilson coefficients included in our analysis. } 
\end{figure}

We present here the first global dimension-6 SMEFT analysis to include top data and operators in a simultaneous combination of the constraints from the Higgs, electroweak, diboson and top sectors. We use a full set of data from LHC Run 2, in particular the latest Higgs Simplified Template Cross Section (STXS) measurements, differential distributions in $WW$ diboson and $Zjj$ measurements, and updated top observables including kinematic distributions, $t \bar t$, single-top and $t \bar t W/Z$ production. In addition to expanding our dataset, improvements over previous fits include a proper computation using {\tt SMEFT@NLO}~\cite{Degrande:2020evl} of the dimension-6 contributions to Higgs gluon fusion in STXS bins and incorporating the full SMEFT dependence in off-shell Higgs to 4 lepton decays~\cite{Brivio:2019myy}. We also provide a self-consistent treatment of the triple-gluon operator at linear order that had been omitted from our previous fit~\cite{Ellis:2014jta} on the basis of strong constraints at quadratic order~\cite{Krauss:2016ely, Hirschi:2018etq, Goldouzian:2020wdq}. 
We discuss two possible options
for the fermion flavour structure, one assuming a flavour-universal symmetry and the other
allowing the coefficients of operators containing third-generation fermions to vary independently through a top-specific flavour symmetry, both of which are explicitly broken only by the dimension-6 operators that induce shifts in Yukawa couplings.

Fig.~\ref{fig:venn} provides a schematic representation of 
the dependences of these datasets on the 34 
dimension-6 operator coefficients in the top-specific flavour scenario
introduced in Section~\ref{sec:SMEFT} and defined in Tables~\ref{tab:operators} and \ref{tab:4F_SU2_2_SU3_3}. 
The overlaps between the different oblongs show explicitly 
how a given operator may contribute to several classes of
measurements. For example, ${\cal O}_{Ht}, {\cal O}_{tW}, {\cal O}_{tB}$ and
${\cal O}_{Qq}^{3,1}$ contribute to
top EW observables, i.e., single-top and $W$ helicity fraction measurements, as well as
$t \bar t V$ observables. In addition
to these observables, ${\cal O}_{Hu}, {\cal O}_{Hd}, {\cal O}_{Hq}^{(3)}$ 
and ${\cal O}_{Hq}^{(1)}$
also contribute to
diboson, electroweak precision and Higgs observables, exemplifying the interplay between the
various data sectors.

Our analysis is based on a newly-developed public code called {\tt Fitmaker} that is to be continuously updated in the future, with the goal of
continuing to provide an adaptable, flexible and extensible framework for global SMEFT fits over the longer term. We have implemented both a fast analytical method for linear-order fits and a
Markov Chain Monte Carlo (MCMC) procedure that is necessary to
incorporate positivity priors in operator coefficients, as appear in 
specific BSM scenarios. 

The global analysis we present here is performed at linear order 
in the SMEFT operator coefficients. It makes manifest the interplay between Higgs and top 
observables, as well as the electroweak and diboson data, and its calculational speed makes possible a broad-band search for
physics beyond the Standard Model.
The linear constraints can be viewed as provisional for operators where quadratic contributions are non-negligible. Nevertheless, keeping those operators in the global fit typically yields conservative marginalised limits and allows one to assess where the impact on other operators can be significant to a first approximation. Our framework is not limited in principle to linear order:
we discuss in the text the possible impact of quadratic effects in the top sector and illustrate in 
an Appendix focusing on Higgs data how our procedure can be used for quadratic fits as well, though a global analysis
at quadratic order is beyond the scope of this paper. We note that the quadratic contributions can be dominant for strongly-coupled new physics, though 
dimension-8 operators would in general be relevant
at the same order in the new physics scale, and could be included relatively economically in our framework to linear order.

The layout of our paper is as follows. Section~\ref{sec:SMEFT} reviews the SMEFT framework we use,
and sets out the flavour-universal scenario with SU(3)$^5$ symmetry and the top-specific scenario with SU(2)$^2 \times$SU(3)$^5$ symmetry that we analyze. Section~\ref{sec:dataset} summarizes
the dataset that we use in the global fit, which is described in more detail in Appendix~\ref{app:datasets}.
Section~\ref{sec:theorypredictions} describes how theory 
predictions are calculated, and
Section~\ref{sec:procedure} sets out our fitting
procedure. This is described in more detail in Appendix~\ref{app:nestedsampling}, including a nested-sampling calculational method that we illustrate in an analysis of Higgs signal strengths to quadratic order in the dimension-6 operator
coefficients. The results from our global linear fit are presented in Section~\ref{sec:results}, where we
display results from the electroweak, Higgs and top sectors separately and in combination. In both the flavour-universal SU(3)$^5$ symmetric and top-specific SU(2)$^2 \times$SU(3)$^5$
scenario, we derive constraints for all individual operators and also constraints including all 
dimension-6 operators and marginalising. 
Our principal results are displayed in Fig.~\ref{fig:all_fit}, with numerical results for the
SU(2)$^2 \times$SU(3)$^3$ top-specific scenario presented in Table~\ref{tab:all_fit}. We include dedicated discussions of sensitivities in ‘Higgs-only’ operator planes that illustrates the impact of the top data, and of the triple-gluon operator coefficient $C_G$. We also analyze correlations between the coefficients of the 34 operators in the top-specific scenario and perform a principal component analysis, identifying the most and least constrained combinations of SMEFT operators. Applications of our analysis to some specific BSM scenarios are presented in
Section~\ref{sec:UV}, including single-field extensions of the SM and a light-stop scenario, and a survey
of fits with contributions from any combination of two, three, four or
five dimension-6 operators, none of which provide any significant evidence for physics beyond the SM. Our conclusions are summarised in Section~\ref{sec:conx}.

\section{SMEFT Framework}
\label{sec:SMEFT}

We describe in this Section the dimension-6 SMEFT framework that we use for our analysis~\footnote{For some reviews of the SMEFT, see, e.g., Refs.~\cite{Willenbrock:2014bja,Falkowski:2015fla,Brivio:2017vri}.}. After reviewing flavour symmetry in the SM, we classify accordingly the dimension-6 operators of the SMEFT~\cite{Faroughy:2020ina} and describe two flavour symmetry assumptions that we adopt for our global fit, namely flavour-universal and top-specific scenarios.

\subsection{Flavour Symmetry in the Standard Model}
\label{sec:SMflavour}

In the absence of the Yukawa interactions, the SM Lagrangian has a U(3)$^5$ 
global symmetry~\cite{Gerard:1982mm,Chivukula:1987py} associated with independent unitary transformations among the flavour 
components of its five types of fermions: the left-handed quark doublets, right-handed charge +2/3 and -1/3 quarks,
left-handed lepton doublets and right-handed charged leptons. This symmetry can be decomposed into 
five SU(3) flavour rotations and five U(1) symmetries: baryon and lepton numbers, 
hypercharge, a Peccei-Quinn symmetry and independent rotations of the 
right-handed leptons: 
\begin{align}
    \mathcal{G} &= SU(3)_q\times SU(3)_u\times SU(3)_d\times SU(3)_l\times SU(3)_e \times U(1)^5 \, .
\end{align}
The fermion representations transform as follows under the SU(3) rotations:
\begin{align}
    q^i&\to U_q^{ij}q^j\,,\,
    u^i \to U_u^{ij}u^j\,,\,
    d^i \to U_d^{ij}d^j\,,\,
    l^i \to U_l^{ij}l^j\,,\,
    e^i \to U_e^{ij}e^j,
\end{align}
with U$_x\in $SU(3)$_x$. However, the SM Yukawa Lagrangian
\begin{align}
    \mathcal{L}_{\text{Yuk.}}=\lambda^{ij}_d(\bar{q}_{i}\, H )d_j
                             +\lambda^{ij}_u(\bar{q}_{i}\,\tilde{ H })u_j
                             +\lambda^{ij}_e(\bar{l}_{i}\, H )e_j+\text{h.c.}
\end{align}
is manifestly not invariant under this set of transformations. The down-type 
Yukawa interactions break SU(2)$_q\times $SU(2)$_d$, the up-type interactions break SU(2)$_q\times $SU(2)$_u$ 
and the lepton Yukawa interactions break SU(2)$_l\times $SU(2)$_e$.

One can recover the complete flavour symmetry in the presence of the Yukawa interactions by 
introducing three spurions that take the places of the Yukawa matrices and 
transform as follows:
\begin{align}
    Y_u\to U_q\,Y_u\,U^\dagger_u\quad ,\quad
    Y_d\to U_q\,Y_d\,U^\dagger_d\quad,\quad
    Y_e\to U_l\,Y_e\,U^\dagger_e \, ,
\end{align}
such that the Yukawa Lagrangian becomes
\begin{align}
    \mathcal{L}_{\text{Yuk.}}=Y^{ij}_d(\bar{q}_{i}\, H )d_j
                             +Y^{ij}_u(\bar{q}_{i}\,\tilde{ H })u_j
                             +Y^{ij}_e(\bar{l}_{i}\, H )e_j+\text{h.c.} \, .
\end{align}
The spurions take background values that can be rotated using 
flavour symmetry to a basis in which the down-type quark and lepton mass terms 
are diagonal and the up-type quark mass term is diagonal up to a factor of the CKM matrix, $V$:
\begin{align}
    \langle Y_d\rangle^{ij} = y_d^{ij}\propto m_d^{ij},\,\,
    \langle Y_e\rangle^{ij} = y_e^{ij}\propto m_e^{ij},\,\,
    \langle Y_u\rangle^{ij} = (V^\dagger y_u)^{ij}\propto (V^\dagger)^{ik} m_u^{kj} \, .
\end{align}
After electroweak symmetry breaking (EWSB), the remaining $V$ factor can be moved to the charged-current gauge 
interaction terms by an independent rotation of the left-handed up-type quark 
field~\footnote{The opposite convention is sometimes used, with a diagonal up-type mass matrix before EWSB.}, $u_L^i\to (V^\dagger)^{ik}u_L^k$.

According to the Minimal Flavour Violation (MFV) hypothesis~\cite{DAmbrosio:2002vsn}, no new 
sources of flavour violation exist beyond the SM Yukawa couplings. It assumes that 
flavour-violating effects in, {e.g.,} higher-dimensional operators, appear 
with an associated Yukawa matrix via spurion insertions. This can be taken as a prior for the expected sizes of the operator coefficients. Often the operators are defined to absorb a Yukawa factor determined by the spurions but here we shall consider the coefficients of the operators that break SU(3)$^5$, such as those that induce shifts in the Yukawas, as explicit breakings of the flavour symmetry, without these spurion insertions. This simplifies the normalisation of the operators to make the interpretation of their coefficients clear and unambiguous. 

\subsection{SMEFT operators}
\label{sec:smeftoperators}

\begin{table}[t] 
\begin{adjustwidth}{-0.28in}{-0.28in}
{\small
\centering
\renewcommand{\arraystretch}{1.0}
\begin{tabular}{||c|c||c|c||c|c||} 
\hline \hline
\multicolumn{2}{||c||}{$X^3$} & 
\multicolumn{2}{|c||}{$\vp^6$~ and~ $\vp^4 D^2$} &
\multicolumn{2}{|c||}{$\psi^2\vp^3$}\\
\hline
$\Op{G}$                 & $f^{ABC} G_\mu^{A\nu} G_\nu^{B\rho} G_\rho^{C\mu} $ &  
$\Op{\vp}$               & $(\vp^\dag\vp)^3$ &
\lgc $\Op{e\vp}$           & \lgc $(\vp^\dag \vp)(\bar l_p e_r \vp)$\\
$\Op{\wt G}$          & $f^{ABC} \wt G_\mu^{A\nu} G_\nu^{B\rho} G_\rho^{C\mu} $ &   
$\Op{\vp\Box}$        & $(\vp^\dag \vp)\raisebox{-.5mm}{$\Box$}(\vp^\dag \vp)$ &
\lgc $\Op{u\vp}$           & \lgc $(\vp^\dag \vp)(\bar q_p u_r \tvp)$\\
$\Op{W}$                 & $\eps^{IJK} W_\mu^{I\nu} W_\nu^{J\rho} W_\rho^{K\mu}$ &    
$\Op{\vp D}$          & $\left(\vp^\dag D^\mu\vp\right)^\star \left(\vp^\dag D_\mu\vp\right)$ &
\lgc $\Op{d\vp}$           & \lgc $(\vp^\dag \vp)(\bar q_p d_r \vp)$\\
$\Op{\wt W}$          & $\eps^{IJK} \wt W_\mu^{I\nu} W_\nu^{J\rho} W_\rho^{K\mu}$ &&&& \\    
\hline \hline
\multicolumn{2}{||c||}{$X^2\vp^2$} &
\multicolumn{2}{|c||}{$\psi^2 X\vp$} &
\multicolumn{2}{|c||}{$\psi^2\vp^2 D$}\\ 
\hline
$\Op{\vp G}$            & $\vp^\dag \vp\, G^A_{\mu\nu} G^{A\mu\nu}$ & 
\lgc $\Op{eW}$               & \lgc $(\bar l_p \sigma^{\mu\nu} e_r) \tau^I \vp W_{\mu\nu}^I$ &
$\Opp{\vp l}{(1)}$      & $(\vpj)(\bar l_p \gamma^\mu l_r)$\\
$\Op{\vp\wt G}$         & $\vp^\dag \vp\, \wt G^A_{\mu\nu} G^{A\mu\nu}$ &  
\lgc $\Op{eB}$               & \lgc $(\bar l_p \sigma^{\mu\nu} e_r) \vp B_{\mu\nu}$ &
$\Opp{\vp l}{(3)}$      & $(\vpjt)(\bar l_p \tau^I \gamma^\mu l_r)$\\
$\Op{\vp W}$            & $\vp^\dag \vp\, W^I_{\mu\nu} W^{I\mu\nu}$ & 
\lgc $\Op{uG}$               & \lgc $(\bar q_p \sigma^{\mu\nu} T^A u_r) \tvp\, G_{\mu\nu}^A$ &
$\Op{\vp e}$            & $(\vpj)(\bar e_p \gamma^\mu e_r)$\\
$\Op{\vp\wt W}$         & $\vp^\dag \vp\, \wt W^I_{\mu\nu} W^{I\mu\nu}$ &
\lgc $\Op{uW}$               & \lgc $(\bar q_p \sigma^{\mu\nu} u_r) \tau^I \tvp\, W_{\mu\nu}^I$ &
$\Opp{\vp q}{(1)}$      & $(\vpj)(\bar q_p \gamma^\mu q_r)$\\
$\Op{\vp B}$            & $ \vp^\dag \vp\, B_{\mu\nu} B^{\mu\nu}$ &
\lgc $\Op{uB}$               &\lgc  $(\bar q_p \sigma^{\mu\nu} u_r) \tvp\, B_{\mu\nu}$&
$\Opp{\vp q}{(3)}$      & $(\vpjt)(\bar q_p \tau^I \gamma^\mu q_r)$\\
$\Op{\vp\wt B}$         & $\vp^\dag \vp\, \wt B_{\mu\nu} B^{\mu\nu}$ &
\lgc $\Op{dG}$               & \lgc $(\bar q_p \sigma^{\mu\nu} T^A d_r) \vp\, G_{\mu\nu}^A$ & 
$\Op{\vp u}$            & $(\vpj)(\bar u_p \gamma^\mu u_r)$\\
$\Op{\vp WB}$           & $ \vp^\dag \tau^I \vp\, W^I_{\mu\nu} B^{\mu\nu}$ &
\lgc $\Op{dW}$               & \lgc $(\bar q_p \sigma^{\mu\nu} d_r) \tau^I \vp\, W_{\mu\nu}^I$ &
$\Op{\vp d}$            & $(\vpj)(\bar d_p \gamma^\mu d_r)$\\
$\Op{\vp\wt WB}$        & $\vp^\dag \tau^I \vp\, \wt W^I_{\mu\nu} B^{\mu\nu}$ &
\lgc $\Op{dB}$               &\lgc  $(\bar q_p \sigma^{\mu\nu} d_r) \vp\, B_{\mu\nu}$ &
\lgc $\Op{\vp u d}$          & \lgc $i(\tvp^\dag D_\mu \vp)(\bar u_p \gamma^\mu d_r)$\\
\hline \hline
\hline\hline
\multicolumn{2}{||c||}{$(\bar LL)(\bar LL)$} & 
\multicolumn{2}{|c||}{$(\bar RR)(\bar RR)$} &
\multicolumn{2}{|c||}{$(\bar LL)(\bar RR)$}\\
\hline
$\Op{ll}$               & $(\bar l_p \gamma_\mu l_r)(\bar l_s \gamma^\mu l_t)$ &
$\Op{ee}$               & $(\bar e_p \gamma_\mu e_r)(\bar e_s \gamma^\mu e_t)$ &
$\Op{le}$               & $(\bar l_p \gamma_\mu l_r)(\bar e_s \gamma^\mu e_t)$ \\
$\Opp{qq}{(1)}$  & $(\bar q_p \gamma_\mu q_r)(\bar q_s \gamma^\mu q_t)$ &
$\Op{uu}$        & $(\bar u_p \gamma_\mu u_r)(\bar u_s \gamma^\mu u_t)$ &
$\Op{lu}$               & $(\bar l_p \gamma_\mu l_r)(\bar u_s \gamma^\mu u_t)$ \\
$\Opp{qq}{(3)}$  & $(\bar q_p \gamma_\mu \tau^I q_r)(\bar q_s \gamma^\mu \tau^I q_t)$ &
$\Op{dd}$        & $(\bar d_p \gamma_\mu d_r)(\bar d_s \gamma^\mu d_t)$ &
$\Op{ld}$               & $(\bar l_p \gamma_\mu l_r)(\bar d_s \gamma^\mu d_t)$ \\
$\Opp{lq}{(1)}$                & $(\bar l_p \gamma_\mu l_r)(\bar q_s \gamma^\mu q_t)$ &
$\Op{eu}$                      & $(\bar e_p \gamma_\mu e_r)(\bar u_s \gamma^\mu u_t)$ &
$\Op{qe}$               & $(\bar q_p \gamma_\mu q_r)(\bar e_s \gamma^\mu e_t)$ \\
$\Opp{lq}{(3)}$                & $(\bar l_p \gamma_\mu \tau^I l_r)(\bar q_s \gamma^\mu \tau^I q_t)$ &
$\Op{ed}$                      & $(\bar e_p \gamma_\mu e_r)(\bar d_s\gamma^\mu d_t)$ &
$\Opp{qu}{(1)}$         & $(\bar q_p \gamma_\mu q_r)(\bar u_s \gamma^\mu u_t)$ \\ 
&& 
$\Opp{ud}{(1)}$                & $(\bar u_p \gamma_\mu u_r)(\bar d_s \gamma^\mu d_t)$ &
$\Opp{qu}{(8)}$         & $(\bar q_p \gamma_\mu T^A q_r)(\bar u_s \gamma^\mu T^A u_t)$ \\ 
&& 
$\Opp{ud}{(8)}$                & $(\bar u_p \gamma_\mu T^A u_r)(\bar d_s \gamma^\mu T^A d_t)$ &
$\Opp{qd}{(1)}$ & $(\bar q_p \gamma_\mu q_r)(\bar d_s \gamma^\mu d_t)$ \\
&&&&
$\Opp{qd}{(8)}$ & $(\bar q_p \gamma_\mu T^A q_r)(\bar d_s \gamma^\mu T^A d_t)$\\
\hline\hline
\multicolumn{2}{||c||}{ $(\bar LR)(\bar RL)$ and $(\bar LR)(\bar LR)$} &
\multicolumn{4}{|c||}{ $B$-violating}\\\hline
\lgc $\Op{ledq}$  & \lgc $(\bar l_p^j e_r)(\bar d_s q_t^j)$ &
\lgc $\Op{duq}$   &  \multicolumn{3}{|c||}{\lgc $\eps^{\alpha\beta\gamma} \eps_{jk} 
 \left[ (d^\alpha_p)^T C u^\beta_r \right]\left[(q^{\gamma j}_s)^T C l^k_t\right]$}\\
\lgc $\Opp{quqd}{(1)}$ & \lgc $(\bar q_p^j u_r) \eps_{jk} (\bar q_s^k d_t)$ &
\lgc $\Op{qqu}$ & \multicolumn{3}{|c||}{\lgc $\eps^{\alpha\beta\gamma} \eps_{jk} 
  \left[ (q^{\alpha j}_p)^T C q^{\beta k}_r \right]\left[(u^\gamma_s)^T C e_t\right]$}\\
\lgc $\Opp{quqd}{(8)}$ & \lgc $(\bar q_p^j T^A u_r) \eps_{jk} (\bar q_s^k T^A d_t)$ &
\lgc $\Op{qqq}$ & \multicolumn{3}{|c||}{ \lgc $\eps^{\alpha\beta\gamma} \eps_{jn} \eps_{km} 
  \left[ (q^{\alpha j}_p)^T C q^{\beta k}_r \right]\left[(q^{\gamma m}_s)^T C l^n_t\right]$}\\
\lgc $\Opp{lequ}{(1)}$ &\lgc  $(\bar l_p^j e_r) \eps_{jk} (\bar q_s^k u_t)$ &
\lgc $\Op{duu}$ & \multicolumn{3}{|c||}{ \lgc $\eps^{\alpha\beta\gamma} 
  \left[ (d^\alpha_p)^T C u^\beta_r \right]\left[(u^\gamma_s)^T C e_t\right]$}\\
\lgc $\Opp{lequ}{(3)}$ & \lgc $(\bar l_p^j \sigma_{\mu\nu} e_r) \eps_{jk} (\bar q_s^k \sigma^{\mu\nu} u_t)$ &
& \multicolumn{3}{|c||}{}\\
\hline\hline
\end{tabular}
} 
\end{adjustwidth}
\caption{\it Dimension-6 operators in the Warsaw basis, adapted from Ref.~\cite{Grzadkowski:2010es}. The grey cells indicate operators that break flavour SU(3)$^5$ explicitly. \label{tab:operators}}
\end{table}

The SM Lagrangian is only the leading approximation, consisting of operators up to dimension 4, to an EFT that also includes higher-dimensional operators. The lepton number-violating Weinberg operator~\cite{Weinberg:1979sa} is the unique one at dimension 5. At dimension 6, we may write the SMEFT Lagrangian as
\begin{equation}
\mathcal{L}_\text{SMEFT} = \mathcal{L}_\text{SM} + \sum_{i=1}^{2499}\frac{C_i}{\Lambda^2}\mathcal{O}_i \, ,
\label{eq:Ldim6}
\end{equation}
where $\Lambda$ is a dimensionful scale and $C_i$ are the dimensionless Wilson coefficients. The dimension-6 operators $\mathcal{O}_i$ in the Warsaw basis are given in Table~\ref{tab:operators} (adapted from Ref.~\cite{Grzadkowski:2010es}). Those with 
only bosonic fields are invariant under flavour symmetry, whereas those 
containing fermions can be classified in terms of their transformation 
properties under SU(3)$^5$. Any off-diagonal flavour entry violates SU(3)$^5$ explicitly, as do 
terms with both a left- and a right-handed field, 
even in their flavour-diagonal entries. For terms involving two fermions, 
these include the Yukawa ($\psi^2\vp^3$) and dipole ($\psi^2 X\vp$) operators as 
well as the right-handed charged-current operator ($\Op{\vp u d}$). The 
four-fermion sector includes the $(\bar LR)(\bar RL)$, 
$(\bar LR)(\bar LR)$ and $B$-violating classes of operators. All other operators containing
fermionic currents are SU(3)$^5$-invariant in the flavour-universal case, 
{i.e.}, when the diagonal entries have a common Wilson coefficient 
and off-diagonal entries vanish.

\subsubsection{Flavour-universal scenario}
\label{sec:flavouruniversalscenario}

Assuming SU(3)$^5$ symmetry reduces the Warsaw basis to the flavour-universal scenario in which only the operators in cells not 
shaded grey in Table~\ref{tab:operators} are 
allowed, with common flavour-diagonal Wilson coefficients and no off-diagonal entries. 
Neglecting CP-violating interactions, one is left with 31 degrees of freedom, of which 16 are relevant for a leading-order fit to electroweak precision, diboson and Higgs data. To this we also add 4 operators that explicitly break the flavour symmetry and affect Higgs physics through a shift of the tau, muon, $b$-quark and top-quark Yukawa couplings. The 20 operators in our ``flavour-universal'' scenario are then 
\begin{mdframed}[style=boxed]
\begin{align}
    \text{EWPO: } &\quad \mathcal{O}_{HWB} \, , \, \mathcal{O}_{HD} \, , \, \mathcal{O}_{ll} \, , \, \mathcal{O}_{Hl}^{(3)} \, , \, \mathcal{O}_{Hl}^{(1)} \, , \, \mathcal{O}_{He} \, , \, \mathcal{O}_{Hq}^{(3)} \, , \, \mathcal{O}_{Hq}^{(1)} \, , \, \mathcal{O}_{Hd} \, , \, \mathcal{O}_{Hu} \, , \, \nonumber \\
    \text{Bosonic: } &\quad  \mathcal{O}_{H\Box} \, , \, \mathcal{O}_{HG} \, , \, \mathcal{O}_{HW} \, , \, \mathcal{O}_{HB} \, , \, \mathcal{O}_{W} \, , \, \mathcal{O}_G \, , \, \nonumber \\
    \text{Yukawa: } &\quad \mathcal{O}_{\tau H} \, , \, \mathcal{O}_{\mu H} \, , \, \mathcal{O}_{bH} \, , \, \mathcal{O}_{tH} \, .
\label{eq:flavunivops}
\end{align}
\end{mdframed}
We have categorised these operators roughly, into sets that are mostly constrained by electroweak precision observables (EWPO), those that can only be constrained at tree-level by Higgs and diboson measurements (bosonic), and operators that induce shifts in the Yukawa couplings (Yukawa).

\subsubsection{Top-specific flavour scenario}
\label{sec:topspecificflavourscenario}

The minimal flavour scenario that singles out top-quark couplings relaxes 
the SU(3)$^5$ symmetry as follows~\cite{AguilarSaavedra:2018nen}:
\begin{align*}
    SU(3)^5 &\to SU(2)^2\times SU(3)^3\\
    &=SU(2)_q\times SU(2)_u\times SU(3)_d\times SU(3)_l\times SU(3)_e \, .
\end{align*}
This allows chirality-flipping interaction terms involving the third-generation 
quark doublet and right-handed up-type fields, notably the 
top-quark Yukawa interaction. 
The following three additional dimension-6 operators in the SMEFT are now allowed:~\footnote{The analogous operator $[\Op{u H }]^{\sss 33} = \Op{t H }$
is already included in the set of Yukawa operators listed in Eq.~(\ref{eq:flavunivops}).}
\begin{equation}
\label{firstfour}
[\Op{uG}]^{\sss 33} \; = \; \Op{tG} \, , \quad
    [\Op{uB}]^{\sss 33} \; = \; \Op{tB} \, , \quad [\Op{uW}]^{\sss 33} \; = \; \Op{tW} \, .
\end{equation}
The flavour-universality conditions on operators in the 
$\psi^2 H ^2D$, $\bar{L}L\bar{L}L$, $\bar{R}R\bar{R}R$ and $\bar{L}L\bar{R}R$
classes that contain $q$ or $u$ are also relaxed.
Schematically,
\begin{align}
    C_{\text{univ.}}\sum_{i=1,2,3}\mathcal{K}^\mu\bar{f}_i\gamma_\mu\,f_i\to 
    \begin{cases}
    &C_3\mathcal{K}^\mu\bar{f}_3\gamma_\mu\,f_3 \, ,\\
    &C_{\text{univ.}}\sum_{j=1,2{\red{(,3)}}}\mathcal{K}^\mu\bar{f}_j\gamma_\mu\,f_j \, ,
    \end{cases}
    \label{eq:c_univ}
\end{align}
where $\mathcal{K}^\mu$ is a combination of other fields.
Here a choice must be made in the second line of Eq.~\ref{eq:c_univ} whether to split the degrees of freedom into 
a fully-universal operator that preserves the full flavour symmetry or an 
operator that respects only the reduced symmetry SU(2)$^2\times $SU(3)$^3$, corresponding to keeping or removing the index in red, respectively. The two are related by a basis rotation. We adopt the second option, since it better separates the degrees of freedom that affect only top measurements from the rest.
The $\psi^2 H ^2D$ class grows to
\begin{align}
\begin{split}
    [\Opp{ H  q}{(1)}]&\to [\Opp{ H  q}{(1)}]^{j,j}\text{ and } [\Opp{ H  q}{(1)}]^{3,3}= \{\Opp{ H  q_i}{(1)},\Opp{ H  Q}{(1)}\} \, , \\
    [\Opp{ H  q}{(3)}]&\to [\Opp{ H  q}{(3)}]^{j,j}\text{ and } [\Opp{ H  q}{(3)}]^{3,3}= \{\Opp{ H  q_i}{(3)},\Opp{ H  Q}{(3)}\} \, , \\
    [\Op{ H  u}]&\to [\Op{ H  u}]^{j,j}\text{ and } [\Op{ H  u}]^{3,3}= \{\Op{ H  u_i},\Op{ H  t\phantom{i}}\} \, , 
\end{split}
\label{threemore}
\end{align}
where $Q$ denotes here the third-generation quark doublet.
Four-fermion operators are split generically into `four-light', `two-heavy-two-light' 
and `four-heavy' flavour components, where `light' and `heavy' denote the first two and the 
third generations, respectively. The four-light degrees of freedom are the same 
as the four-fermion operators of the SU(3)$^5$ scenario, except that they are flavour-universal 
over three generations for $d$, $l$ and $e$, and only the first two 
generations for $q$ and $u$.
The classification of the additional four-fermion operators under this generalisation is 
slightly more involved, due to the permutation symmetries on the flavour 
indices as well as the equivalence of certain degrees of freedom via Fierz identities. 
This is discussed in Ref.~\cite{AguilarSaavedra:2018nen}, where a `dim6top' basis is chosen for the operators 
involving top fields with LHC top physics observables in mind.
The new operators are shown in Table~\ref{tab:4F_SU2_2_SU3_3} with their 
definitions in terms of the Warsaw basis coefficients. 

A total of 31 new CP-conserving degrees of freedom are introduced 
by this relaxation of the flavour symmetry.
However, our analysis is only sensitive to a subset of these, for two main reasons. First, our chosen dataset does not constrain a number of the operators allowed by this flavour assumption. These include all flavour-universal four-light fermion operators, which are constrained by electron-positron collider data and numerous low-energy scattering and decay experiments (see Ref.~\cite{Falkowski:2017pss} for a recent compilation of constraints), as well as high-energy Drell-Yan and dijet observables at hadron colliders~\cite{Domenech:2012ai, Farina:2016rws, Dawson:2018dxp, Alte:2018xgc, Fuentes-Martin:2020lea, Ricci:2020xre}. Furthermore, it is also effectively blind to the two-heavy quark two-lepton class of operators listed in Table~\ref{tab:4F_SU2_2_SU3_3}. Although many of them mediate the same final states as those selected by, \emph{e.g.}, $t\bar{t}V$ measurements, they do so in the absence of a resonant intermediate $W$ or $Z$ boson decaying into the lepton pair. So far, searches have implemented selections to enhance this resonant contribution, and are therefore not sensitive to the non-resonant phase space populated by the operators in question. The second important feature of our analysis is that it is restricted to the linear, $\mathcal{O}(1/\Lambda^2)$, level in the EFT expansion. This restricts the sensitivity to the set of operators that interfere appreciably with the dominant SM amplitudes for the processes of interest. This is not the case for the six neutral-current mediating, two-heavy two-light operators in the upper left section of Table~\ref{tab:4F_SU2_2_SU3_3}. These operators mediate $q\bar{q}\to t\bar{t}$ production in the colour-singlet channel, which does not interfere at LO with the 
strongly-dominant SM QCD contribution. In contrast, the corresponding charged-current operator affects single-top quark production that, being an EW process in the SM, does have such an interference term. Finally, we also omit the four-heavy operators that would mainly be constrained by $t\bar{t}b\bar{b}$ measurements and four-top production searches. These data have been shown to be  largely sensitive, at present, to the quadratic EFT contributions~\cite{Degrande:2010kt,DHondt:2018cww,Hartland:2019bjb, Banelli:2020iau}, and our analysis would not yield meaningful bounds in these directions. We therefore only include in our analysis 8 two-heavy two-light quark degrees of freedom: the colour-singlet, charged-current operator and seven neutral, colour-octet operators. 

To summarise, the 34 operators relevant for our leading-order, linear fit in the top-specific flavour scenario are the 20 listed in Eqs.~(\ref{eq:flavunivops}) plus the 14 discussed above (three in (\ref{firstfour}), three more in (\ref{threemore}) and eight two-light two-heavy quark operators): 
\begin{mdframed}[style=boxed]
\begin{align}
    \label{eq:topspecificops}
    \text{EWPO: } &\quad \mathcal{O}_{HWB} \, , \, \mathcal{O}_{HD} \, , \, \mathcal{O}_{ll} \, , \, \mathcal{O}_{Hl}^{(3)} \, , \, \mathcal{O}_{Hl}^{(1)} \, , \, \mathcal{O}_{He} \, , \, \mathcal{O}_{Hq}^{(3)} \, , \, \mathcal{O}_{Hq}^{(1)} \, , \, \mathcal{O}_{Hd} \, , \, \mathcal{O}_{Hu} \, , \, \nonumber \\
    \text{Bosonic: } &\quad  \mathcal{O}_{H\Box} \, , \, \mathcal{O}_{HG} \, , \, \mathcal{O}_{HW} \, , \, \mathcal{O}_{HB} \, , \, \mathcal{O}_{W} \, , \, \mathcal{O}_G \, , \, \nonumber \\
    \text{Yukawa: } &\quad \mathcal{O}_{\tau H} \, , \, \mathcal{O}_{\mu H} \, , \, \mathcal{O}_{bH} \, , \, \mathcal{O}_{tH} \, , \nonumber \\
    \text{Top 2F: } &\quad \mathcal{O}_{HQ}^{(3)} \, , \, \mathcal{O}_{HQ}^{(1)} \, , \, \mathcal{O}_{Ht} \, , \, \mathcal{O}_{tG} \, , \, \mathcal{O}_{tW} \, , \, \mathcal{O}_{tB} \, , \nonumber \\
    \text{Top 4F: } &\quad \mathcal{O}_{Qq}^{3,1} \, , \, \mathcal{O}_{Qq}^{3,8} \, , \, \mathcal{O}_{Qq}^{1,8} \, , \, \mathcal{O}_{Qu}^{8} \, , \, \mathcal{O}_{Qd}^{8} \, , \, \mathcal{O}_{tQ}^{8} \, , \, \mathcal{O}_{tu}^{8} \, , \, \mathcal{O}_{td}^{8} \, . 
\end{align}
\end{mdframed}
These are grouped into top operators involving two (top 2F) and four (top 4F) heavy fermions, respectively.

\begin{table}[t]
\centering
{\small
\renewcommand{\arraystretch}{1.0}    
\centering
  \begin{tabular}{|p{0.7cm}p{6.1cm}|p{0.7cm}p{4.1cm}|}
  \hline
  $\mathcal{O}_i$& $C_i$ Definition
  &
  $\mathcal{O}_i$& $C_i$ Definition
  \tabularnewline
  \hline
  \multicolumn{4}{|l|}{\emph{4 quark (2 heavy 2 light)}}
  \tabularnewline
  \hline
   \lgc $\Opp{Qq}{1,1}$& 
   \lgc $\sum\limits_{\sss i=1,2}\left([\Cpp{qq}{(1)}]^{\sss ii 33}
    +\frac{1}{6}[\Cpp{qq}{(1)}]^{\sss i33i}
    +\frac{1}{2}[\Cpp{qq}{(3)}]^{\sss i33i} \right)$ 
   &
   $\Opp{Qq}{1,8}$&
   $\sum\limits_{\sss i=1,2}\left([\Cpp{qq}{(1)}]^{\sss i33i}
   +3[\Cpp{qq}{(3)}]^{\sss i33i}\right)$ 
   \tabularnewline
    $\Opp{Qq}{3,1}$&
    $\sum\limits_{\sss i=1,2}\left([\Cpp{qq}{(3)}]^{\sss ii 33}
    +\frac{1}{6}[\Cpp{qq}{(1)}]^{\sss i33i}
    -\frac{1}{6}[\Cpp{qq}{(3)}]^{\sss i33i}\right)$
   &
   $\Opp{Qq}{3,8}$&
   $\sum\limits_{\sss i=1,2}\left([\Cpp{qq}{(1)}]^{\sss i33i}
    -[\Cpp{qq}{(3)}]^{\sss i33i}\right)$ 
   \tabularnewline
   \lgc $\Opp{tu}{1}$&
   \lgc $\sum\limits_{\sss i=1,2}\left([\Cp{uu}]^{\sss ii33}
    +\frac{1}{3}[\Cp{uu}]^{\sss i33i}\right)$ 
   &
   $\Opp{tu}{8}$&
   $\sum\limits_{\sss i=1,2}2[\Cp{uu}]^{\sss i33i}$
   \tabularnewline
   \lgc $\Opp{td}{1}$&
   \lgc $\sum\limits_{\sss i=1,2(,3)}[\Cpp{ud}{(1)}]^{\sss 33ii}$ 
   &
   $\Opp{td}{8}$&
   $\sum\limits_{\sss i=1,2(,3)}[\Cpp{ud}{(8)}]^{\sss 33ii}$
   \tabularnewline
   \lgc $\Opp{tq}{1}$&
  \lgc  $\sum\limits_{\sss i=1,2}[\Cpp{qu}{(1)}]^{\sss ii33}$ 
   &
   $\Opp{tq}{8}$&
   $\sum\limits_{\sss i=1,2}[\Cpp{qu}{(8)}]^{\sss ii33}$ 
   \tabularnewline
  \lgc  $\Opp{Qu}{1}$&
  \lgc  $\sum\limits_{\sss i=1,2}[\Cpp{qu}{(1)}]^{\sss 33ii}$ 
   &
   $\Opp{Qu}{8}$&
   $\sum\limits_{\sss i=1,2}[\Cpp{qu}{(8)}]^{\sss 33ii}$ 
   \tabularnewline
   \lgc $\Opp{Qd}{1}$&
   \lgc $\sum\limits_{\sss i=1,2(,3)}[\Cpp{qd}{(1)}]^{\sss 33ii}$ 
   &
   $\Opp{Qd}{8}$&
   $\sum\limits_{\sss i=1,2(,3)}[\Cpp{qd}{(8)}]^{\sss 33ii}$ 
   \tabularnewline
   \hline
   \hline
   \multicolumn{4}{|l|}{\emph{4 quark (4 heavy)}}
   \tabularnewline
   \hline
   \lgc $\Opp{QQ}{1}$&
  \lgc  $2[\Cpp{qq}{(1)}]^{\sss 33 33}
    -\frac{2}{3}[\Cpp{qq}{(3)}]^{\sss 3333}$ 
   &
   \lgc $\Opp{QQ}{8}$&
  \lgc  $8[\Cpp{qq}{(3)}]^{\sss 3333}$ 
  \tabularnewline
   \lgc $\Opp{Qt}{1}$&
  \lgc  $[\Cpp{qu}{(1)}]^{\sss 33 33}$ 
   &
  \lgc  $\Opp{Qt}{8}$&
  \lgc  $[\Cpp{qu}{(8)}]^{\sss 33 33}$ 
  \tabularnewline
   \lgc $\Op{tt}$&
  \lgc  $[\Cpp{uu}{(1)}]^{\sss 33 33}$ 
   &
   &
   \tabularnewline
   \hline
   \hline
   \multicolumn{4}{|l|}{\emph{2 heavy 2 lepton}}
   \tabularnewline
   \hline
   \lgc $\Opp{Ql}{-(1)}$&
   \lgc $\sum\limits_{\sss i=1,2,3}[\Cpp{lq}{1}]^{\sss ii 33}-[\Cpp{l q}{3}]^{\sss ii 33}$ 
   &
   \lgc $\Opp{tl}{(1)}$&
   \lgc $\sum\limits_{\sss i=1,2,3}[\Cp{lu}]^{\sss ii 33}$
   \tabularnewline
   \lgc $\Opp{Ql}{3(1)}$&
   \lgc $\sum\limits_{\sss i=1,2,3}[\Cpp{l q}{3}]^{\sss ii 33}$ 
   &   
   \lgc $\Opp{te}{(1)}$&
   \lgc $\sum\limits_{\sss i=1,2,3}[\Cp{eu}]^{\sss ii 33}$ 
  \tabularnewline
   \lgc $\Opp{Qe}{(1)}$&
   \lgc $\sum\limits_{\sss i=1,2,3}[\Cp{eQ}]^{\sss ii 33}$ &
   &
  \tabularnewline
  \hline
  \end{tabular}
  } 
  \caption{\it
  \label{tab:4F_SU2_2_SU3_3}
  Four-fermion operators containing at least one third-generation bilinear in
  the `dim6top' basis~\cite{AguilarSaavedra:2018nen} assuming an SU(2)$^2\times $SU(3)$^3$ flavour symmetry. The relations of the corresponding Wilson coefficients with those of the Warsaw basis are also shown. The shaded entries indicate operators that are not included in our analysis because significant constraints cannot be obtained from the chosen dataset at leading order and linear level in the EFT expansion, as discussed in the text.
  }
\end{table}

\section{Dataset description}
\label{sec:dataset}

In this Section we describe the data used in our global fit.  We summarise here the main categories of data and refer the reader to Appendix~\ref{app:datasets} for a complete list of the observables that have been implemented in {\tt Fitmaker}, together with their source references.

The most precise electroweak measurements, other than the $W$ mass, remain those from LEP and the SLC~\footnote{We note that global SMEFT fits would benefit greatly from a future $Z$-pole run~\cite{Ellis:2015sca, deBlas:2019wgy, Abada:2019zxq}.}. 
The Higgs boson discovery at the LHC enabled the possibility of a closed global SMEFT fit to a complete set of dimension-6 operators for the first time. Higgs physics has since progressed rapidly to include more channels and sub-categories beyond signal strengths. In particular, the STXS categorisations of the various Higgs production sub-channels provide further sensitivity to different directions in the fit, as illustrated, for example, in Fig.~\ref{fig:STXS} below for the case of gluon fusion and described further in the next section.

The higher energies at the LHC also allow certain measurements of diboson and dilepton final states to become competitive with LEP~\cite{Zhang:2016zsp, Grojean:2018dqj, Farina:2016rws, Dawson:2018dxp, Alte:2018xgc, Fuentes-Martin:2020lea, Ricci:2020xre}, enable complementary probes of higher-dimensional operators~\cite{Henning:2018kys, Falkowski:2020znk}, and, moreover, give access to top physics with higher statistics than ever before, including the previously unreachable $t\bar{t}W/Z/H$ and other, rare production processes such as four-top production~\cite{Degrande:2010kt,DHondt:2018cww,Hartland:2019bjb, Banelli:2020iau}. More operators, under less restrictive flavour assumptions, can then be included in a global SMEFT fit. This is particularly motivated since the top quark is often expected to be more sensitive to BSM physics. 

The following is a summary of the different categories of observables that we consider---see Appendix~\ref{app:datasets} for more details and references.
We build our selected dataset by combining statistically independent measurements, including correlation information by means of published covariance/correlation matrices, when available~\footnote{See Ref.~\cite{Bissmann:2019qcd} for a study of the impact of correlations in global fits.}.
In general, for LHC data, this amounts to a single ATLAS and CMS measurement in a particular final state for each LHC run. When multiple measurements, {e.g.}, differential distributions, are reported, a single one is chosen based on maximising the sensitivity of our fit.  

\begin{itemize}
\item The set of electroweak precision observables (EWPOs) include the pseudo-observables measured on the $Z$ resonance by LEP and SLD, together with the $W$ boson mass measurements by CDF and D0 at the Tevatron and ATLAS at the LHC: 
\begin{align}
    \text{EWPO: } \quad \{\Gamma_Z, \sigma^0_\text{had.}, R_l^0, A_{FB}^l, A_l, R_b^0, R_c^0, A_{FB}^b, A_{FB}^c, A_b, A_c, M_W \} \, .
\end{align}
{\it We include a total of 14 electroweak measurements.}

\item For diboson measurements, we include the $W^+W^-$ measurements of total cross-sections at different energies and angular distributions at LEP, the fiducial differential cross-section in leading lepton $p_T$ by ATLAS at the LHC, and ATLAS and CMS fiducial differential cross-section measurements of the $Z$-boson $p_T$ in leptonic $W^\pm Z$ production. We also incorporate the differential distribution in $\Delta \phi_{jj}$ for the $Zjj$ measurement given by ATLAS, which we include in the diboson category because it is sensitive to related physics. 
{\it We include a total of 118 diboson measurements.}

\item The Higgs dataset at the LHC includes the combination of Higgs signal strengths by ATLAS and CMS for Run 1, and for Run 2 both signal strengths and STXS measurements are used. ATLAS in particular provide the combined stage $1.0$ STXS for $4l, \gamma\gamma, WW^*, \tau^+\tau^-$ and $b\bar{b}$, while for CMS we use the signal strengths of $4l$, $\gamma \gamma$,  $WW^*$, $\tau^{+} \tau^{-}$, $b \bar{b}$ and $\mu^{+} \mu^{-}$. 
We also include the $Z\gamma$ signal strength from ATLAS and a differential $WW^*$ cross-section measurement from CMS. 
{\it We include a total of 72 Higgs measurements.} 

\item The top data consists of differential distributions in various $t\bar{t}$ channels and cross-section measurements of top pair production in association with a $W/Z$ boson or a photon (the $t \bar t V$ dataset), as well as various single top differential and inclusive cross-section measurements, for both Runs 1 and 2. 
{\it We include a total of 137 top measurements.}
\end{itemize}

{\it Overall, we include a total of 341 measurements in our analysis.}\\

If not already given in such a form, each measurement is converted into a corresponding `signal strength', $\mu$, defined as the ratio of the observed value to the best available theory prediction, usually quoted in the experimental publication. Differential data is taken from the publication and its associated entry in {\tt HEPdata}, where available, using absolute differential 
cross section measurements, $\vec{\sigma}_{\text{abs}}$, and their associated 
covariance matrices, $ \mathbf{\Sigma}_{\text{abs}}$. If only normalised 
differential cross sections ($\vec{\sigma}_{\text{norm}}$) are published, they 
are converted to absolute ones using the best available measurement of the inclusive cross section for that process in the same channel. 
Covariance matrices are then updated to reflect the 
correlations between the bins induced by the common rescaling of the total 
cross section, $\sigma_{\text{tot}}\pm\delta\sigma_{\text{tot}}$. 
The absolute differential measurement and its covariance matrix are then
\begin{align*}
    \vec{\sigma}_{\text{abs}}&=\vec{\sigma}_{\text{norm}}\sigma_{\text{tot}}\, ,\\
   \mathbf{\Sigma}_{\text{abs}} &= 
   \mathbf{\Sigma}_{\text{norm}}\sigma_{\text{tot}}^2 +
   \delta\sigma_{\text{tot}}^2 \vec{\sigma}_{\text{norm}}\otimes \vec{\sigma}_{\text{norm}} \, .
\end{align*}
Where available, \verb|fastnlo| tables~\cite{Czakon:2017dip,Czakon:2019bcq,Czakon:2019yrx} were used to obtain NNLO QCD predictions for the differential $t\bar{t}$ data.
The SM theoretical errors are taken to be uncorrelated and the relative signal strength covariance matrix  is obtained
by adding the relative experimental and theoretical covariances as follows:
\begin{equation}
    \mathbf{\Sigma}_{\mu} = \left(
    \frac{\mathbf{\Sigma}_{\text{exp}}}
         {\vec{\sigma}_{\text{exp}}\otimes\vec{\sigma}_{\text{exp}}}
         +
    \text{diag}(\vec{\delta}_{\text{th}}/\vec{\sigma}_{\text{th}})^2
    \right)(\vec{\mu}\otimes\vec{\mu}), \; \; {\rm where} \; \; \; \vec{\mu} \equiv \frac{\vec{\sigma}_{\text{exp}}}{\vec{\sigma}_{\text{th}}} \, .
\end{equation}
This corresponds to adding the relative experimental and theory uncertainties in 
quadrature. The observables are stored in the \verb|Fitmaker| database in \verb|json| format, together with metadata and information about how each signal strength was obtained.

\section{SMEFT Predictions}
\label{sec:theorypredictions}
\subsection{General strategy}
The SMEFT predictions for all of the included observables were computed using the code {\tt MadGraph5\_aMC@NLO} together with the {\tt SMEFTsim}~\cite{Brivio:2017btx} and/or the {\tt SMEFT@NLO}~\cite{Degrande:2020evl} UFO models. These are used to extract the linear contribution, $a_i^{\sss X}$, of a given Wilson coefficient, $C_i$, to a physical quantity, $X$, such as a production cross-section, partial or total decay width, or asymmetry:
\begin{align}\label{eqn:muX}
    \mu_{\sss X} \equiv \frac{X}{X_{SM}} =  1 + \sum_i a_i^{\sss X} \frac{C_i}{\Lambda^2}  + \mathcal{O}\left(\frac{1}{\Lambda^4}\right).
\end{align}
The $a^X_i$ can usually be obtained with a single, high statistics Monte-Carlo (MC) run for each coefficient.
In some cases, non-linear contributions from Wilson coefficients can arise in MC predictions due to the $W$-mass shift and modifications of total widths of intermediate particles. The former is a consequence of using the EW $\{ \aEW, \GF, \mz \}$ input scheme, while the latter may modify branching ratios of narrow resonances such as the Higgs, $W$, $Z$ or top. When such states are produced on shell, factorising a given process into production and decay via the narrow-width approximation (NWA) allows the decay contributions to be computed separately, then added to the prediction. If these effects cannot be factorised, as in the case of $W$-mass modifications or off-shell vector bosons in Higgs decays, the $a^X_i$ are obtained by generating the predictions over a range of coefficient values and numerically fitting for the linear dependence. This is done, in particular, for Higgs production in association with a $W$-boson (WH), vector boson fusion (VBF), $pp \to WW\to \ell\nu\ell\nu$ and $pp \to W^\pm Z\to \ell^+\ell^-\ell^\pm\nu$.

Predictions are obtained at Leading Order (LO) in perturbation theory throughout, which, in almost all cases, corresponds to tree-level computations. Unless stated otherwise, we generate and analyse our events at the parton level and apply analysis-specific selection criteria to obtain the relevant fiducial regions of phase space. We use the following values in the aforementioned EW input scheme:
\begin{equation}
\begin{gathered}
\aEW^{-1} = 127.95,\quad  
 \GF = 1.16638\times 10^{-5}\,\text{GeV}^{-2} \, ,\\
 \mz = 91.1876\,\text{GeV},\quad m_{\sss H} = 125.09\, \text{GeV},\quad
 m_t = 173.2\, \text{GeV} \, .
\end{gathered}
\end{equation}
All other fermions are taken to be massless, which implies the use of five-flavour scheme PDFs, for which we use the default {\tt NNPDF23\_nlo\_as\_0119} sets~\cite{Ball:2012cx} provided by  {\tt MadGraph5\-\_aMC@NLO}. The one exception is when the lighter fermions appear as Higgs boson decay products, when we assume that they interact with the Higgs
via their Yukawa couplings, $y_f\equiv 2^{\frac{3}{2}} m_f \sqrt{\GF}$, with masses taken to be
\begin{align}
     m_\mu=0.106\,\text{GeV}\,, \quad m_\tau = 1.77\,\text{GeV}, \quad m_c = 0.907\,\text{GeV}, \quad \text{and}  \quad m_b = 3.237\,\text{GeV} \, .
\end{align}
The latter two have been run up to the Higgs mass scale. In some cases, the 3rd generation-specific operators independently modify $b$-quark initiated contributions to EW processes such as VBF and diboson. We do not take these contributions into account as they are highly suppressed by the $b$ PDFs. We do not assign a theory uncertainty to our predictions of the SMEFT contributions, assuming that they will be subdominant with respect to other uncertainties such as the baseline SM theory predictions. We also neglect other theoretical uncertainties inherent to the SMEFT framework itself, such as omitting quadratic dimension-6 or linear dimension-8 contributions and other higher-order effects (see, e.g., Refs.~\cite{Englert:2014cva, Hays:2020scx, Horne:2020pot, Keilmann:2019cbp, Hays:2018zze, Baglio:2020oqu} for discussions of these and related uncertainties). We note that operator mixing from RGE running and loops can also induce extra constraints~\cite{Cirigliano:2016nyn, Aoude:2020dwv, Bissmann:2019gfc, Silvestrini:2018dos} and the effects of including SMEFT operators in parton distribution functions are also starting to be investigated~\cite{Carrazza:2019sec}.

\subsection{Higgs production}
\label{subsec:higgs_production}
We computed LO predictions in the full parameter space of our basis for the five main Higgs production modes: gluon-fusion (ggF), VBF, associated production with a $W$ or $Z$ ($WH,ZH$), and associated production with a top quark pair, $t\bar{t}H$. Some results are taken from Ref.~\cite{Ellis:2018gqa}, after being cross checked by independent computations.   Predictions in STXS bins are compared to and found to be in agreement with the predictions presented in Refs.~\cite{ATLAS:2020naq,ATLAS:2019dhi}.
The only one-loop calculations that we employ in our analysis involve the Higgs coupling to gluons, for which the LO contribution in the SM arises at one-loop level. Despite being a loop-induced coupling it mediates the $gg \to H$ Higgs boson production mode, which is dominant at hadron colliders. Since many Higgs measurements are very sensitive to this production mechanism and the associated, $gg\to H(+\text{jets})$ processes, we include as leading effects in the SMEFT both the tree-level contribution from $\Cp{HG}$ and the leading effects from the operators that modify the top-loop contribution to the SM coupling: $\Cp{tH}$,  $\Cp{tG}$, and $\Cp{H\Box}$. One final operator, $\Cp{G}$, modifies the gluon self-interaction allowing for a contribution to gluon fusion Higgs production in association with one or more jets, which we also include at one-loop order in this channel for the first time. The computations of linear contributions involve extracting the interference between loop diagrams of the SM with tree-level diagrams from $\Cp{HG}$ as well as loop diagrams with a single operator insertion, and are made possible by {\tt SMEFT@NLO}. We use a fixed renormalisation and factorisation scale of $\mh$ in all such computations. 

\begin{figure}[h!] 
\centering
\includegraphics[width=1.0\textwidth]{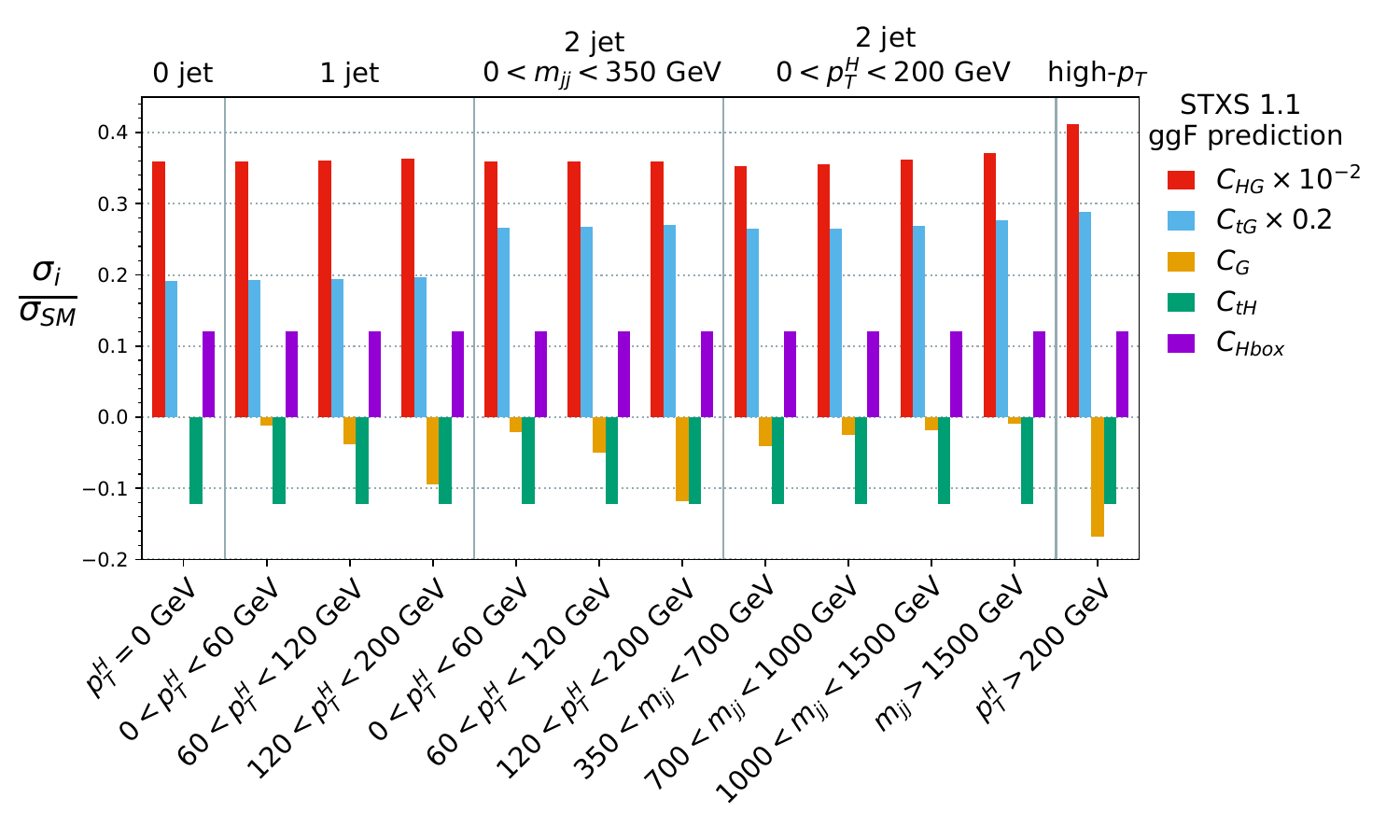}\caption{\it Illustration of the effects of selected SMEFT operators on representative gluon-fusion simplified template cross-sections $\sigma_i$ relative to the corresponding SM cross-sections $\sigma_{SM}$, for $C_i/\Lambda=1$ TeV$^{-2}$. The various Higgs $p_T$ and $m_{jj}$ bins provide complementary sensitivities and hence discriminating power between the operators.
\label{fig:STXS}}
\end{figure}

Figure~\ref{fig:STXS} illustrates the predictions we obtain for a
selection of stage 1.1 STXS gluon fusion bins~\cite{Berger:2019wnu},
highlighting the potential additional discriminating power offered by
the inclusion of $H$+jet(s). 
These predictions were obtained by parton-level generation of Higgs
production in association with one or two additional jets. Specific
predictions for, {e.g.}, the 0-jet gluon-fusion bin with Higgs 
$p_T$ > 10 GeV or the `3-jet like' $\geq 2$-jet bins with 
$p_T(Hjj)> 25$ GeV would require a matching/merging procedure 
interfaced with
parton showering that goes beyond the level of sophistication of our
analysis. Instead, we take the same dependence on
the coefficients as we find for the associated parton-level bin
($p_T^H=0$ and $p_T(Hjj)=0$). This corresponds to assuming that the
main effect on the population of the non-zero $p_T$ bins will come from
the parton shower, which does not depend on the EFT coefficients.
Comparing our results with merged sample analysis of Ref.~\cite{ATLAS:2020naq} (Tables 10-14), we find this assumption to be
excellent for the two $0$-jet bin, which have almost identical linear
coefficients. Furthermore, the relevant 2-jet bins are compatible with
the `parton shower only' assumption within about $10$\%. We note that
the comparison of individual coefficients between the two analyses is
not completely possible due to the different STXS binnings and EW input
schemes used; additional information on the MC generation would be
needed for a detailed cross-check. Nevertheless, we compare numbers for
each operator where reasonable, and find that our $C_{HG}$
contributions agree within 10--20\%, while $C_{tG}$ displays larger
differences on the order of 20--60\%. The other operators do not induce
kinematics-dependent effects, contributing overall rescalings for which
we also find good agreement with Ref.~\cite{ATLAS:2020naq}.

If one were to use only Higgs signal strength data, they would be limited to the sensitivity shown in the left-most `0-jet' entry, and
measurements of gluon fusion would exhibit a degeneracy in the four
relevant coefficients. Instead, allowing for associated production with
jets offers sensitivity to $\Cp{G}$ and, most importantly, breaks the
aforementioned degeneracy by exploiting the different energy
dependences of the five operators. Besides $\Op{H\Box}$ and $\Op{tH}$,
there are 3 additional operators that globally rescale gluon fusion
rates by shifting the top Yukawa interaction: $\Cp{HD}$,
$\Cpp{Hl}{(3)}$ and $\Cp{ll}$. We do not show their impact because
their contributions can simply be obtained from that of $\Op{H\Box}$ by
multiplicative factors of $-\tfrac{1}{4}, -1$ and $\tfrac{1}{2}$,
respectively. Furthermore, they are severely constrained by EWPO, to
the point where they are not expected to affect gluon fusion (or
$t\bar{t}H$).

Another relevant loop-induced process in Higgs production is $gg\to ZH$,  which, while being formally an NNLO QCD contribution to $ZH$, accounts for a significant portion of the inclusive cross section~\cite{Brein:2003wg}, and has a harder $p_T$ spectrum, which could be especially relevant for EFT interpretations~\cite{Mimasu:2015nqa}. Recent STXS definitions include dedicated bins for this contribution, although we are not aware of any explicit extraction of the cross sections. Notably,  this process was shown to be sensitive to top quark interactions with the $Z$, which are relatively poorly measured elsewhere~\cite{Bylund:2016phk,Englert:2016hvy}. However, we leave to future work
the inclusion of this loop-induced sensitivity to EW top operators, also present in other processes such as $gg\to VV$~\cite{Azatov:2016xik}, and EW Higgs production and decay~\cite{Vryonidou:2018eyv}.

\subsection{Gauge boson and Higgs decays}

Many of the measurements included in our fit involve the on-shell production and decay of SM gauge bosons and/or the Higgs boson. As mentioned above, we employ the NWA to factorise production and decay, such that, at the linear level in the SMEFT, the modification to the cross section of a given process is a combination of the $a^X_i$ for production, the partial width to the decay channel and the total width of the parent particle. The total width shifts of the $W$ and $Z$ bosons, have been determined as follows, where $\Lambda$ has been fixed to 1 TeV:
\begin{align}
\label{fig:WZwidth}
\begin{split}
        \frac{\Gamma_Z}{\Gamma_Z^{SM}} &= 1 - 0.05\,\Cp{HWB} - 0.041\,\Cp{HD} + 0.082\, \Cp{ll}-0.13 \,\Cpp{Hl}{(3)} -0.012 \,\Cpp{Hl}{(1)} -0.012 \,\Cp{He}\\
        & +0.077 \,\Cpp{Hq}{(3)} +0.0078 \,\Cpp{Hq}{(1)} +0.016 \,\Cp{Hu}-0.012 \,\Cp{Hd} +0.021 \,\Cpp{HQ}{(3)} +0.021\, \Cpp{HQ}{(1)} \, ,
\end{split}\\
\frac{\Gamma_W}{\Gamma_W^{SM}} &=1 -0.14\, \Cp{HWB} -0.065 \,\Cp{HD} +0.10\, \Cp{ll} -0.16  \,\Cpp{Hl}{(3)} +0.081  \,\Cpp{Hq}{(3)} \, .
\end{align}

Higgs boson decays present a richer structure, due to the importance of four-fermion decay modes that are mediated in the SM by the $H$ couplings to the $W$ and $Z$. Although the Higgs mass is too small for both gauge bosons to be on-shell, it is often assumed that at least one of the gauge bosons is, simplifying the decay to a three-body, $H\to V f^\prime\bar{f}$, final state with the NWA applied to on-shell vector boson, $V=W,Z$. This neglects certain interference effects between, {e.g.}, neutral-current and charged-current mediated four-fermion decays. The SMEFT introduces tree-level $H\gamma\gamma$, $H\gamma Z$ and $Hgg$ interactions, all of which can contribute to four-fermion decay modes, and degrade the accuracy of the NWA. In practice, however, experiments often make invariant mass cuts around $\mz$ in, {e.g.}, $h\to 4\ell$ analyses, that could largely mitigate this effect.

Ref.~\cite{Brivio:2019myy} performs an in-depth calculation and analysis of  Higgs decays to four fermions beyond the NWA, from which we take our predictions. We also include the contributions to $h\to g g$ mediated by operators that modify the top-quark loop contribution. These have the same relative impact as they do on $g g\to H$, discussed in the previous section~\footnote{However, for reasons of consistency,
we do not include the contributions to $h\to \gamma \gamma$ and $Z\gamma$ mediated by operators that modify the top-quark loop contribution,
which are formally of the same order as other NLO electroweak corrections that we do not include in general, and would be similarly relevant for, \emph{e.g.}, Z-pole data.}. The results of Ref.~\cite{Brivio:2019myy} are given for the flavour-universal $SU(3)^5$ scenario, which is more restrictive than our top-specific one, singling out operators affecting the left-handed $b$-quark couplings to the $Z$, $\Cpp{HQ}{(1)}$ and $\Cpp{HQ}{(3)}$.  In order to account for four-fermion decays involving $b$ quarks, it was necessary to adapt these results, complemented with some predictions from {\tt SMEFT@NLO}.  The contributions to the total Higgs width were found to be
\begin{align}
\begin{split}
     \frac{\Gamma_H}{\Gamma_H^{SM}} &=1 +0.18\, \Cp{HWB} +0.018\, \Cp{HD} +0.045 \,\Cp{ll} +0.11 \,\Cp{H\Box} +1.7 \,\Cp{HG} -0.075 \,\Cp{HW} \\
     &-0.093\, \Cp{HB} -0.089 \,\Cpp{Hl}{(3)} - 0.000059\, \Cp{Hl}^{(1)} - 0.000039 \,\Cp{He} -0.000051\,\Cp{Hd} \\
     &+0.0037\, \Cpp{Hq}{(3)} -0.00025 \Cp{Hq}^{(1)} +0.000055 \,\Cp{Hu} -0.00014 \,\Cpp{HQ}{(3)} +0.00038\, \Cpp{HQ}{(1)}\\ &-0.73\, \Cp{\tau H} -0.0057 \,\Cp{tH} -4.0 \,\Cp{bH} -0.043 \,\Cp{\mu H} +0.044 \,\Cp{tG} \, .
\end{split}
\end{align}
In any case, the only experimentally accessible four-fermion decay modes are the leptonic ones, so quark current operators are only practically relevant via their effect on the total width, which is clearly very small. For comparison, we computed the Higgs decays using the NWA for $h\to VV^\ast\to 4f$ processes, noting that the two approaches can have significantly different predictions for some operators. However, we do not find any appreciable differences in the results of the global fit, indicating that Higgs decays are not the primary source of constraints for the operators whose predictions are sensitive to whether or not this approximation is used.  Finally, we note that a recent EFT interpretation of Higgs production measurements in the $h\to 4\ell$ channel~\cite{Aad:2020mkp} pointed out significant acceptance differences in the kinematical selection between the SM and the EFT due to the additional off-shell photon contributions from $\Cp{HW}$, $\Cp{HB}$ and $\Cp{HWB}$. We do not take any acceptance corrections into account in our analysis, but also do not expect these to have a large impact on the global fit, given the fact that our results are unaffected by whether or not we use the NWA.

\subsection{Diboson data}
We have obtained predictions for the fiducial signal strengths in bins of leading-lepton $p_T$ for the ATLAS $WW$~\cite{Aaboud:2019nkz} analysis, the $Z$-boson $p_T$ for the CMS and ATLAS $WZ$ analyses~\cite{Sirunyan:2019bez,Aaboud:2019gxl}, and in bins of $\Delta \phi_{jj}$ for the ATLAS $Zjj$ analysis~\cite{Aad:2020sle}, following the general strategy outlined above. For the $WW$ and $WZ$ analyses, we find good agreement of the total SM fiducial cross-section, whereas in the  $Zjj$ cases we have validated our analysis by comparison with the binned signal strengths for the operator coefficients reported by ATLAS and CMS. In particular, as pointed out in Ref.~\cite{Aad:2020sle}, we find that this channel is the most sensitive to the interference term that is linear in the triple gauge boson operator $\mathcal{O}_W$.  
For LEP we used the $WW$ results of Ref.~\cite{Berthier:2016tkq} that are provided for the total and differential cross-sections at different centre of mass energies. We note that this analysis uses a restricted set of angular distribution bins to mitigate the effects of unknown correlations in those bins. 

\subsection{Top data}
Since we only make use of parton-level unfolded data, all top predictions are generated with stable top quarks. The data assume SM-like decay chains for the top quarks, and we do not take into account the small modifications to the $W$ boson branching fractions in top decays due to current operators. For top production in association with a gauge or Higgs boson, we do take into account the modified branching fraction of the associated boson in the measured decay channel (usually leptonic for $W$ and $Z$). However, this choice is not expected to have a significant impact on the results, given that top data is mainly sensitive to top-related operators, with those that modify gauge or Higgs boson decays being better constrained elsewhere. Much of the data we use overlaps with the data used in the top sector fit of Ref.~\cite{Hartland:2019bjb}, in which predictions are also obtained with {\tt SMEFT@NLO}, but including parton shower effects. We make use of the linear, LO parts of these in our work~\footnote{We thank the authors of Ref.~\cite{Hartland:2019bjb} for sharing the predictions with us.} and generate new, parton-level samples for these measurements, as including parton showering was not found to be very significant in determining the $a^X_i$. Our work also includes some top quark asymmetry measurements, namely those of the forward-backward and charge asymmetries from the Tevatron and LHC, respectively.
Splitting a measured cross section, $\sigma$, into two regions, $\sigma_F$ and $\sigma_B$,
the asymmetry, $A$, is defined as $A\equiv(\sigma_F-\sigma_B)/(\sigma_F+\sigma_B)$. The linearised contribution of a given operator is proportional to the difference between its relative contributions to the cross sections in the two regions, $a^{\sss \sigma_F}_i$ and  $a^{\sss \sigma_B}_i$,
\begin{align}
    a^i_{A} = \frac{1-A_{SM}^2}{2A_{SM}}(a_i^{\sss \sigma_F}-a_i^{\sss \sigma_B}) \, ,
\end{align}
highlighting how asymmetry measurements may be useful in breaking parameter space degeneracies in total cross section measurements that would, instead, be sensitive to the sum, $a_i^{\sss \sigma_F}+a_i^{\sss \sigma_B}$. We use the latest NNLO QCD + NLO EW theory predictions for the top asymmetries from Ref.~\cite{Czakon:2017lgo}. The impact of the asymmetries is quantified in Section~\ref{sec:top}.

\section{Fitting procedure}
\label{sec:procedure}

We perform a $\chi^2$ fit for a vector of observables, $\Vec{y}$, with covariance matrix~\footnote{In the case of a non-symmetric covariance matrix $\bf{\tilde{V}}$, we symmetrise it by defining ${\bf V}^{-1} = \frac{1}{2}\left({\bf\tilde{V}}^{-1} + ({\bf\tilde{V}}^{-1})^T\right)$. We have verified in a numerical fit to Higgs data that this does not affect significantly the results. }, ${\bf V}$, and theory predictions for those observables, $\Vec{\mu}(C_i)$, using a $\chi^2$ function defined as
\begin{equation}
\chi^2(C_i) = \left(\Vec{y} - \Vec{\mu}(C_i)\right)^T {\bf V}^{-1} \left(\Vec{y} - \Vec{\mu}(C_i)\right) \, .
\label{eq:chi2}
\end{equation}
The predictions are functions of the dimension-6 operator coefficients $C_i$, as defined in Eq.~(\ref{eq:Ldim6}), and are truncated at the linear level so as to include only the interference term with the SM. The quadratic dependence on $C_i$ is generically of the same order as linear interference terms with coefficients of dimension-8 operators, though exceptions exist in some specific UV completions~\cite{Liu:2016idz}. For example, in a UV completion with a single particle and a single coupling the quadratic dimension-6 contributions can be larger than the linear dimension-6 ones in the strong-coupling regime, though not more generally. The importance of their effect is therefore a model-dependent question. We note also that sensitivity of a linear fit  to quadratic contributions is an indicator for a possible breakdown in the regime of validity of the SMEFT, so care must be taken in the interpretation of the fit~\footnote{This issue is illustrated in the context of Higgs measurements in Appendix~\ref{app:nestedsampling}.}.

The least-squares estimators $\hat{C}_i$ that extremise the $\chi^2$ function can be obtained analytically in the case of a linear fit (see, e.g., Ref.~\cite{Zyla:2020zbs} for a review). We may write the linear theory prediction in terms of a matrix ${\bf H}$ that characterises the modification of the SM predictions $\Vec{\mu}^\text{SM}$ at linear order:
\begin{align}
    \mu_\alpha(C_i) = \mu^\text{SM}_\alpha + {\bf H}_{\alpha i} C_i  \, .
\end{align}
A summation over repeated indices is implied; the index $\alpha$ ranges over the number of observables and $i$ ranges over the number of dimension-6 coefficients. Solving $\partial\chi^2 / \partial C_i = 0$ gives the best fit values as
\begin{equation}
\hat{\Vec{C}} = \left({\bf H}^T {\bf V}^{-1} {\bf H}\right)^{-1} {\bf H}^T {\bf V}^{-1} (\Vec{y} - \vec{\mu}^\text{SM}) \equiv {\bf F}^{-1} \Vec{\omega}  \, .
\end{equation}
It is convenient to define the symmetric Hessian matrix ${\bf F}$, also known as the Fisher information matrix, and the $\chi^2$ gradient vector $\Vec{\omega}$ as
\begin{equation}
{\bf F} \equiv {\bf H}^T {\bf V}^{-1} {\bf H} \quad , \quad \Vec{\omega} \equiv {\bf H}^T {\bf V}^{-1} (\Vec{y} - \vec{\mu}^\text{SM}) \, ,
\end{equation}
in terms of which the $\chi^2$ function Eq.~(\ref{eq:chi2}) can be written as 
\begin{equation}
\chi^2(C_i) = \chi^2_\text{SM} - 2 \Vec{C}^T \Vec{\omega} + \Vec{C}^T {\bf F} \Vec{C}
            = \chi^2_\text{min} + \big(\Vec{C}-\hat{\Vec{C}}\big)^{T}\mathbf{F} \big(\Vec{C}-\hat{\Vec{C}}\big)\, ,
\end{equation}
where ${\bf F} \equiv{\bf U^{-1}}$ is the inverse of the covariance matrix of the least-squares estimators, $\hat{\Vec{C}}$.

Splitting the coefficients into $\Vec{C} = \{\Vec{C}_A, \Vec{C}_B\}$, we may profile over a subset of coefficients $\Vec{C}_A$ to obtain the least-squares estimators $\hat{\Vec{C}}_B$ for the remaining coefficients $\Vec{C}_B$. For this purpose, the Fisher information matrix may be decomposed into the sub-matrices
\begin{equation}
{\bf F} = \left( \begin{matrix} {\bf F}_A & {\bf F}_{AB} \\ {\bf F}_{AB}^T & {\bf F}_B \end{matrix} \right) \, ,
\end{equation}
and the gradient vectors as $\Vec{\omega} = \{ \Vec{\omega}_A, \Vec{\omega}_B \}$. The profiled best fit values are then given by
\begin{equation}
\hat{\Vec{C}}_B = \left({\bf F}_B - {\bf F}_{AB}^T {\bf F}_A^{-1} {\bf F}_{AB} \right)^{-1} \left( \Vec{\omega}_B - {\bf F}_{AB}^T {\bf F}_A^{-1} \Vec{\omega}_A \right) \, . 
\end{equation}
In cases where a prior on the coefficients needs to be imposed, for example when the magnitude-squared of couplings cannot go negative, as when matching to specific UV models in Section~\ref{sec:UV}, or when including quadratic dependences on the coefficients, this analytic method may no longer be used. A numerical MCMC method using {\tt MultiNest} has therefore also been implemented in {\tt Fitmaker}, as described in Appendix~\ref{app:nestedsampling}.

\section{Global results}
\label{sec:results}


\subsection{Higgs, Diboson and Electroweak fit}
\label{sec:EWH}

The main emphasis of this Section is on the improvements in the Higgs data since the
Run~1 and early Run~2 data that were analysed in the SMEFT framework in~\cite{Ellis:2018gqa}. However, as
has been emphasised previously, e.g., in~\cite{Ellis:2014dva, Ellis:2014jta},
there is considerable overlap between the sets
of operators whose coefficients are constrained by both electroweak and Higgs data~\footnote{We show later in Fig.~\ref{fig:margnoCG} the effect on the marginalised fit of removing the LEP EWPO and $WW$ datasets.}, as visualised in
Fig.~\ref{fig:venn}.
Therefore we present in this Section results from a joint fit to the 
combined Higgs, diboson and electroweak data, including the 20 operators of relevance listed in Eq.~\ref{eq:flavunivops}. We recall that the latter are dominated by data from LEP, with the most important LHC contribution coming from an ATLAS measurement of $M_W$.

As already mentioned, this joint fit is carried out to linear order in the dimension-6 SMEFT operator coefficients, neglecting
quadratic dimension-6 contributions to the LHC measurements and linear dimension-8 contributions,
which, as discussed above, are {\it a priori} of similar order in the scale $\Lambda$ of high-mass BSM physics.
Details of the fit procedure are described for the analytic method in Section~\ref{sec:procedure} and a numerical MCMC method in Appendix~\ref{app:nestedsampling}. We use mainly the former, but have verified in representative cases that fit results do not depend significantly on the method used. Appendix~\ref{app:nestedsampling} also discusses the importance of effects that are quadratic in the dimension-6
operators, and we refer
the interested reader to~\cite{Hays:2018zze} for a discussion of possible dimension-8 effects in
Higgs measurements. We emphasise again the importance of using, as well as total Higgs production and
decay rates, kinematic measurements of Higgs production as encapsulated in STXS measurements,
due to the different $p_T$ dependences of dimension-6 contributions to production amplitudes, whose relative 
importances are generally enhanced at higher $p_T$~\cite{Ellis:2018gqa}, as seen in Fig.~\ref{fig:STXS}.

\begin{figure}[h!] 
\centering
\vspace{-7mm}
\includegraphics[width=1.0\textwidth]{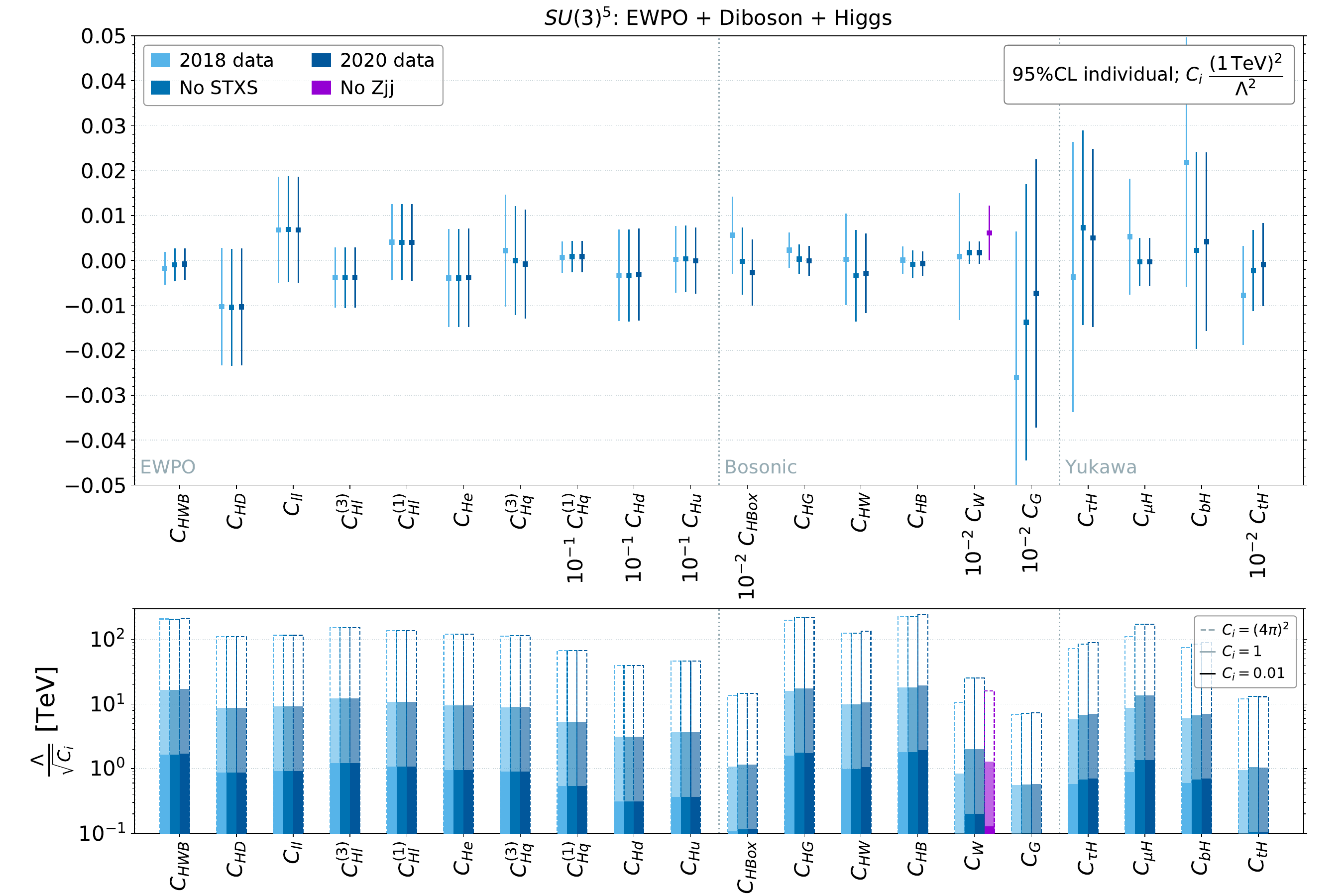}\\
%
\includegraphics[width=1.0\textwidth]{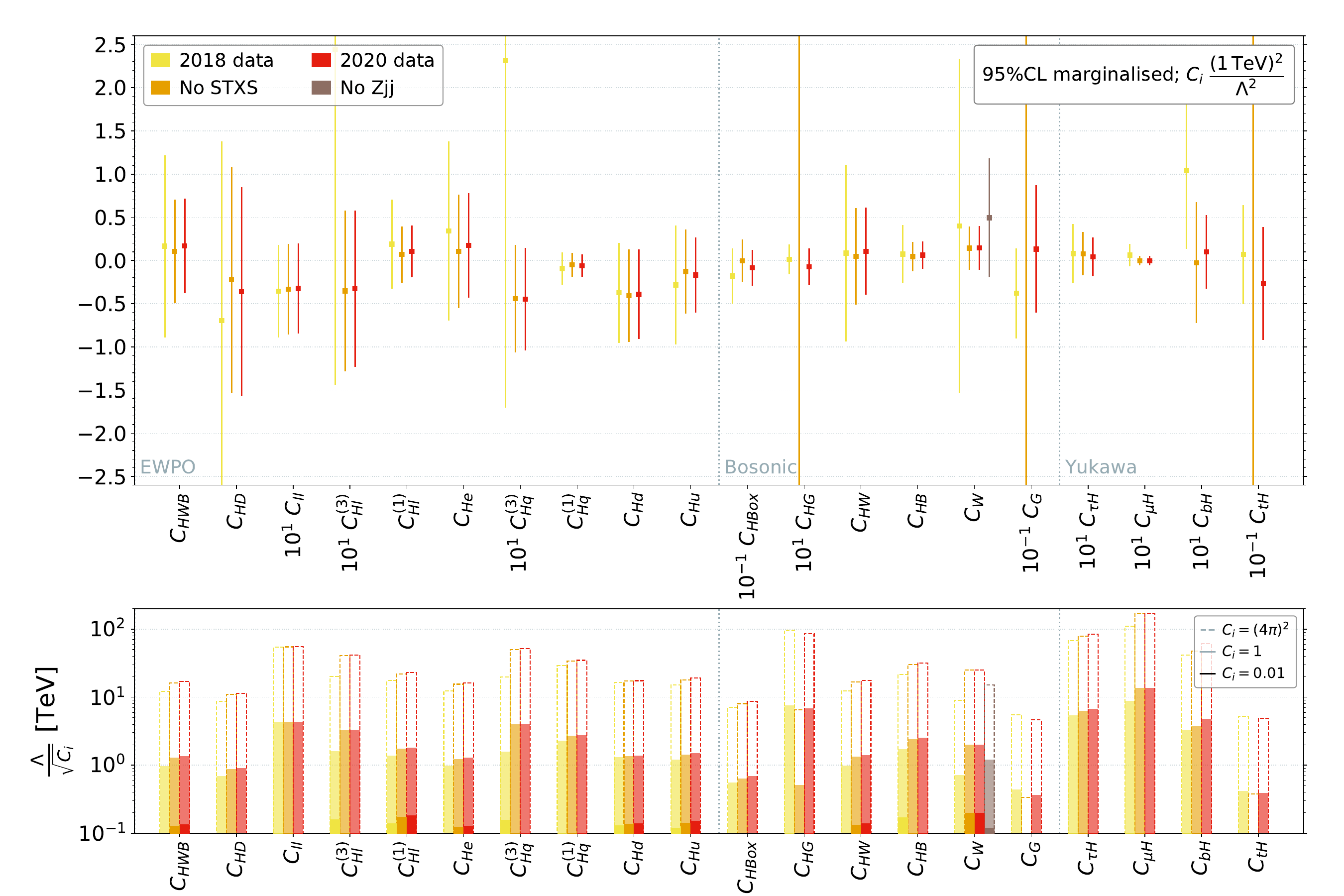}\\
\vspace{-5mm}
\caption{\it Constraints on the indicated individual and marginalised operator coefficients 
$C_i (1~{\rm TeV})^2/\Lambda^2$ (top and third panels) and the corresponding scales $\Lambda$ for the indicated values of the ${C_i}$ at the 95\% confidence level
(second and bottom panels), from a combined linear fit to the Higgs, diboson and electroweak precision observables.
In the top two panels, the bars show the 95\% CL
ranges from the LHC Run~1 and early Run~2 data (light blue), current data without using the STXS measurements (intermediate blue), and current data 
including STXS using either the on-shell vector boson approximation or the full $1\to4$ matrix elements for the 4-fermion Higgs decay modes taken from Ref.~\cite{Brivio:2019myy} (dark blue). In the bottom two panels, the corresponding marginalised results are indicated by yellow, orange and red bars, respectively. We also show in purple in the individual case (grey in the marginalised case) the effect of dropping the ATLAS $Z +$ jets measurement.}
\label{fig:flavour_universal_EWdiBH}
\end{figure}

The top panel in Fig.~\ref{fig:flavour_universal_EWdiBH} 
shows the 95\% CL intervals for the coefficients 
$C_i (1~{\rm TeV})^2/\Lambda^2$
of the 20 dimension-6 SMEFT operators contributing to the joint Higgs, diboson and precision electroweak data analysis when each operator is analysed individually.
We note that certain coefficients have been scaled as indicated by the labels on the x-axis. In particular, the Yukawa operator $C_{tH}$, the triple-gauge boson operators $C_W$ and $C_G$, and the Higgs-only operator $C_{H\Box}$ have all been rescaled by 2 orders of magnitude to appear on the same scale. Dotted grey vertical lines separate the sets of operators that contribute mostly to electroweak precision observables (EWPO), that contribute mostly to Higgs and diboson measurements (Bosonic), and that modify the Yukawa couplings (Yukawa). These categories are indicated as guides to aid the reader; however, as discussed in more detail later, there are correlations between these sectors. 

The two right-most bars for each operator are from fits
using all the available Higgs and diboson data from Run~2 of the LHC together with the precision electroweak
constraints, differing only in whether they include the STXS constraints. The intermediate blue bars demonstrate the effect of replacing the combined ATLAS STXS dataset with the latest combined ATLAS signal strengths, in which case there are slightly weaker constraints for some of the operator coefficients. These results do not differ significantly whether predictions in the on-shell approximation are used, where the $H \to 4f$ process is assumed to originate from an underlying $H \to VV^*\to4f$ decay with one gauge boson taken to be on-shell, or whether the SMEFT dependence of the full $H\to 4f$ matrix elements is used (see~\cite{Brivio:2019myy} and Section~\ref{sec:procedure}). For comparison,
the left-most bar (light blue) is from a fit in which only the Higgs and diboson data analysed in~\cite{Ellis:2018gqa} are used. In the case of $C_W$, we also show (in purple) the effect of dropping the ATLAS $Zjj$ measurement~\cite{Aad:2020sle}, which does not impact significantly the constraints on the other operator coefficients. 

In general, the individual fit ranges for the first 10 operator coefficients of Fig.~\ref{fig:flavour_universal_EWdiBH} counting from the left, 
i.e., up to and including $C_{Hu}$, are very similar in fits including all the LHC Run~2 data 
to those found using the earlier set of Higgs, diboson and
electroweak data. This is because the precision electroweak
data provide the strongest constraints on these individual operator coefficients.
The impacts of the Higgs and diboson data are more apparent for the rest of the operator coefficients,
i.e., from $C_{H\Box}$ rightwards, particularly for the Yukawa operators that have benefited from improved sensitivity of the Higgs couplings to the tau and bottom. The relative constraining power of datasets, as measured by the Fisher Information, is given in Table~\ref{tab:individual} in the Appendix, and confirms the points discussed above. It also quantifies the importance of the $Zjj$ data (84\%) in pinning down $C_W$, compared to $W^\pm Z$ (13\%) and LEP 2 $W^+W^-$ (3\%)

Using the same colours, the bars in the second panel show the 95\% CL lower limits on the $\Lambda_i$ on logarithmic scales in units of TeV, for different values of ${C_i}$. These reaches are estimated by taking half the width of the 95\% CL ranges of $C_i/\Lambda_i^2$ as the typical sensitivity of the measurement. Here and in subsequent analogous panels, the darker (lighter) coloured
histograms are for $C_i = 0.01 (1)$ and the histograms with dashed
outlines are for the strong-coupling perturbativity limit $C_i = (4 \pi)^2$.
In general, the $\Lambda_i$ scales would be modified by a factor $\sqrt{C_i}$, which would depend on whether the Wilson coefficient is induced by strongly- or weakly-coupled new physics, at tree- or loop-level.

The corresponding 95\% CL constraints for the marginalised case, where we include simultaneously all operators in the analysis and then profile the likelihood over all coefficients except one, as described in Section~\ref{sec:procedure}, are shown in
the lower pair of panels in Fig.~\ref{fig:flavour_universal_EWdiBH}. The yellow, orange and red bars from left to right are the fits to the old data, the new data without the STXS measurements, and including them, respectively. There is again no significant difference for different treatments of $h \to 4f$. The marginalised constraints are weaker overall than the individual constraints, as the fit is allowed to explore all possible variations in the space of coefficients. We also note that the STXS measurements play a key role in the marginalised constraints for some operators, e.g., $C_{HG}, C_G$ and $C_{tH}$. As discussed in Section~\ref{subsec:higgs_production}, removing them causes a degeneracy in the parameter space that prevents meaningful constraints in these directions. We show in dark brown the impact on $C_W$ of dropping the $Zjj$ constraint, which still does not significantly affect the other operators. The more traditional diboson measurements suffer from suppressed SM interference at high energy due to helicity selection~\cite{Azatov:2016sqh}. This is particularly so for $W^+W^-$, while $W^\pm Z$ appears to retain some sensitivity. This is why the bound without $Zjj$ changes significantly, becoming dominated by the $W^\pm Z$ and/or LEP 2 data, with the high mass $W^+W^-$ distributions from the LHC yielding no significant improvement. The $Zjj$ observable is therefore extremely useful for constraining anomalous gauge boson self-interactions at linear level, overcoming the non-interference issue and accessing the leading contributions in the SMEFT expansion. 

We see that most of the 95\% CL ranges are reduced when the full Run~2
data are included, some quite substantially, the only exceptions being $\Cp{G}$ and $\Cp{HG}$. This occurs despite the individual constraints improving in both parameter space directions. We attribute the slight worsening of the marginalised bounds to the presence of a highly boosted $H\to b\bar{b}$ measurement in the 2018 data~\cite{Sirunyan:2017dgc} that selects Higgs $p_T$ > 450 GeV, which is significantly higher than the highest, $p_T$ > 200 GeV of the stage 1.0 STXS bins used in our 2020 dataset. Removing this observable degrades the 2018 bounds below our most recent results. We expect this sensitivity to be recovered once the stage 1.2 measurements, which probe a similarly high $p_T$ region, are incorporated. {We find $\chi^2/{\rm dof} = 0.94$ ($p = 0.72$) for our flavour-universal global fit, to be compared with $\chi^2/{\rm dof} = 0.93$ ($p = 0.76$)
for the SM.}
Among the 20 operators considered in the fit to
the Higgs, diboson and electroweak data, the most weakly
constrained operator
coefficients are all constrained so that $\Lambda/\sqrt{C_i} \gtrsim 500 \, (400)$~GeV
in the individual (marginalised) fits, suggesting that the linear SMEFT
treatment may be adequate in this sector~\cite{Contino:2016jqw}.

\FloatBarrier

\subsection{Top fit}
\label{sec:top}

\begin{figure}[t] 
\centering
\includegraphics[width=\textwidth]{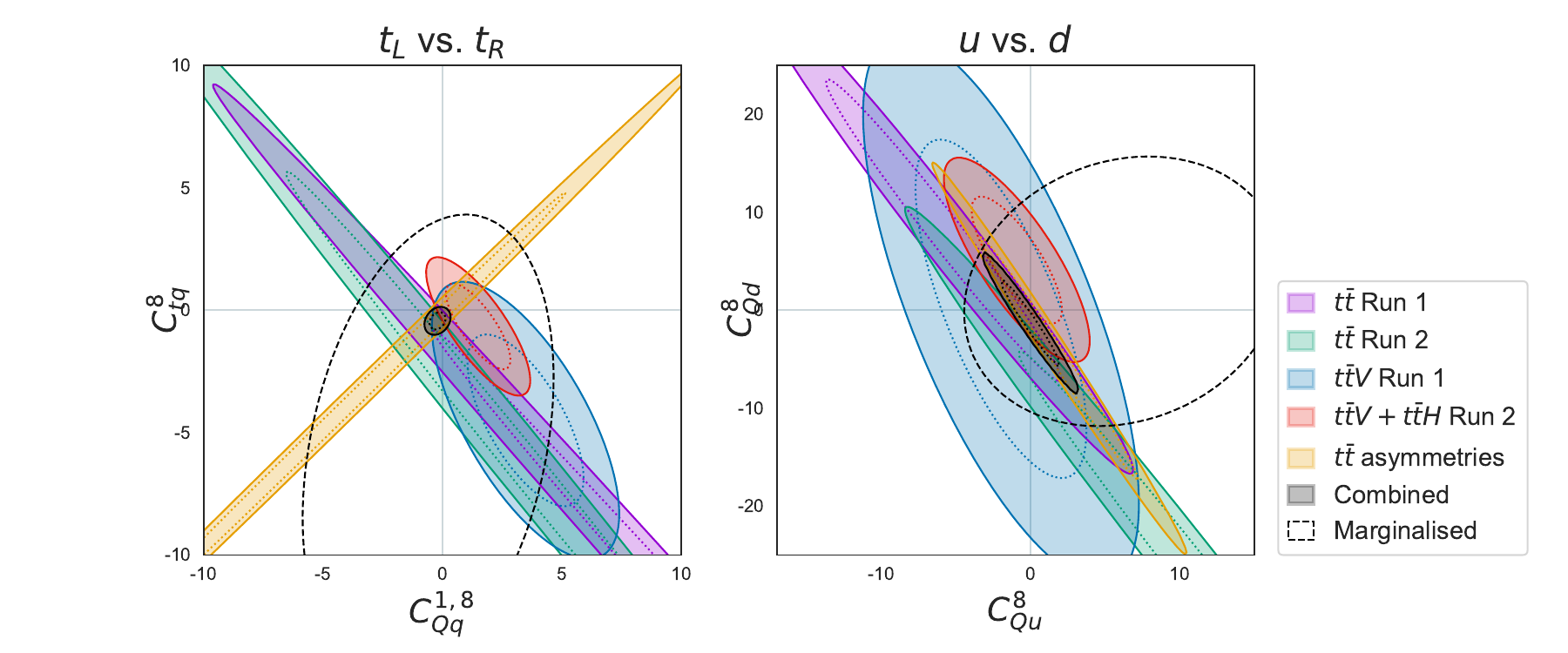}
\caption{\it \label{fig:top_breakdown_mtt} 
\it Left panel: Breakdown of the impacts of various top datasets on the 2-dimensional subspace of the four-fermion operators ($C_{Qq}^{1,8}, C_{tq}^{8}$), setting all other operator coefficients to zero. Right panel: Similar breakdown of the impacts of various top datasets on the 2-dimensional subspace of the four-fermion operators ($C_{Qu}^{8}, C_{Qd}^{8}$), setting all other operator coefficients to zero.
}
\end{figure}

\begin{figure}[t!] 
\centering
\includegraphics[width=1.0\textwidth]{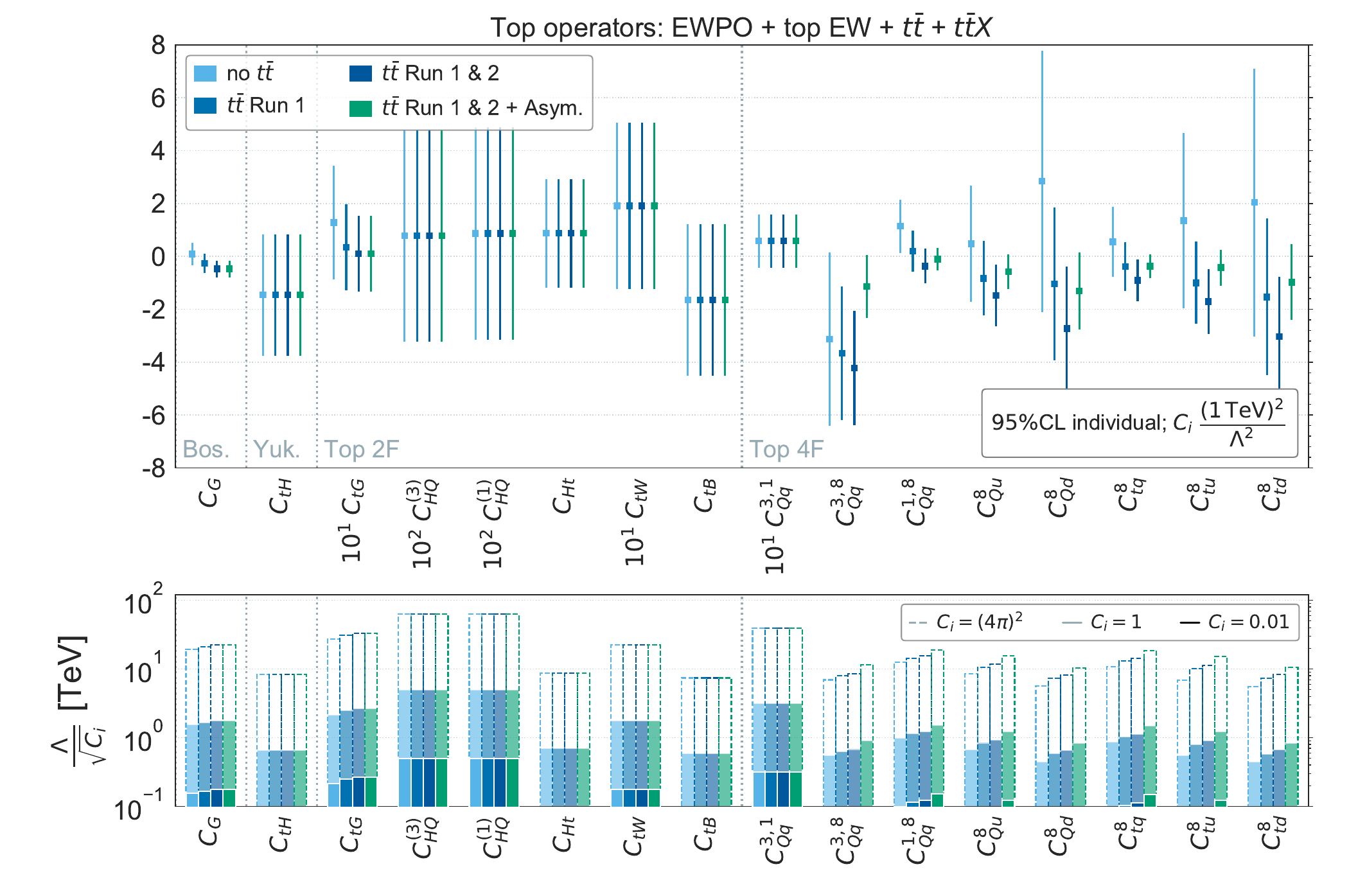}
\includegraphics[width=1.0\textwidth]{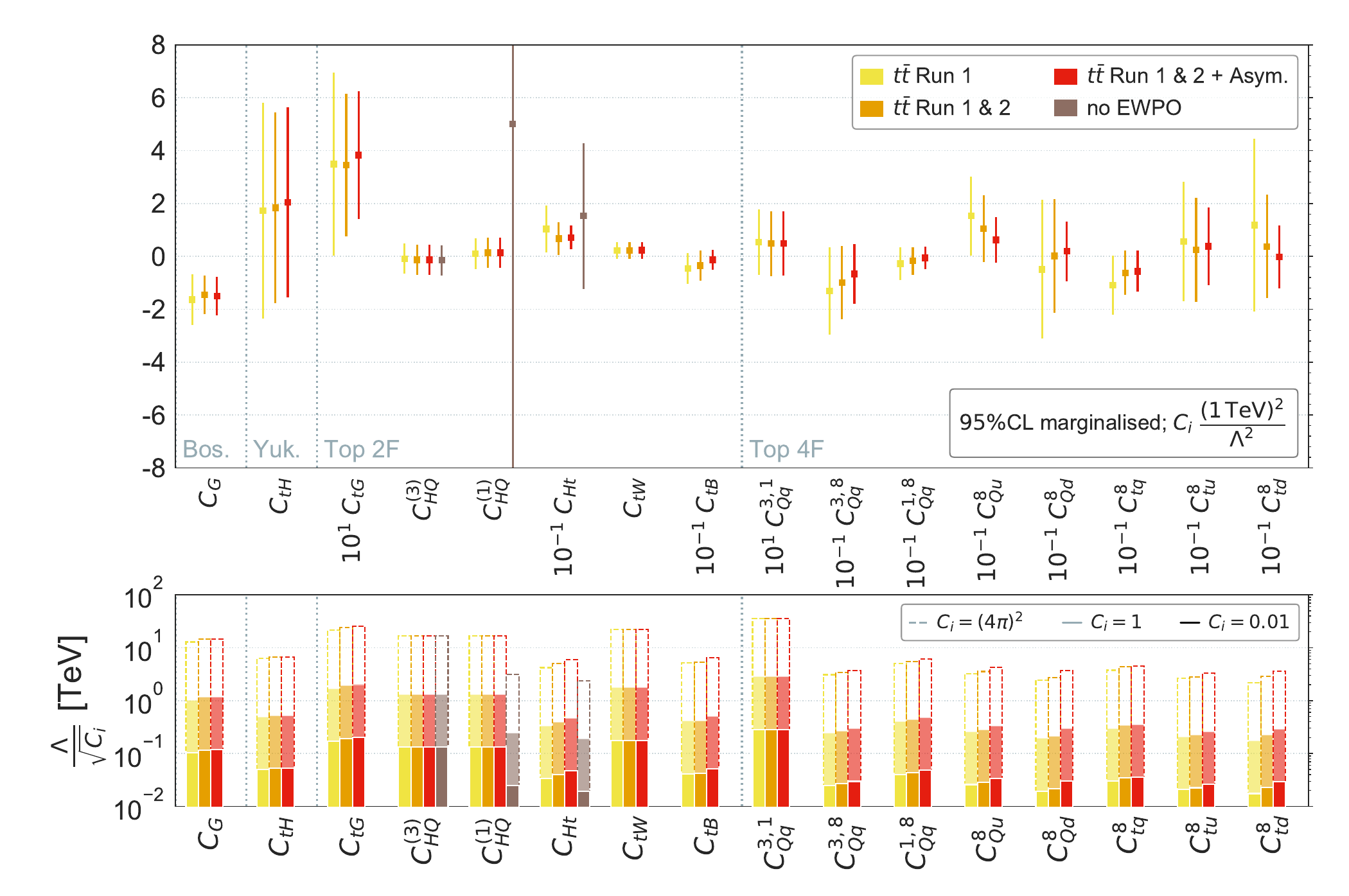}
\caption{\it \label{fig:top_asym_impact} Constraints on the indicated individual and marginalised operator coefficients at the 95\% confidence level (upper and lower figures, respectively), from
a combined linear fit to the top data and electroweak precision observables. The impact of $t\bar{t}$ data is highlighted by the evolution of the constraints starting from no $t\bar{t}$ data (light blue/yellow) adding Run 1 $t\bar{t}$ total and differential cross-section data (blue/pink), the corresponding Run 2 $t\bar{t}$ data (purple/orange), and finally $t\bar{t}$ asymmetry measurements $A_{FB}$ from the Tevatron and $A_C$ from the LHC (green/red).}
\end{figure}

We consider now the information provided by top quark data, in a dedicated fit to a subset of relevant `top quark operators', {i.e.}, those involving at least one top quark field (the top Yukawa operator, and the top 2F and top 4F categories in Equation~\ref{eq:topspecificops}) plus the triple gluon operator, $\Cp{G}$, which globally affects most QCD-induced processes. We also include EW precision observables, which close a single blind direction ($\Cpp{HQ}{(1)}+\Cpp{HQ}{(3)}$) affecting the left-handed $Zb\bar{b}$ coupling, as well as the latest $t\bar{t}H$ signal strength measurements from the two CMS and ATLAS Higgs combination papers included in the final fit, to constrain the top Yukawa operator.
Overall, the $\chi^2/n_{\text{dof}}$ for the SM is 0.92. However, there are a number of observables that exhibit some tension with the SM predictions, such as the recent 13 TeV measurement of the differential $p_T$ distribution in $t$-channel single top production~\cite{Sirunyan:2019hqb} ($\chi^2/n_{\text{d.o.f}}=5.3$), $t\bar{t}W$ cross section measurements by CMS at 8 TeV~\cite{Khachatryan:2015sha} and ATLAS at 8 and 13 TeV~\cite{Aad:2015eua,Aaboud:2019njj} ($\chi^2/n_{\text{d.o.f}}=2.6,2.1$ and 1.9), and CMS $t\bar{t}$ differential distributions at 8~TeV ($\tfrac{d\sigma}{dm_{t\bar{t}}y_{t\bar{t}}}$ in the l+jets channel) and 13~TeV ($\tfrac{d\sigma}{dm_{t\bar{t}}}$ in the dilepton channel), both with $\chi^2/n_{\text{d.o.f}}=1.6$. These tensions may lead to a preference for non-zero Wilson coefficients, 
though this depends on whether other, more consistent observables also constrain the operators in question.

Fig.~\ref{fig:top_asym_impact} shows the 95\% CL intervals for the coefficients individually (top two panels) and after marginalisation (bottom two panels). The pale blue bars in the top two panels are obtained without using any $t \bar t$ data, the intermediate blue bars are obtained including the Run~1 $t \bar t$ total and differential cross-section data, the dark blue bars include also the corresponding Run~2 $t \bar t$ data, and finally the green bars include also the $t \bar t$ asymmetry measurements $A_{FB}$ from the Tevatron and $A_C$ from the LHC. When removing the $t\bar{t}$ data entirely, a closed fit is not possible, so marginalised constraints do not exist. The corresponding colours for the latter three sets of data (Run 1 $t\bar{t}$, including Run 2, and including $A_{FB}$ asymmetries) in the bottom two panels are yellow, orange and red. We also show in dark brown the impact of removing electroweak precision observables for the two operators most affected, $\Cpp{HQ}{(1)}$ and $\Cp{Ht}$. As previously, the constraints on the scales $\Lambda/\sqrt{C_i}$ are estimated by taking half the width of the 95\% CL ranges of $C_i/\Lambda_i^2$ for $C=1$ as the typical sensitivity of the measurement. We note that many of the constraints on the quantities $C_i (1 {\rm TeV})^2/\Lambda^2$ in the individual and marginalised cases differ by an order of magnitude, as shown in the top and third panels of 
Fig.~\ref{fig:top_asym_impact}, 
and that the corresponding constraints on the scales $\Lambda/\sqrt{C_i}$ are typically a factor $\sim 3$ stronger in the individual analysis, as seen in
the second and bottom panels. 

The impact of the tension with the SM for the aforementioned observables can be seen from the individual constraints of Fig.~\ref{fig:top_asym_impact}. We see that $\Cp{tH}$ and the `top 2F' category appear consistent with the SM.  Although it disagrees the most with the SM, the single top differential $p_T$ data does not lead to significant pulls for the operators that can affect it, $\Cpp{Qq}{3,1}$, $\Cpp{HQ}{3}$ and $\Cp{tW}$. This is due to the fact that single operators are not able to improve the fit significantly, given that the disagreement predominantly comes from the lowest $p_T$ bin, combined with the relatively good agreement of most other single-top data. Instead, the effect of the $t\bar{t}W$ data can be seen in the individual constraints without $t\bar{t}$ data, that cause the deviation of the best-fit values for $\Cpp{Qq}{1,8}$ and $\Cpp{Qq}{3,8}$ away from zero in the positive and negative directions, respectively. This is consistent with a relative minus sign in the linear dependence of the $t\bar{t}W$ process on these two operators. On the other hand, $t\bar{t}$ data depend on 9 of the operators in question, namely all of the `top 4F' category except $\Cpp{Qq}{3,1}$, as well as $\Cp{tG}$ and $\Cp{G}$. Gradually adding the differential $t\bar{t}$ data draws the coefficients towards negative values, resulting in particularly large pulls, especially for $\Cp{G}$. Finally, the $t\bar{t}$ asymmetry observables add an orthogonal constraint on the four-fermion operators that restore consistency with the SM in this sector. Since $\Cp{G}$ does not produce an angular asymmetry in the $t\bar{t}$ matrix element, its significant, non-zero best-fit value remains. Considering now the marginalised results, we see that more $t\bar{t}$ data and the inclusion of asymmetries has a significant impact on the global sensitivity, even indirectly affecting sensitivity to EW top quark couplings $\Cp{Ht}$ and $\Cp{tB}$ by constraining the allowed four-fermion contributions to $t\bar{t}Z/\gamma$. We also show the importance of EW precision observables in closing the parameter space for neutral top quark couplings, by noting the large significant degradation of the limits in this space when removing them from the fit. All of the potentially large pulls are washed out by the marginalisation, except for $\Cp{G}$ and $\Cp{tG}$, which lie 4.1 and 3.2 standard deviations away from zero, respectively.

The interplay between different $t \bar t$ measurements in constraining
the four fermion sector is shown in Fig.~\ref{fig:top_breakdown_mtt}, inspired by a similar analysis in Ref.~\cite{Brivio:2019ius}. Two pairs of operators are selected, setting all other coefficients to zero: $C_{Qq}^{1,8}$ and $C_{tq}^{8}$, which couple to left- and right-handed top quarks, respectively, in the left panel, and $C_{Qu}^{1,8}$ and $C_{Qd}^{8}$, which couple left handed top quarks to the up and down quarks, respectively, in the right panel. We see explicitly the
complementarity between the $t \bar t$ cross-section measurements,
which strongly constrain one linear combination of $C_{Qq}^{1,8}$ and $C_{tq}^{8}$, and the asymmetry measurements, which constrain an orthogonal direction. The combination of these measurements constrains each of
$C_{Qq}^{1,8}$ and $C_{tq}^{8}$ quite tightly, though less so in
the marginalised case, indicated by the dashed lines. The fact that forward-backward asymmetries are sensitive to the chiral structure of the $t\bar{t}$ matrix element explains why they excel at distinguishing modified interactions for left- and right-handed tops but less so the isospin of the initial state quark ($u$ or $d$). 
In this case the combination of $t \bar t$ measurements constrains a highly correlated combination of the coefficients, though each is only weakly constrained in the marginalised fit, as seen in the bottom two panels of Fig.~\ref{fig:top_asym_impact}. 
The large differences between their individual and marginalised limits that we observe indicate strong correlations among the top quark four fermion operators constraints and that the overall marginalised sensitivity is set by the less precise $t\bar{t}X$ data.

Overall, the best-constrained coefficients in Fig.~\ref{fig:top_asym_impact}
are $C_{tG}$, $C_{HQ}^{(3)}$, $C_{HQ}^{(1)}$, $C_{Qq}^{3,1}$ and $C_{G}$, with one direction being driven by electroweak precision observables. The large
negative values of $C_G$ can be traced back 
to $t \bar t$ differential cross-section measurements, and we discuss a specific example in Section~\ref{sec:CG} below, where we also consider the possibility that
$\Cp{G}$ is very small, as suggested on the basis of a quadratic analysis
of multijet data. Four-fermion operator coefficients are less well constrained, with scales $\Lambda$ between 
$800-1500$~GeV in the individual analysis and $300-500$~GeV 
in the marginalised analysis when the corresponding $C_i = 1$,
in which case the validity of the global SMEFT interpretation for these operators could be
questioned for weakly-coupled UV completions, given that some of the $t\bar{t}$ data extends up to TeV
energies. We therefore expect the differences between our top data analysis and those performed at quadratic level to be especially significant for the top quark four-fermion operators, as shown in Ref.~\cite{Hartland:2019bjb}. 

The neutral top quark operator coefficients $\Cp{Ht}, \Cp{tB}$ are also particularly hard to constrain. Production of $t\bar{t}Z/\gamma$ and $tZ$ are the main handles we have on these couplings, and these are still not so well measured and only beginning to produce differential data. The $t\bar{t}\gamma$ differential distributions in photon $p_T$ turn out to provide the best handle on $\Cp{tB}$.
Unfortunately, $\Cp{Ht}$ does not predict any effects that grow in energy in either of these processes,
and $\Cp{tB}$ has a suppressed interference with the SM, meaning that one does not expect spectacular
gains from differential measurements, especially in a linear
analysis~\cite{Bylund:2016phk,Degrande:2018fog}.  Other rare EW top processes, such as $t\bar{t}Wj$ and
$tWZ$ have been shown to be sensitive to such unitarity-violating behaviour and will therefore provide
useful constraining power, once they are measured at the LHC ~\cite{Dror:2015nkp,Maltoni:2019aot}.
However, in all cases we see that the dashed histograms extend beyond a TeV even in the marginalised case, indicating that the SMEFT analysis should be a good approximation in the strong-coupling limit $C_i = (4 \pi)^2$.

\begin{figure}[t] 
\vspace{-5mm}
\centering
\includegraphics[width=1.0\textwidth]{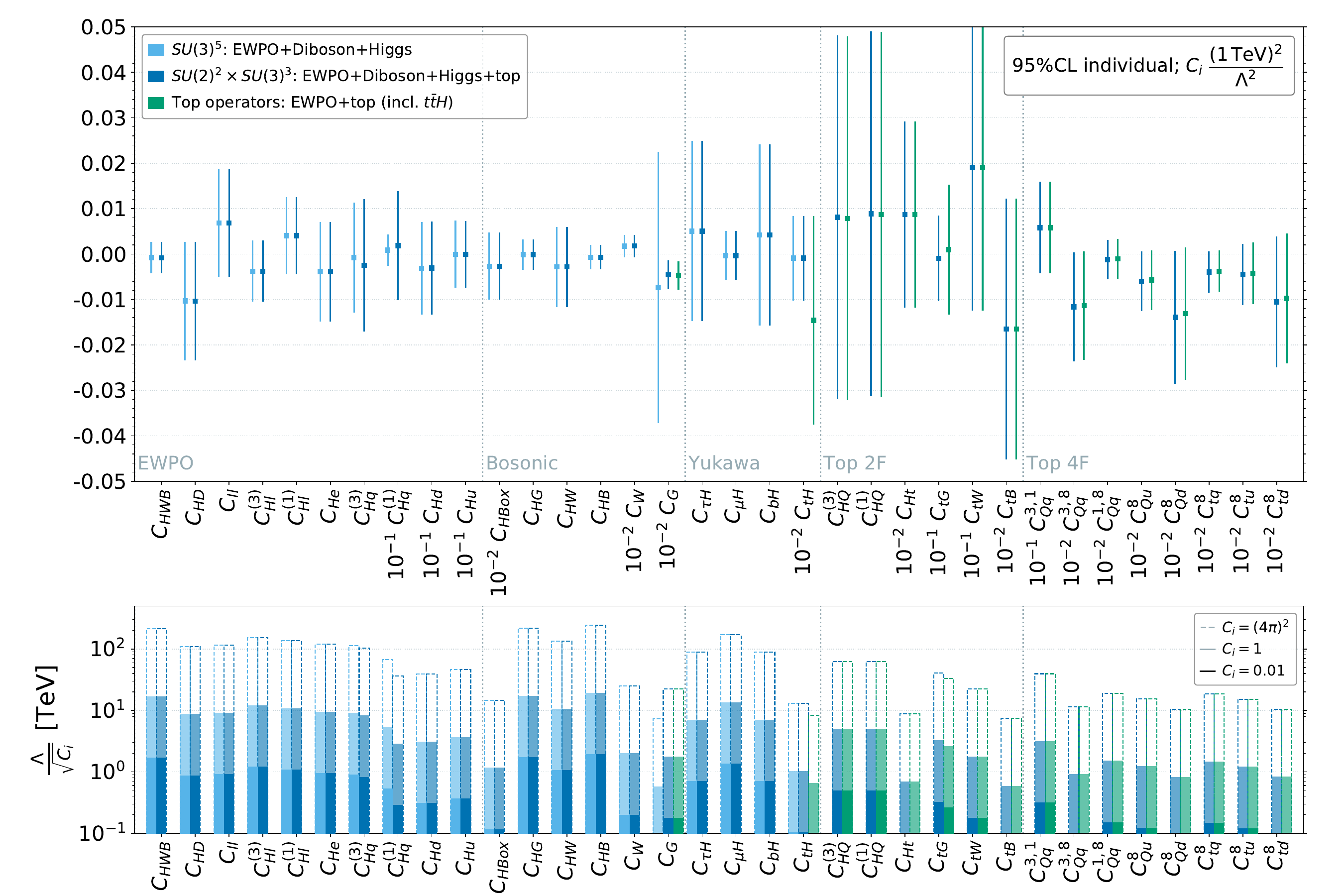}\\
\includegraphics[width=1.0\textwidth]{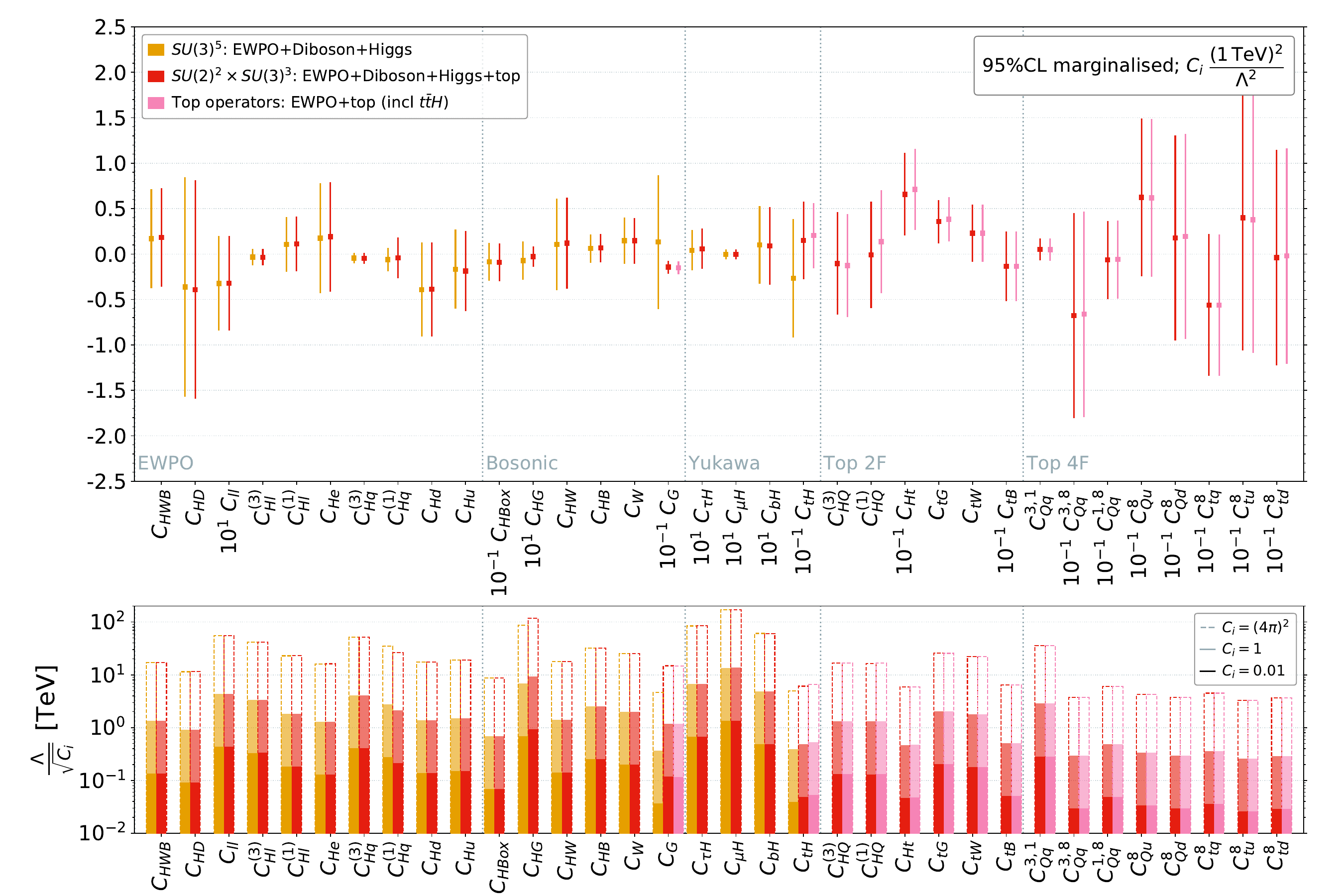}
\vspace{-7mm}
\caption{\it \label{fig:all_fit} Results from the global fit
to the electroweak, diboson, Higgs and top data in the top-specific
SU(2)$^2\times$SU(3)$^3$ scenario. Top two panels: fits to the
individual operators, showing the 95\% CL ranges for the operator
coefficients $C_i$ normalising the new physics scale $\Lambda$ to
1~TeV, and the ranges for the scales $\Lambda$ for different values of the $C_i$.
Bottom two panels: similar, but marginalising over the other
operators. In the top panels, fit results found using only electroweak, diboson
and Higgs data are shown in light blue, those found in the combination with the
top data are shown in dark blue, and those using only top data are shown in green.
In the bottom panels, the corresponding marginalised fit results are shown in yellow, red and pink.}
\end{figure}

\FloatBarrier

\subsection{Combined Top, Higgs, Diboson and Electroweak Fit}

As we discussed in Section~\ref{sec:SMEFT}, there are different possible treatments of the flavour
degrees of freedom within the SMEFT framework. Specifically, in this paper we
assume either an SU(3)$^5$
symmetry in the operator coefficients (broken by the Yukawa operators) or allow this symmetry to be broken to an SU(2)$^2 \times$SU(3)$^3$ symmetry by the
coefficients of third-generation fermions in a
top-specific flavour scenario. 

We find that, among the 20 flavour-universal 
operators in the SU(3)$^5$ scenario,
only the marginalised constraints on $C_G, C_{HG}$ and $C_{tH}$
are improved significantly by the additional 
top measurements beyond the constraints
that are already provided by the
electroweak, diboson and Higgs data, while the sensitivity to $C_{Hq}^{(1)}$ and $C_{Hq}^{(3)}$ decreases, since their third-generation flavour components have been separated into separately constrained degrees of freedom. This indicates the robustness of the fit despite the increase in the number of parameters. However, there are significant correlations between the datasets. 
Accordingly,
in the following we focus attention on the results 
from a simultaneous global fit to the 34 operators 
of the top-specific flavour scenario that was described in Section~\ref{sec:topspecificflavourscenario}. 

\subsubsection{Operator sensitivities in individual and marginalised fits}

Figs.~\ref{fig:all_fit} shows the results from this combined fit to all the available Higgs, electroweak, diboson and top data, switching on one operator at a time (top two panels) and marginalising over all other coefficients (bottom two panels), respectively. In
each case, the upper panel shows the 95\% CL ranges for the operator coefficients $C_i$ normalising the corresponding new physics scales to 1~TeV. As indicated on the x-axis labels, certain operator coefficients have been rescaled for the sake of convenience. The bars in the lower panels show the 95\% CL lower limits on the $\Lambda_i$ on logarithmic scales in units of TeV. As previously, these reaches are estimated by taking half the width of the 95\% CL ranges of $C_i/\Lambda_i^2$ for $C=1$ as the typical sensitivity of the measurement. 

The differences in the constraints on the 20 operators entering the flavour-independent SU(3)$^5$ fit between including top data or not are small in the individual case (dark vs light blue), except for $C_G$. For a more detailed breakdown of the relative constraining power of different datasets on each individual coefficient, we refer the reader to Table~\ref{tab:individual}. Marginalising widens the ranges allowed by the fit, but the effect of marginalising over a larger set of coefficients - 34 compared to 20 in the top-specific flavour scenario compared to the flavour-universal case (red vs yellow), introduces noticeable differences for only a few operator coefficients, namely $C_G, C_{HG}$ and $C_{tH}$. The differences in the constraints on the top operators (shown in green in the individual case and pink in the marginalised case) when the electroweak, diboson and Higgs data are included in the fit are generally small, apart in the case of the Yukawa operator, $\Cp{tH}$ and the top chromomagnetic dipole operator, $\Cp{tG}$. The loop-level constraints from Higgs production via gluon fusion are clearly very powerful in  constraining these operator individually, but this sensitivity is diluted by marginalisation, which allows the other operators affecting this process to float.

Overall, the data are sufficient for a closed fit with no flat directions, {and we find $\chi^2/{\rm dof} = 0.81$ ($p = 0.99$) for our top-specific global fit}. The sensitivities to the scale of new physics in the operator coefficients in the individual case are generally several hundred GeV or more for $C_i = 1$. We also note that the $C_i$ for the Yukawa operators would be expected to contain a Yukawa factor in the MFV hypothesis, as discussed in Sec.~\ref{sec:SMEFT}, with the bounds then weakened appropriately. The scale sensitivities in the
marginalised case still reach a TeV for $C_i = 1$ for most of the electroweak precision observables set and some of the operators in the bosonic and top categories, falling to $\sim 300$~GeV for some of the other top operators.  We emphasise, however, that specific UV completions each generate only a subset of operators, for which they have correspondingly improved reaches. The individual and marginalised fits can therefore be taken as optimistic and pessimistic sensitivity estimates, respectively, with realistic cases living somewhere between the two.

\subsubsection{Sensitivities in `Higgs-only' operator planes}
\label{sec:Higgsonly}

\begin{figure}[t] 
\centering
\includegraphics[width=0.8\textwidth]{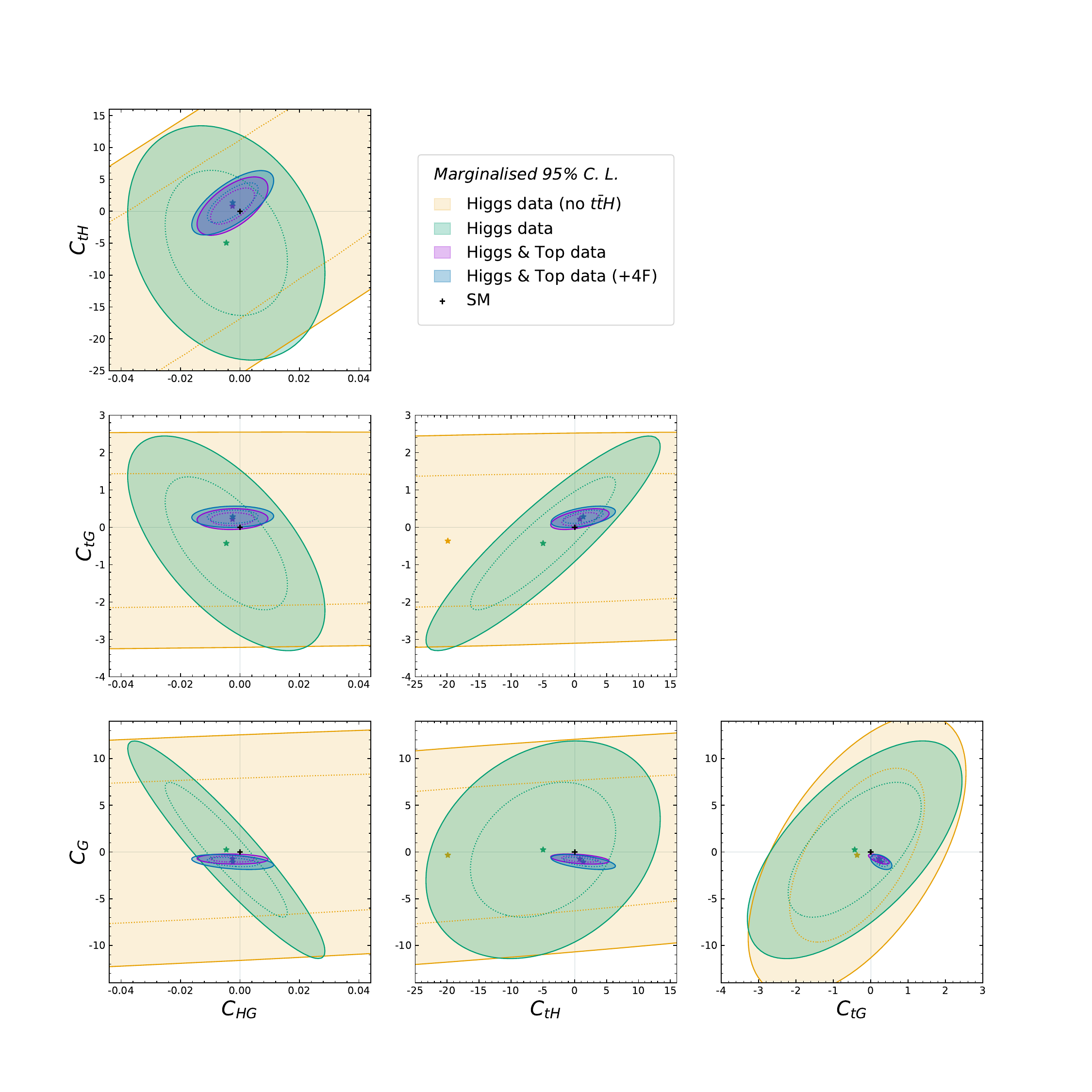}
\caption{\it \label{fig:interplay1} Constraints on the indicated pairs of operator coefficients 
at the 95\% confidence level, marginalised over the remaining degrees of freedom in the `Higgs only' operator set. The green and mauve shaded areas correspond to combined linear fits to Higgs data and Higgs + top data, respectively. The blue ellipses indicate the marginalised constraints from Higgs + top data after introducing top-quark four-fermion operators into the fit, and the yellow ellipses are obtained from a fit dropping the $t \bar t H$ data.}
\end{figure}

\begin{figure}[t] 
\centering
\includegraphics[width=0.8\textwidth]{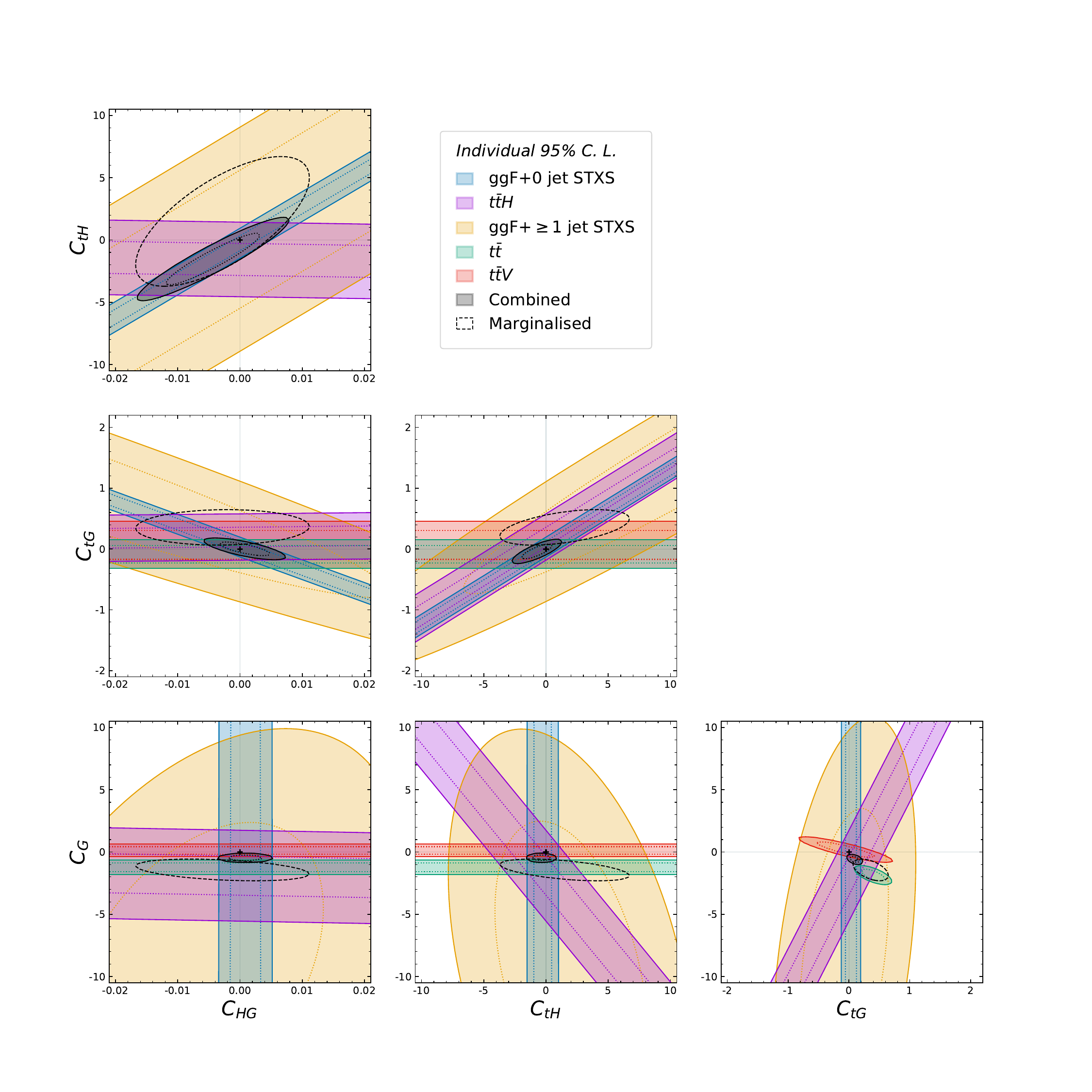}
\caption{\it \label{fig:interplay2} Constraints on the indicated pairs of operator coefficients 
at the 95\% confidence level, setting the other operator
coefficients to zero. The shaded regions correspond to 
linear fits to
Higgs signal strengths and 0 jet STXS bins (blue), 
$t\bar{t}H$ signal strengths (mauve),  $\geq 1$ jet STXS 
bins (orange) $t\bar{t}$ data (green), $t\bar{t}V$ data (red) 
and their combination (grey). The dashed ellipses show the constraints obtained by marginalising over the remaining Wilson coefficients of the full fit.}
\end{figure}

In order to assess the potential impact of the interplay between top and Higgs data, we may consider the following subset of `Higgs-only' operators: 
\begin{equation}
    \{\Cp{H\Box}, \Cp{HG},\Cp{HW}, \Cp{HB}, \Cp{tH}, \Cp{bH}, \Cp{\tau H}, \Cp{\mu H} \}
\label{eq:Higgsonlyops}
\end{equation}
together with $\Cp{G}$ and $\Cp{tG}$, which do not modify Higgs interactions directly but can impact gluon fusion. Performing a fit to this subset, Fig.~\ref{fig:interplay1} displays the result for the 95\% CL constraints
when top data are combined with
Higgs data in planes showing different pairs of
the operator coefficients $\Cp{HG}, \Cp{tG}, \Cp{tH}$
and $\Cp{G}$, marginalised over the other coefficients in~(\ref{eq:Higgsonlyops}). This is the relevant set of operators in which the interplay between Higgs and top physics is most evident, taking place in the gluon fusion and $t\bar{t}$ associated Higgs production processes. It is well known that there is a degeneracy in gluon fusion between $\Cp{HG}$ and $\Cp{tH}$ that prevents it from being used as a robust indirect constraint on the top Yukawa coupling, or conversely, heavy coloured particles that couple to the Higgs. This strongly motivates the direct measurement of the top Yukawa, via $t\bar{t}H$. In Fig.~\ref{fig:interplay1}, the yellow ellipses show results for Higgs data, without $t\bar{t}H$, while the green ellipses show the sensitivity for Higgs data including it, indicating how a relative flat direction in this plane (top left panel) is lifted by the inclusion of $t\bar{t}H$.  However, despite not being directly related to Higgs couplings, both $\Cp{tG}$ and $\Cp{G}$ can also affect Higgs(+jet) production in gluon fusion. Thankfully, these can be constrained by top data, particularly $t\bar{t}$ (and multi-jet data at order $\Lambda^{-4}$, for the latter). 
The results for the combination of Higgs with top data are shown as mauve ellipses, where we see in each plane a very substantial reduction of the area allowed at the 95\% CL. The difference between the two sensitivities underlines the strength and importance of this data in indirectly pinning down BSM interactions of the Higgs, where now the $\Cp{tG}$ and $\Cp{G}$ directions are squeezed down by an order of magnitude. We also see that several (anti)correlations between pairs of operator coefficients are suppressed when top data are included,
most noticeably in the $(C_{HG}, C_G)$ plane. However, using top data to constrain only two operators is not in keeping with the global spirit of SMEFT interpretations, especially given the large number of degrees of freedom discussed in Sec.~\ref{sec:topspecificflavourscenario} that could potentially dilute its power to bound $\Cp{tG}$ and $\Cp{G}$. We address this question by increasing our operator subset to include the 7 four-fermion operators that impact $t\bar{t}$ production, with the new marginalised constraints shown as blue ellipses. Surprisingly, there is little further change when
adding the four-fermion operators, indicating a very limited dilution effect and underlining the robustness of the complementarity of top data in indirectly constraining Higgs couplings. This is especially encouraging given the fact that, as discussed in Section~\ref{sec:top}, our constraints on this set of operators are significantly weakened by the linear approximation used in our analysis, allowing for larger marginalisation effects than a quadratic-level fit would.

Fig.~\ref{fig:interplay2} displays the constraints on the same
pairs of operator coefficients at the 95\% confidence level when
the coefficients of other operators are set to zero, with more fine-grained information on the constraints provided by the different datasets. The shaded regions are the results of linear fits to Higgs signal strengths and 0 jet STXS bins (blue), $t \bar t H$ signal strengths (mauve), $\geq 1$ jet STXS bins (orange), $t \bar t$ data (green), $t \bar t V$ data (red) and their combination (grey).
The dashed ellipses show the constraints in the corresponding
parameter planes when marginalising over the remaining Wilson coefficients of the full fit, as shown in
Fig.~\ref{fig:interplay1}. As was to be expected, the
constraints obtained when the other operator coefficients are
set to zero are significantly stronger. The complementarity between $ggF$, $t\bar{t}H$ and $t\bar{t}$ is again evident, with $t\bar{t}V$ data also providing some additional information on $\Cp{G}$. We also see that $\geq 1$ jet STXS bins have not yet reached the level of precision needed to offer significant complementary information in this parameter space. However, we expect this to improve as increasingly fine-grained STXS binnings are measured.

\subsubsection{The triple-gluon operator $C_G$}
\label{sec:CG}

The operator $\mathcal{O}_G$ consists of triple-gluon field strengths~\cite{Simmons:1990dh, Dreiner:1991xi, Cho:1994yu}
and so affects any observables sensitive to jets~\cite{Ghosh:2014wxa, Hirschi:2018etq, Goldouzian:2020wdq, Krauss:2016ely}. This includes many of the Higgs and top processes in our global fit, as shown in Fig.~\ref{fig:correlationmatrixfull} that we discuss in the next Section, where we see sizeable correlations of $C_G$ with the operators $C_{HG}, C_{tH}, C_{tG}, C_{Qq}^{3,8}, C_{Qq}^{1,8}, C_{Qu}^{8}, C_{tq}^{8}, C_{tu}^{8}$ and $C_{td}^{8}$, spanning both the Higgs and top sectors. Therefore, the gluonic operator complicates the SMEFT interpretation of the measurements in these sectors.

\begin{figure}[h!] 
\centering
\includegraphics[width=0.6\textwidth]{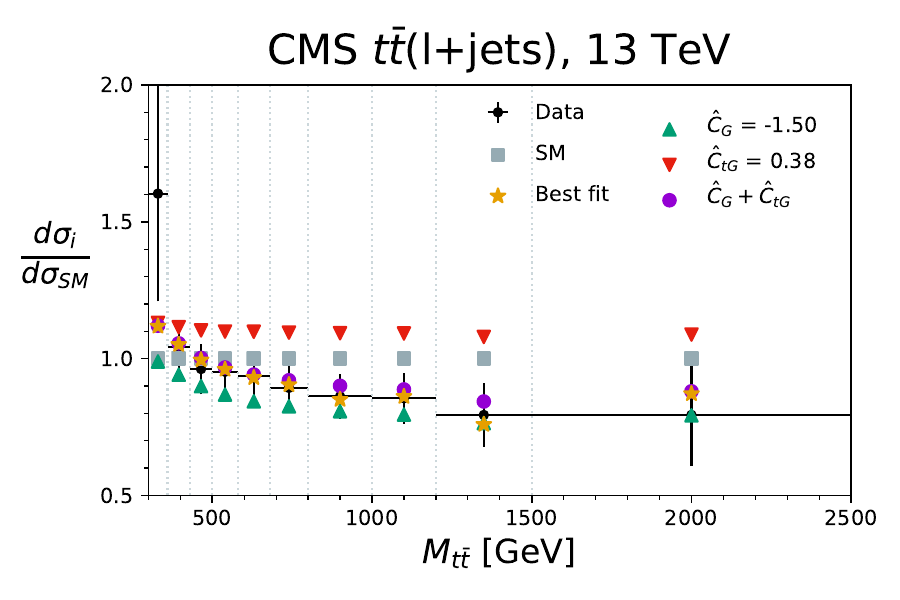} 
\caption{\it \label{fig:mtt_measurement} 
\it The measurement of the $t\bar{t}$ invariant mass distribution in the lepton+jets channel at 13 TeV by the CMS experiment~\cite{Sirunyan:2018wem}
compared to to the SM prediction at the NNLO QCD + NLO EW level~\cite{Czakon:2017wor}. Also shown are predictions corresponding to the best-fit values for $\hat C_G$ (green upward triangles), $\hat C_{tG}$ (red downward triangles), their combination (purple circles) and the global best-fit point in the full parameter space (orange stars).
}
\end{figure}

It has been argued in Refs.~\cite{Krauss:2016ely, Hirschi:2018etq, Goldouzian:2020wdq} that a very strong constraint on $C_G$ is provided by multijet data that are not included
in our default data set, and that one can set $C_{G} = 0$ when analysing electroweak, Higgs or top data. However, this strong constraint relies on quadratic contributions to multijet observables, whereas our global fit is made to linear order. The linear contributions of $C_G$ are small since the amplitude involving $C_G$ in gluon-gluon scattering does not interfere with the SM, and in quark-gluon scattering it does so only proportionally to the quark masses. This has made it a challenge to constrain in past studies~\cite{Simmons:1990dh, Dreiner:1991xi, Cho:1994yu}. However, the top sector, with its large quark mass, provides an opportunity to recover sensitivity at linear order to $C_G$, as studied most recently in Ref.~\cite{Bardhan:2020vcl}. Table~\ref{tab:individual} indicates that $t\bar{t}$ (43\%) and $t\bar{t}V$ data (56\%) provide the entirety of the individual sensitivity to $\Cp{G}$. This is confirmed by comparing Figs.~\ref{fig:flavour_universal_EWdiBH}, \ref{fig:top_asym_impact} and  \ref{fig:all_fit}, which also emphasise that these bounds are robust when marginalising over the other operators in both the Higgs and top sectors.

The $C_G$ fit also shows the strongest pull away from zero, with a significance $\sim 3$  and $4 \, \sigma$ for the best fit in the individual and marginalised cases, respectively, as shown in Fig.~\ref{fig:all_fit}. This effect is due to $t\bar{t}$ differential data, an example of which is given by the 13-TeV invariant mass distribution data shown in Fig.~\ref{fig:mtt_measurement}. We see there that the 
$m_{t \bar t}$ dependence of the cross-section data, normalised to the SM prediction, denoted by black points, differs quite 
significantly from that of the SM, represented by grey boxes at a value of 1, and it is this 
discrepancy that pulls $C_G$ into negative territory, as shown by the green triangles, plotting the best-fit $\Cp{G}$ contribution. We see that the agreement with the data is obtained by an interplay between $\Cp{G}$ and $\Cp{tG}$, whose best fit prediction, shown by the red triangles, improves the agreement in the low mass bins while not too significantly affecting the high mass region. The sum of the best-fit $\Cp{G}$ and $\Cp{tG}$ predictions, shown by the purple circles, coincides with the global best-fit prediction, shown by the orange stars, demonstrating that the fit to the data is obtained primarily by this interplay.
We emphasise,
however, that the significance of this effect could be reduced if
there were some important contribution to the $m_{t \bar t}$
distribution close to threshold that has not been included in the
SM calculation. Moreover, such a large pull away from the SM in this case is not meaningful, in view of the potentially important quadratic contributions from $C_G$. What the linear fit demonstrates is the size of the linear constraint on $C_G$, which is not known {\it a priori}, and its dependence on other operators. Significant indirect effects of $C_G$ on the other operators may also then be questioned.

\begin{figure}[t] 
\centering
\includegraphics[width=0.65\textwidth]{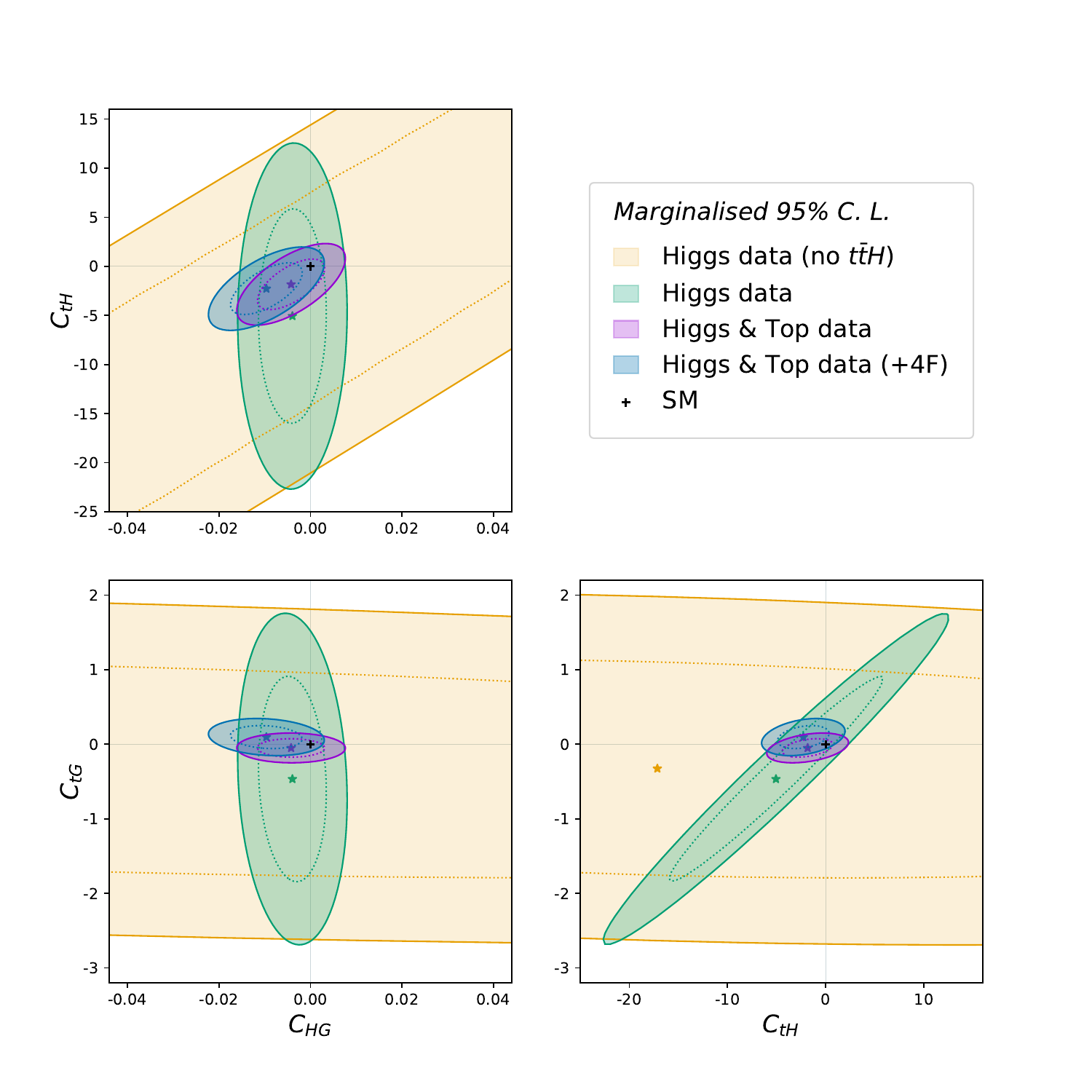}\\
\caption{\it \label{fig:interplay3} Same as Fig~\ref{fig:interplay1}, but setting $\Cp{G}=0$.
}
\end{figure}

\begin{figure}[h!] 
\centering
\includegraphics[width=1.0\textwidth]{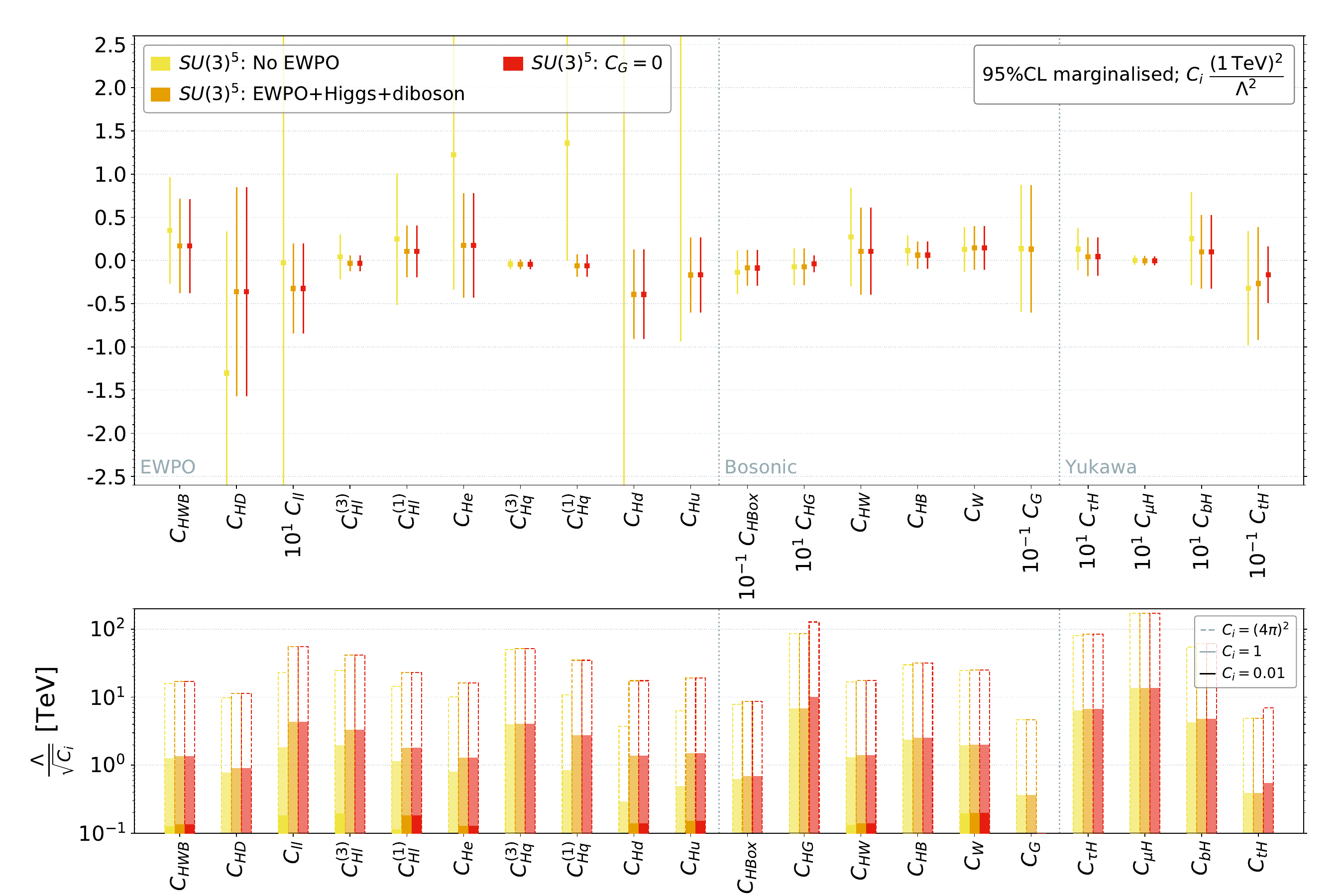}\\
\includegraphics[width=1.0\textwidth]{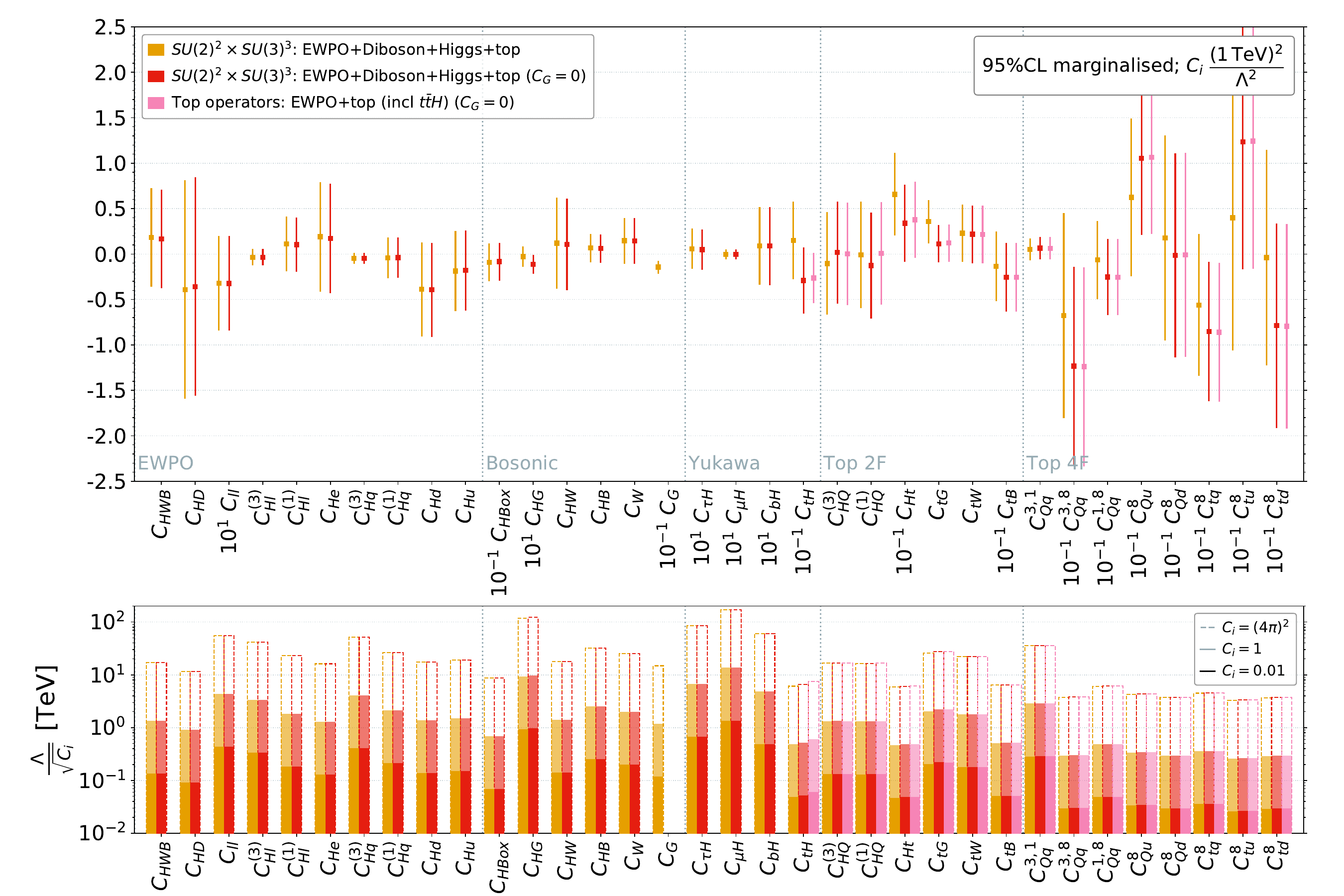}
\caption{\it Comparison of the constraints on the indicated marginalised operator coefficients 
$C_i (1~{\rm TeV})^2/\Lambda^2$ (top and third panels) and the corresponding scales $\Lambda$ for the indicated values of the ${C_i}$ at the 95\% confidence level
(second and bottom panels), found in a combined linear fit to the Higgs,
diboson and electroweak precision observables (top two panels) and
including in addition top data (bottom two panels), including 
$C_G$ in the fit (orange) and setting $C_G = 0$ (red). Also displayed in yellow in the top two panels is a fit without LEP (EWPO and $WW$) measurements. 
}
\label{fig:margnoCG}
\end{figure}

Accordingly, we have investigated the consequences of assuming that 
$C_G$ can be better constrained by including 
dedicated QCD multi-jet data, and have analysed the effects of setting 
$C_{G} = 0$ in our study of the impact of top data on Higgs coupling measurements, as well as in our marginalised global fits in the flavour-universal SU(3)$^5$ and the top-specific SU(2)$^2 \times$SU(3)$^3$ scenarios. Fig.~\ref{fig:interplay3} shows that the global space of constraints is changed compared to  Fig.~\ref{fig:interplay1}, especially when only including Higgs data (green ellipse), with improved sensitivity (see Fig~\ref{fig:margnoCG}) and a different pattern of correlations. Once the top data is included, the overall sensitivity is improved, and the best-fit point before including the four-fermion operators (mauve ellipse) is closer to the SM. The relative impact of adding the four-fermion operators (blue ellipse) is more noticeable when $\Cp{G} = 0$ than when it is non-zero. When $\Cp{G} = 0$, the tension with the SM in the $t\bar{t}$ data pulls the four-fermion operators and $\Cp{tG}$. These operators also affect $t\bar{t}H$, and $\Cp{tG}$ also modifies gluon fusion, leading to a cascade of shifts in the ranges of these operators.  
Moving to the global results, we see in the top two panels of Fig.~\ref{fig:margnoCG} that this 
constraint has little effect on the 95\% ranges we
find in our marginalised SU(3)$^5$ fit, except that
the ranges of $C_{HG}$ and $C_{tH}$ are reduced
noticeably when we set $C_G = 0$~\footnote{For completeness, we also
show in the top two panels of Fig.~\ref{fig:margnoCG} the effects of
dropping the EWPOs (and LEP $WW$) from the  marginalised SU(3)$^5$ fit
(yellow bars). As could be expected, there are large effects on the
constraints on the operators that 
contribute most to the EWPOs, but quite small effects in the 
bosonic and Yukawa sectors.}, as expected from the previous discussion of Fig.~\ref{fig:interplay3}. In the SU(2)$^2 \times$SU(3)$^3$
(bottom two panels of Fig.~\ref{fig:margnoCG})
there are shifts in the central values of several top operator 
coefficients, with the four fermion operators moving further away from the SM to absorb the aforementioned discrepancies in top data. Overall, no significant reductions in the ranges of any operator
coefficients are observed, and thus our fit results are relatively insensitive to the treatment
of $C_G$.

\FloatBarrier

\subsubsection{Correlation matrix and principal component analysis}
\label{sec:correlations}

\begin{figure}[p] 
\centering
\includegraphics[width=1.0\textwidth]{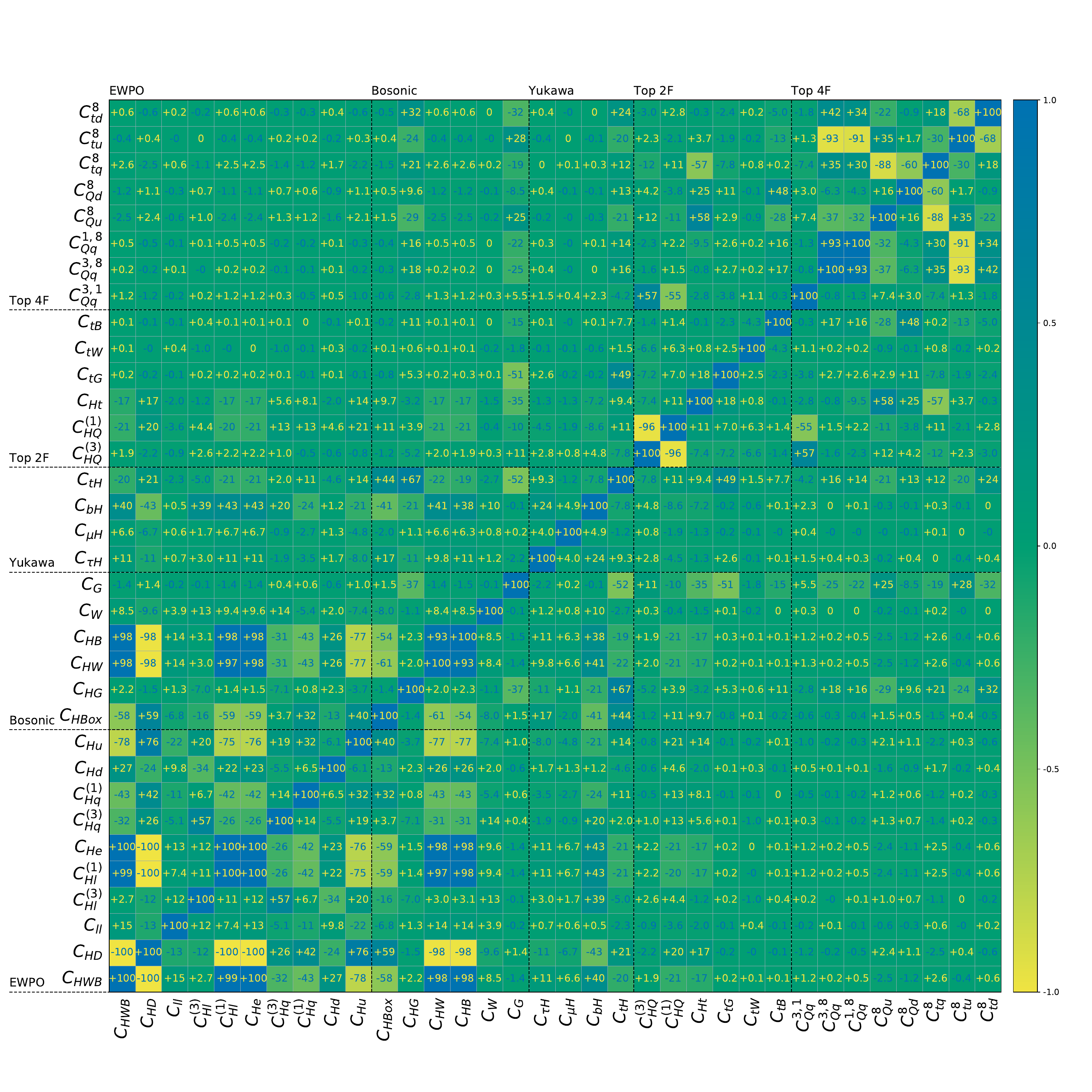}
\caption{\it The full $34 \times 34$ correlation matrix for
the marginalised top-specific fit. The operators are grouped into those
affecting primarily electroweak precision observables, 
bosonic observables,
Yukawa measurements and top electroweak measurements, as well
as top-quark four-fermion operators. The entries in the
correlation matrix are colour-coded according to the 
indicated magnitudes of the correlation coefficients.
}
\label{fig:correlationmatrixfull}
\end{figure}

The full correlation matrix for the top-specific marginalised fit is shown in Fig.~\ref{fig:correlationmatrixfull}, 
colour-coded and labelled in percentages. Some of these correlations can be explained intuitively by the simple fact 
that two operators contribute to the same observable, while others occur more indirectly, 
through a chain of dependencies that is difficult to trace through the inversion of the Fisher information matrix.
We see that there are substantial (anti) correlations between the coefficients of 
operators in the EWPO and Bosonic sectors, and there are also
many (anti)correlations within the top 4F sector.
On the other hand, there is only one large off-diagonal entry
in the top 2F sector, namely a negative correlation between $C^{(3)}_{HQ}$ and $C^{(1)}_{HQ}$. 
Along with $\Cp{Ht}$, that is mildly correlated with these two, 
these operators affect $Zt\bar{t}$ couplings. As previously discussed, 
one linear combination, $C^{(3)}_{HQ}+C^{(1)}_{HQ}$ modifies the $Z b\bar{b}$ coupling, 
and hence has a very strong LEP constraint, while  $\Cp{Ht}$ and the other combination of $C^{(3)}_{HQ}$ and $C^{(1)}_{HQ}$ can only be 
probed in EW top processes. The bottom Yukawa operator ($\Cp{bH}$) exhibits 
some correlations with those of the top ($\Cp{tH}$) and the tau ($\Cp{\tau H}$), 
and in the bosonic sector some moderate correlations are observed, notably 
between $\Cp{HW}$, $\Cp{HB}$ and $\Cp{H\Box}$, and between $C_{HG}$ and $C_{G}$ 
(as expected from the discussion in Sec.~\ref{sec:Higgsonly}).
Turning to correlations between operators in different
sectors, we note substantial  (negative) correlations between$C_{bH}$ in the Yukawa sector and $C_{HWB},( C_{HD}), C_{Hl}^{(3)}, C_{Hq}^{(3)}$ and $C_{He}$
in the EWPO sector, and $(C_{H\Box}), C_{HW}$ and $C_{HB}$ in the bosonic sector respectively, as well as
substantial positive (negative) correlations between $C_{tH}$ in the
Yukawa sector and $C_{H\Box}, C_{G}$ and $(C_{HG})$ in the Bosonic sector.
Finally, there are several large (anti)correlations between
operators in the top 4F and top 2F sectors, namely
$C^{3,1}_{Qq}$ and $C_{HQ}^{(3)}$ (positive),
$C^{3,1}_{Qq}$ and $C_{HQ}^{(1)}$ (negative),
$C^8_{Qu}$ and $C_{Ht}$ (negative), and $C^8_{tq}$ and 
$C_{Ht}$ (positive). Overall, there are  22 correlation
coefficients with magnitude $\ge 0.2$ between operators in a
top sector on the one hand and in a Yukawa, bosonic or 
electroweak sector on the other hand. These and the top sector
may not be talking to each other very loudly, but they are starting to whisper to each other.

\begin{figure}[p] 
\centering
\includegraphics[angle=0,width=0.9\textwidth]{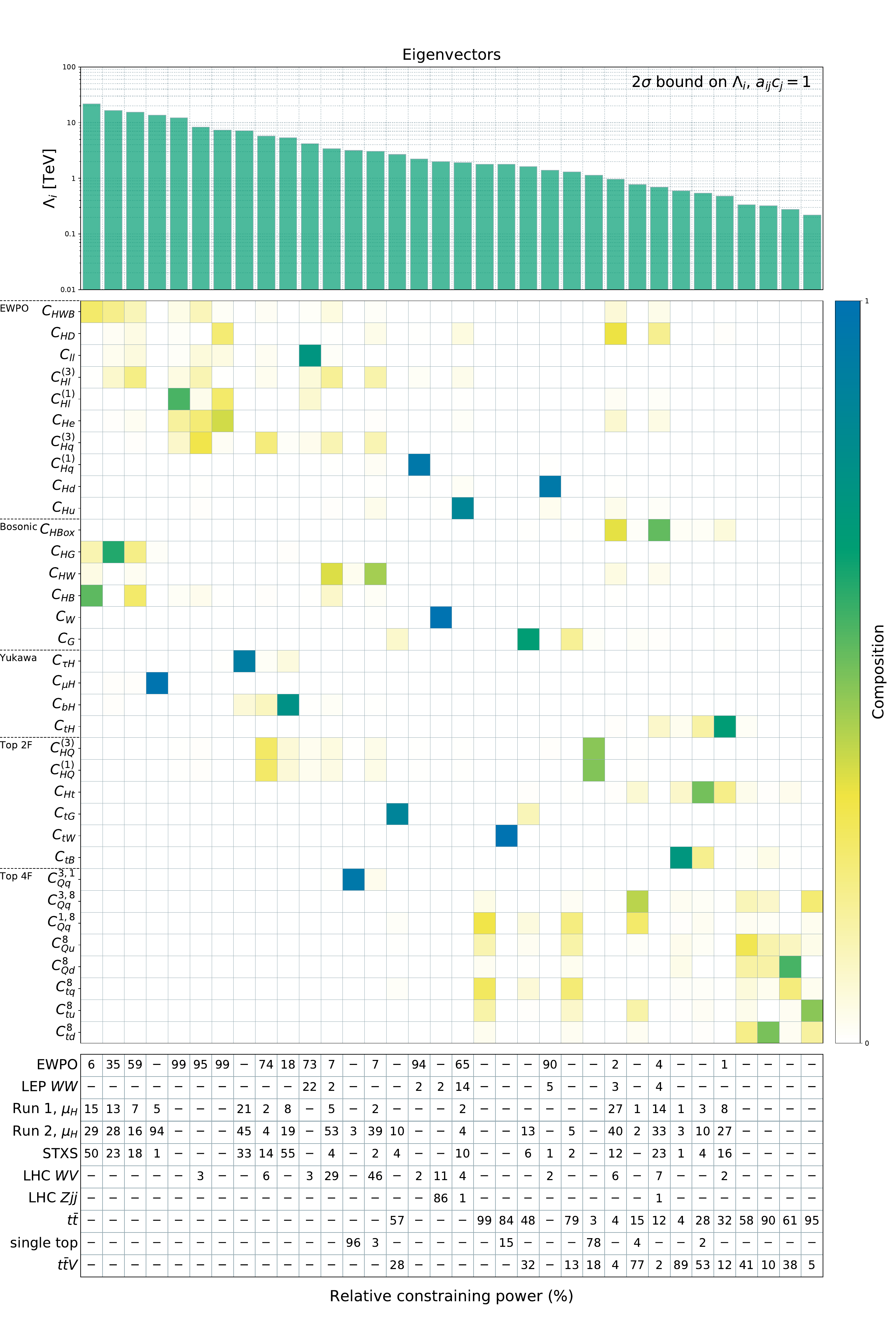}
\vspace{-5mm}
\caption{\it \label{fig:eigenvectors} Constraints on the eigenvectors of the global likelihood function.
The top panel shows the strengths of the global constraints on the eigenvectors.
The rows in the centre panel correspond to the different
operator coefficients, the rows correspond to the different eigenvectors, and the
colour-coded squares represent the moduli-squared of the operator components to the
eigenvectors. The bottom panel shows the constraining powers of the electroweak precision data,
LEP diboson data, Higgs coupling strength measurements from Runs 1 and 2, STXS
, LHC diboson and $Zjj$, $t\bar{t}$, single-top and $t\bar{t}W/Z/\gamma$ measurements, respectively. Instances where there is no significant constraining
power are indicated by {\rm ``-''}. Explicit expressions for the eigenvectors are given in Table~\ref{tab:eigenvectors}
}
\end{figure}

In the Gaussian approximation to the global likelihood that is used here,
it is also informative to diagonalise the constraints on the operator coefficients in
the orthonormal eigenvector basis and perform a principal component analysis. This tells us which directions are most constrained in the fit, and what operators contribute to those directions. We display the constraints on the eigenvectors
graphically in Fig.~\ref{fig:eigenvectors}: the rows 
in the centre panel correspond to the different
operator coefficients, the columns correspond to the different eigenvectors, and the
colour-coded squares represent the moduli-squared of the operator components in the
eigenvectors. The latter are ordered such that the strengths of the global constraints decrease
from left to right, as seen in the top panel of Fig.~\ref{fig:eigenvectors},
where the 95\% CL bounds on the scales $\Lambda$ are calculated assuming 
that the linear combination of operator coefficients making up that particular eigenvector is set to unity. The bottom panel
tabulates the respective relative constraining powers of the 
electroweak precision data,
LEP diboson data, Higgs coupling strength measurements from 
Runs 1 and 2, STXS measurements, LHC diboson and $Zjj$ measurements,
$t \bar t$ measurements, single top measurements and 
$t \bar t V$ measurements. These are defined as the relative contribution of each dataset to the corresponding entry of the diagonalised Fisher information. Entries where there is no
significant constraining power are indicated by ``-''. 

We see that the largest component of the 
best-constrained eigenvector is ${\cal O}_{HB}$,
and that the limit on its scale exceeds 20~TeV, 
with the most important contribution
coming from the STXS measurements, followed by Higgs signal
strength measurements. The scales of four more eigenvectors
are constrained beyond the 10-TeV level, with the most
important contributions coming from the 
electroweak precision measurements as well as STXS and Higgs
signal strength measurements. 
They can broadly be associated with the powerful sensitivity we have obtained in constraining the $H\gamma\gamma$ and $Hgg$
interactions. The first  12 eigenvectors are constrained by a mixture of EWPO and Higgs data, showing that these two sets are providing complementary and competitive bounds in the multi-TeV range. 

We next see a particularly strong constraint coming entirely from single top data on $\Cpp{Qq}{3,1}$ alone. We note that other operators contribute along that eigenvector direction, which is given by $-0.98 C_{Qq}^{3,1} - 0.17 C_{HW} +0.08 C_{HQ}^{(3)} $, but with too small a magnitude to be visibly coloured~\footnote{For completeness, the numerical values of the eigenvector components are provided in Table~\ref{tab:eigenvectors}.}. However, this is partly responsible for the large (anti-)correlations between $\Cpp{Qq}{3,1}$ and $C_{HQ}^{(3)} \, (C_{HQ}^{(1)})$ shown in Fig.~\ref{fig:correlationmatrixfull}, and $C_{Qq}^{3,1}$ also appears in other directions with a small contribution. Several other examples of  relatively isolated operators can be found across the figure, identified by the columns dominated by a single, very dark spot. Here the eigen-directions nearly coincide with a particular operator, such that the rest of the fit should be relatively independent of whether these are included or not. Specific examples are $\Cp{\mu H}$, which is constrained in isolation by the $H\to\mu\mu$ signal strength, and $\Cp{tW}$, which is constrained mainly by $W$-helicity fraction measurements in $t\bar{t}$ data. We also see that $Zjj$ mostly constrains $\Cp{W}$ with not much effect on the rest of the fit. The relation between measurements and constraints on operators can be indirect, illustrating the complementarity between the different datasets: for example, the LHC $WW$ and $WZ$ diboson data are responsible for 46\% of the constraining power along the eigenvector direction principally aligned with $C_{HW}$, despite the lack of $C_{HW}$ dependence in diboson data. However, their inclusion helps to close directions of limited sensitivity in the fit to Higgs and electroweak data that are then better able to constrain {\it e.g.} $C_{HW}$. 

The least strongly constrained eigenvector is predominantly
${\cal O}^8_{tu}$, with a scale bounded just
above 200~GeV when the operator coefficients are normalised to unity, mainly
by $t \bar t$ data. It is followed by three  more eigenvectors with 
scales $\sim 300$~GeV, whose principal components are top operators. 
The most important constraints on these eigenvectors are
in the top sector, principally from the $t \bar t$
 and $t \bar t V$ data. As discussed in
Section~\ref{sec:top}, while the validity of the SMEFT may be questioned
when the operator coefficients are normalised to unity, it should be
reliable for all top operators in the strong-coupling limit. 
Also, we expect the SMEFT to be valid for the better-constrained 
eigenvectors even for unit-normalised coefficients, since these
eigenvectors have relatively small
top operator components, as seen in the middle panel of
Fig.~\ref{fig:correlationmatrixfull}.

\FloatBarrier

\section{Constraints on UV Completions}
\label{sec:UV}

So far our approach has been to use the SMEFT framework to combine all relevant data inputs and perform a global fit to the coefficients of the dimension-6 operators characterising possible modifications of the SM Lagrangian at leading order in the momentum expansion. The result of this fit, summarised in Fig.~\ref{fig:all_fit}, provides a model-independent way to evaluate the compatibility of the SM with the available experimental data.
This SMEFT analysis provides information on the level of 
new physics contributions compatible with the current data. We presented in the bottom panel of Fig.~\ref{fig:all_fit} limits on the dimensionful parameters
$\Lambda/\sqrt{C_i}$, which can be interpreted as
constraints on the possible scale of New Physics 
compatible with current measurements. The individual limits correspond to bounds on a single operator assuming all others are zero, while the marginalised limits allow all other coefficients to vary. These can respectively be taken as optimistic and pessimistic estimates of the sensitivity, and we expect realistic models to generate some intermediate subset.

In this Section we go a step further in the interpretation of our fit and explore how specific UV completions of the SM Lagrangian are constrained by current measurements. In any given model, 
the global analysis we have presented is often not directly applicable, as typical models may generate more than just one of the dimension-6 SMEFT operators, but not all of them. Moreover, when a specific model contributes to more
than a single operator, these contributions are often related, corresponding to a smaller subset of independent parameters.

To illustrate these model-dependent effects we have considered several model interpretations. In Section~\ref{sec:treeUV} we discuss models in which SMEFT operators are induced at 
the tree level, and in Section~\ref{sec:UVpatterns} we discuss classes of UV completions that share similar SMEFT patterns. We then analyze supersymmetric models with TeV-scale 
stops in which SMEFT operators are induced at the one-loop level in Section~\ref{sec:SUSYUV}.
Finally, in Section~\ref{sec:multiple} we present results from a survey of the pulls for
all fits with non-vanishing coefficients for combinations of 2, 3, 4 and 5 operators.

\subsection{Simple tree-level-induced SMEFTs} \label{sec:treeUV}

We first study the implications of our analysis for single-field extensions of the SM that contribute to the SMEFT at tree level. This exercise updates the one
presented in~\cite{Ellis:2018gqa} and is based on the dictionary provided in~\cite{deBlas:2017xtg}~\footnote{We note that these one-parameter extensions of the SM have been included among the BSM benchmark proposals made by the LHC Higgs Working Group~\cite{Marzocca:2020jze}.}.

  \begin{table}[h!]
\begin{center}
\resizebox{\columnwidth}{!}{%
\begin{tabular}{|c |c |c |c |c |c || c |c |c |c |c | c|} 
 \hline
Name & Spin & SU(3) & SU(2) & U(1) & Param. & Name & Spin & SU(3) & SU(2) & U(1) & Param. \\ [0.5ex] 
 \hline\hline
 $S$ & 0 & 1 & 1 & 0 & ($M_S$, $ \kappa_{S}$) &  $\Delta_{1}$ & $\frac{1}{2}$ & 1 & 2 & $- \frac{1}{2}$  & ($M_{\Delta_1}$,$\lambda_{\Delta_{1}}$)  \\ 
 \hline
 $S_{1}$ & 0 & 1 & 1 & 1 & ($M_{S_1}$,$y_{S_{1}}$) & $\Delta_{3}$ & $\frac{1}{2}$ & 1 & 2 & $- \frac{1}{2}$ & ($M_{\Delta_3}$,$\lambda_{\Delta_{3}}$)\\ 
 \hline
  $\varphi$ & 0 & 1 & 2 & $\frac{1}{2}$ & ($M_\varphi$,$ Z_{6} \cos \beta$) & $\Sigma$ & $\frac{1}{2}$ & 1 & 3 & 0  & ($M_\Sigma$,$\lambda_{\Sigma}$)\\ 
 \hline
  $\Xi$ & 0 & 1 & 3 & 0 & ($M_\Xi$,$\kappa_{\Xi}$)& $\Sigma_{1}$ & $\frac{1}{2}$ & 1 & 3 & -1 & ($M_{\Sigma_1}$,$\lambda_{\Sigma_{1}}$)\\ 
 \hline
   $\Xi_{1}$ & 0 & 1 & 3 & 1 & ($M_{\Xi_1}$,$\kappa_{\Xi_1}$) & $U$ & $\frac{1}{2}$ & 3 & 1 & $ \frac{2}{3}$ & ($M_{U}$,$\lambda_{U}$) \\ 
 \hline
$B$ & 1 & 1 & 1 & 0 & ($M_B$,$\hat  g_{H}^{B}$) & $D$ & $\frac{1}{2}$ & 3 & 1 & $- \frac{1}{3}$ & ($M_{D}$,$\lambda_D$) \\ 
 \hline
$B_{1}$ & 1 & 1 & 1 & 1 & ($M_{B_1}$,$g_{B_1}$)&$Q_{1}$ & $\frac{1}{2}$ & 3 & 2 & $ \frac{1}{6}$  &  ($M_{Q_1}$,$\lambda_{Q_1}$)\\ 
 \hline
  $W$ & 1 & 1 & 3 & 0 & ($M_W$,$ \hat{g}_{H}^{W}$)& $Q_{5}$ & $\frac{1}{2}$ & 3 & 2 & $- \frac{5}{6}$ & ($M_{Q_5}$,$\lambda_{Q_5}$)\\ 
 \hline
  $W_{1}$ & 1 & 1 & 3 & 1 & ($M_{W_1}$,$\hat{g}_{W_{1}}^{\phi}$)& $Q_{7}$ & $\frac{1}{2}$ & 3 & 2 & $\frac{7}{6}$ & ($M_{Q_7}$,$\lambda_{Q_7}$)\\ 
  \hline
   $N$ & $\frac{1}{2}$ & 1 & 1 & 0 & ($M_N$,$\lambda_{N}$)& $T_{1}$ & $\frac{1}{2}$ & 3 & 3 & $- \frac{1}{3}$ &  ($M_{T_1}$,$\lambda_{T_{1}}$)\\ 
  \hline
    $E$ & $\frac{1}{2}$ & 1 & 1 & -1 & ($M_E$,$\lambda_{E}$)& $T_{2}$ & $\frac{1}{2}$ & 3 & 3 & $\frac{2}{3}$  & ($M_{T_2}$,$\lambda_{T_2}$)\\ 
  \hline
  $T$ & $\frac{1}{2}$ & 3 & 1 & $\frac{2}{3}$ & ($M_T$,$s_{L}^{t}$)& $TB$ &
  $\frac{1}{2}$ & 3 & 2 & $\frac{1}{6}$ & ($M_{TB}$,$s^{t,b}_{L}$)\\ [1ex]
\hline
\end{tabular}
}
\caption{\it Single-field extensions of the SM constrained by our analysis. \label{table:fields}}
\end{center}
\end{table}

We list in Table~\ref{table:fields} the SU(3)$\times$SU(2)$\times$U(1)  quantum numbers and couplings
of the new fields considered here. We assume flavour-universal couplings in all cases except $T$ and $TB$, in which the
new fields couple only to the third generation.  These two models are taken from \cite{Dawson:2020oco}.  We consider only the renormalisable contributions from each single field extension.  In the case of the models $B$ and $W$ (a $Z'$ and $W'$ respectively) we consider only their couplings to the Higgs doublet and set all fermion couplings to zero. 
In addition to evaluating the constraints on these single-field extensions, we also consider the following
two combinations of the fields in Table~\ref{table:fields} that yield single-parameter models via cancellations (see Ref.~\cite{Marzocca:2020jze}), 
i.e., models that depend on only a single coupling ($\lambda$ or $g_{H}$), as well as a mass $M$:
{\it 1)} Quark bidoublet model: $\{Q_{1}, Q_{7} \}$ with equal masses $M$ and equal couplings  $\lambda$ to the top quark, and {\it 2)} Vector-singlet pair model: $\{B, B_{1} \}$ with equal masses $M$ and Higgs couplings proportional to $g_{H}$.
We exhibit in Tables~\ref{tab:single-field_models} and ~\ref{table:doubleop} the contributions made at tree level by
exchanges of each of these fields to the SMEFT coefficients~\footnote{In general, the coloured, vector-like fermions contribute at one-loop
order to $C_{HG}$. We include this contribution for the 2-parameter model $TB$, and verify in a representative example (the $T$ field) that 
it has a negligible effect on the single-parameter model constraints.}. The numbers shown
in the Tables should each be multiplied by the appropriate squared coupling factors
and divided by the square of the mass scale $M$~\footnote{We do not provide limits on the two-parameter model $\Xi_1$, which has a complex coupling $g_{\Xi_1}$, but note that it behaves similarly to models $\Xi$ and $S_1$. }.

\begin{table}[h]
{
\begin{center}
\begin{tabular}{| c || c |c|c|c|c|c|c|c|c|c|c|c|c|}
\hline 
 \lgr    Model &    $C_{HD}$ &  $C_{ll}$ &    $C_{Hl}^{3}$ &     $C_{Hl}^{1}$ &     $C_{He}$ &   $C_{H \Box}$ &     $C_{\tau H}$ &     $C_{tH}$ &     $C_{bH}$  \\
\hline \hline
           $S$ &      &    &       &           &     &    $-\frac{1}{2}$ &       &       &        \\
          \hline
        $S_{1}$ &      &  1 &       &        &       &             &       &       &         \\
       \hline
       $\Sigma$ &      &    &   $\frac{1}{16}$ &  $\frac{3}{16}$ &          &       &    $\frac{y_{\tau}}{4}$ &       &       \\
       \hline
      $\Sigma_{1}$ &     &   &   $-\frac{1}{16}$ &  $-\frac{3}{16}$ &          &      &  $\frac{y_{\tau}}{8}$ &      &      \\
      \hline
           $N$ &      &    &   $-\frac{1}{4}$ &     $\frac{1}{4}$ &       &              &       &       &         \\
           \hline
           $E$ &      &    &   $-\frac{1}{4}$ &    $-\frac{1}{4}$ &          &       &      $\frac{y_{\tau}}{2}$ &       &        \\
           \hline
     $\Delta_{1}$ &      &      &    &     &    $\frac{1}{2}$ &       &      $\frac{y_{\tau}}{2}$ &       &        \\
     \hline
      $\Delta_{3}$ &      &        &    &     &  $-\frac{1}{2}$ &       &      $\frac{y_{\tau}}{2}$ &       &       \\
      \hline
         $B_{1}$ &    $1$ &   &      &        &    &    $-\frac{1}{2}$ &    $-\frac{y_{\tau}}{2}$ &    $-\frac{y_{t}}{2}$ &    $-\frac{y_{b}}{2} $  \\
         \hline
 $\Xi$ &   $-2$ &    &       &        &     &     $\frac{1}{2}$ &     $y_{\tau}$ &    $y_{t}$ &     $y_{b}$  \\
                   \hline
         $W_{1}$ &  $-\frac{1}{4}$&   &       &    &    &  $-\frac{1}{8}$ &  $-\frac{y_{\tau}}{8}$ &  $-\frac{y_{t}}{8}$ &  $-\frac{y_{b}}{8} $  \\
         \hline
          $\varphi$ &     &   &      &       &      &      &    $-y_{\tau}$ &    $-y_{t}$ &    $-y_{b}$  \\
   \hline
       \small{$\{B,B_{1} \}$} &     &   &         &    &    &     $-\frac{3}{2}$ &      $-$ $y_{\tau}$ &      $-$ $y_{t}$ &     $-$ $y_{b}$  \\
   \hline
       \small{$\{Q_{1}, Q_{7} \}$} &     &   &      &    &    &      &      &      $y_{t}$ &      \\
\hline \hline
\lgr Model &       $C_{Hq}^{3}$ &     $C_{Hq}^{1}$ &  $(C_{Hq}^{3})_{33}$ &     $(C_{Hq}^{1})_{33}$ &   $C_{Hu}$ &        $C_{Hd}$ &     $C_{tH}$ &     $C_{bH}$ & \\
\hline \hline
          $U$ &       $-\frac{1}{4}$ &     $\frac{1}{4}$ &    $-\frac{1}{4}$ &     $\frac{1}{4}$  &    &            &     $\frac{y_{t}}{2}$ &    &  \\
          \hline
          $D$ &    $-\frac{1}{4}$ &    $-\frac{1}{4}$ &    $-\frac{1}{4}$ &    $-\frac{1}{4}$    &    &            &      &     $\frac{y_{b}}{2} $ & \\
          \hline
          $Q_{5}$ &        &       &   &  &         &   $-\frac{1}{2}$    &      &     $\frac{y_{b}}{2} $  & \\
          \hline
          $Q_{7}$ &        &       &   &    &   $\frac{1}{2}$     &      &     $\frac{y_{t}}{2}$ &     & \\
          \hline
          $T_{1}$ &    $-\frac{1}{16}$ &  $-\frac{3}{16}$ &   $-\frac{1}{16}$ &  $-\frac{3}{16}$    &          &      &    $\frac{y_{t}}{4}$ &   $\frac{y_{b}}{8} $  &\\
          \hline
         $T_{2}$ &   $-\frac{1}{16}$ &   $\frac{3}{16}$ &  $-\frac{1}{16}$ &   $\frac{3}{16}$    &         &      &   $\frac{y_{t}}{8}$ &    $\frac{y_{b}}{4} $ &\\
         \hline
              $T$ &    &   &  $-\frac{1}{2} \frac{M_{T}^{2}}{v^{2}}$ &   $ \frac{1}{2} \frac{M_{T}^{2}}{ v^{2}}$    &    &       &   $y_{t} \frac{ M_{T}^{2}}{v^{2}}$ &  &\\
         \hline
\end{tabular}
\end{center}
} 
\caption{\it Operators generated at tree level by the single-field extensions 
listed in the first column.  Each extension depends on a single coupling (see Table~\ref{table:fields})
as well as a new physics mass-scale $M$.  The coefficients of the operators are 
each proportional to the squares of the corresponding coupling $\lambda$ 
by the corresponding entry in the Table and divided by $M^2$.  $y_{t}$, $y_{b}$ and $y_{\tau}$ denote the top, bottom and tau Yukawa couplings respectively, $v$ denotes the electroweak scale and $\alpha_{s}$ denotes the strong coupling.}
\label{tab:single-field_models}
\end{table}

\begin{table}[h]
\begin{center}
\begin{tabular}{|c|c|c|c|c|c| }
\hline \hline
\lgr Model &    $C_{HD}$ &      $C_{H \Box}$ &     $C_{\tau H}$ &     $C_{tH}$ &     $C_{bH}$\\
\hline \hline
$B$& $-2 a^{2}$ &  $-\frac{1}{2} (a^{2} - b^{2})$ & $-ab y_{\tau}$ & $-ab y_{t}$ & $-ab y_{b}$ \\
\hline
$W$& $\frac{1}{2} b^{2}$ &  $-\frac{1}{8} (3 a^{2} + b^{2})$ & $-\frac{1}{4} y_{\tau} (a + b)^{2}$ & $-\frac{1}{4} y_{t} (a+b)^{2} $ & $-\frac{1}{4} y_{b} (a+b)^{2}$ \\
\hline \hline
\lgr Model &         $C_{tH}$ &     $C_{bH}$ &     $C_{Ht}$ & $C_{HG}$ &     \\
\hline \hline
$TB$& $\frac{M_{TB}^{2}}{v^{2}} y_{t} a^{2}$ &  $\frac{M_{TB}^{2}}{v^{2}}  y_{b} b^{2}$ & $-\frac{M_{TB}^{2}}{v^{2}} a^{2}$ &  $-\frac{M_{TB}^{2}}{v^{2}}  \frac{\alpha_{s} (0.65)  }{8  \pi} b^{2}$ &  \\
\hline
\end{tabular}
\end{center}
\caption{\it Operator coefficients generated by the tree-level single-field models 
 $B$, $W$ and $TB$, which each depend on two couplings $a$ and $b$, with
 $y_{t}$, $y_{b}$ and $y_{\tau}$ denoting the top, bottom and tau Yukawa couplings respectively, $v$ denoting the electroweak scale and $\alpha_{s}$ denoting the strong coupling. The coefficients of all operators are proportional to the corresponding entries
in the Table and divided by $M^2$. \label{table:doubleop}}
\end{table}

\begin{figure}[t!] 
\centering
\includegraphics[width=0.98\textwidth]{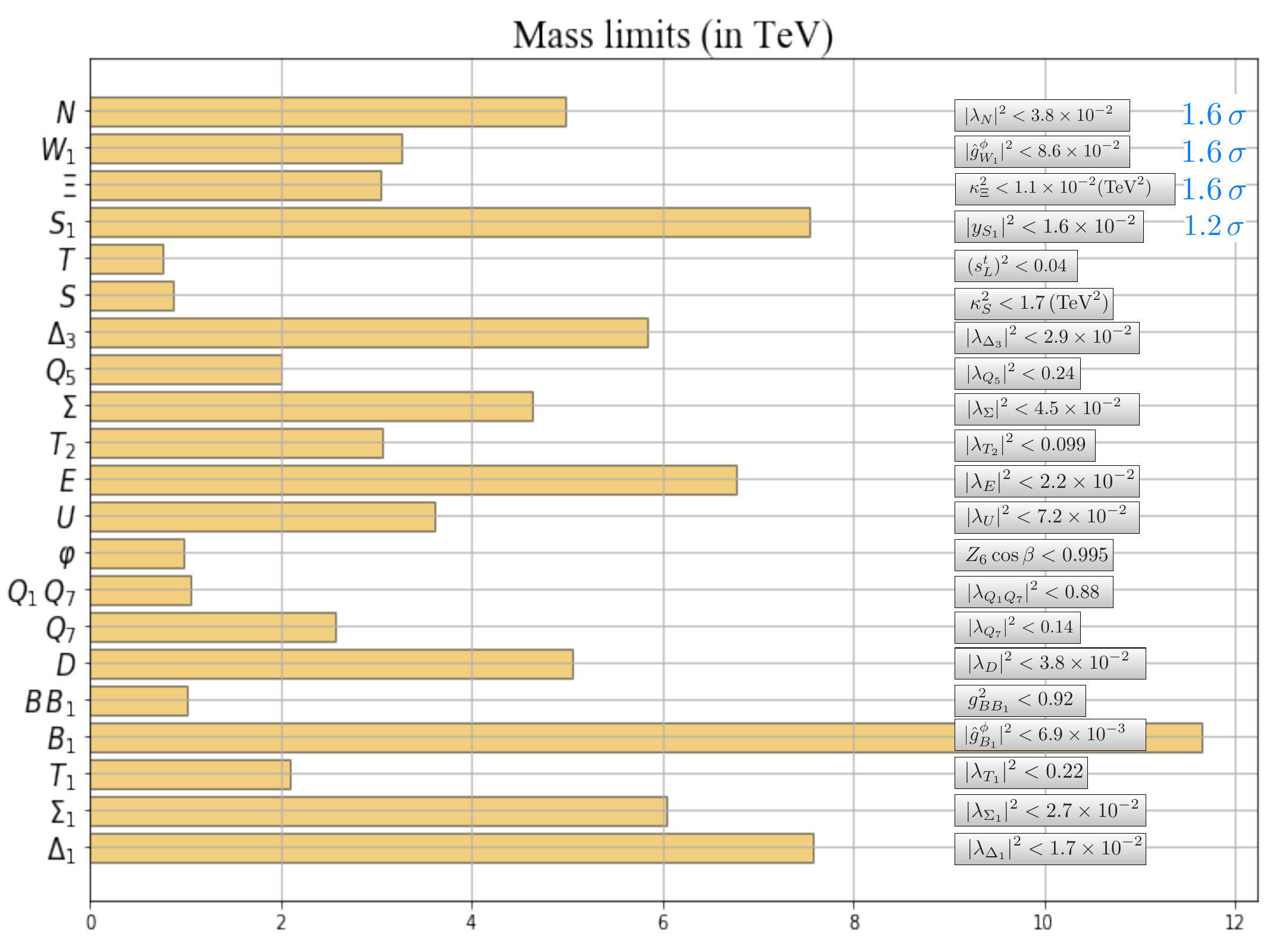}
\caption{\it The horizontal bars show the mass limits (in TeV) at the 95\% CL for the models described in Table~\ref{table:fields},
setting the corresponding couplings to unity. The coupling limits obtained when setting the mass to 1 TeV are shown in grey boxes. We also note in light blue the pulls that exceed 1-$\sigma$.}
\label{fig:1Dlimits}
\end{figure}

We show in Fig.~\ref{fig:1Dlimits} the results from 
our global fit for all of the one-parameter single-field extensions of 
the SM.  In each of these models we constrain a positive quantity: $|\lambda|^{2}$. The constraints in Fig.~\ref{fig:1Dlimits} are found using the numerical MCMC fitter described in Appendix~\ref{app:nestedsampling}.  This method allows us to incorporate the constraint $|\lambda|^{2}>0$ as a Heaviside prior $\pi(|\lambda|^{2}<0)=0$. The 2-$\sigma$ constraints on the mass scales
in TeV units, assuming that the corresponding couplings
are set to unity, are shown as horizontal bars~\footnote{In the case of the $T$ (vector-like quark) model, the mass limit has been obtained using the relation $s^t_L \simeq \lambda v/\sqrt{2} M_T$ and setting $\lambda=1$~\cite{Dawson:2020oco}.}.  We note
that most of these limits exceed 1~TeV for unit coupling and do not depend on kinematic distributions probing this region,
in which case the SMEFT approach is self-consistent. The SU(2)-singlet VLQ top-partner model ($T$) and $S$ are the most poorly constrained, with $m_{S,T}>$900, 770 GeV.  
We also show in grey boxes the corresponding
bounds on the squared couplings, assuming a mass scale of 1~TeV. Most of the bounds are $< 1$, justifying a tree-level treatment.
We also list all the pulls that exceed 1-$\sigma$, 
which is significant.

We can compare the mass limits for these models with the naive scale limits shown in Fig.~\ref{fig:all_fit}.  In a model-independent SMEFT analysis, one  allows all the EFT operators to vary simultaneously. On the other hand, in specific models not all EFT coefficients are generated and those that do appear are related to each other in such a way that the number of free parameters of the model is matched to the number of independent EFT operators generated by the model.
As an example, we discuss the following set of single-parameter models: 
\begin{center}
$\Sigma$, $\Sigma_1$, $N$ and $E$ \, , 
\end{center}
which all span the same types of SMEFT operators. They are all  characterized by non-zero values for the following
set of operators involving electroweak precision lepton observables:
\begin{center}
$C_{H\ell}^{1,3} \neq 0$ \, ,
\end{center}
whereas the other operator coefficients are zero, or very mildly constrained (e.g., $C_{\tau H}$, which is $\propto y_{\tau}$). Interpreted in terms of these models, the global SMEFT fit
leads to mass limits  of the order of 5 TeV for unit couplings, or corresponding coupling limits of ${\cal O} (10^{-1})$ for TeV resonances.  Each of these particles generates a different relation between $C_{H\ell}^{1}$ and $C_{H\ell}^{3}$, leading to slightly different limits, e.g., for the model with a new neutral fermion  $N$ one expects $C_{H\ell}^{1}=-C_{H\ell}^{3}=|\lambda_N|^2/4 m_N^2$.
The sizes of the mass limits justify the SMEFT approach,
and the small couplings for masses of a TeV justify
working at tree level.

The constraints on particles beyond the SM would be
weaker if their effects were not tree-level but loop-induced. 
These are typical of  extensions of the SM where 
couplings to new states have to be in pairs, as would be
the case if they carry a new conserved quantum number. 
We discuss in Section~\ref{sec:SUSYUV} one particularly interesting example with such loop-induced
effects, namely stops in an $R$-parity-conserving
supersymmetric model.

\subsection{Tree-level SMEFT patterns}\label{sec:UVpatterns}

As already commented, and displayed in Table~\ref{tab:single-field_models}, simple extensions of the SM exhibit specific patterns in the operators they generate. Many operators have vanishing coefficients and those that are non-zero are often related.  For example, model $B_{1}$ of Table \ref{table:fields} generates the bosonic operators $C_{H \Box}$, $C_{HD}$ and Yukawa operators such as $C_{tH}$.  These three operators are related by $C_{H \Box} = C_{tH} = -\frac{1}{2} C_{HD}$.  Similarly, model $W_{1}$ generates a pattern $C_{H \Box} = C_{tH} = \frac{1}{2} C_{HD}$.  
Motivated by these patterns, we study the results of our fit in four subspaces of the SMEFT:

\begin{center}
Boson-specific: ($ C_{HD}, \, C_{H\Box}, \, C_{tH}$) \, , \\
     
Lepton-specific: ($ C_{He}, \, C_{H\ell}^{(1,3)}, \, C_{\ell\ell}$)  \, ,   \\          

Quark-specific: ($ C_{Hu}, C_{Hd}, \, C_{Hq}^{(1,3)}, \, C_{tH}$) \, , \\  

Top-specific: (($C_{Hq}^{(1)})_{33}$, $(C_{Hq}^{(3)})_{33}$, $C_{HG}$, $C_{bH}$, $C_{tH}, C_{Ht}$) \, .   

\end{center}

\begin{figure}[t!] 
\centering
\includegraphics[width=0.84\textwidth]{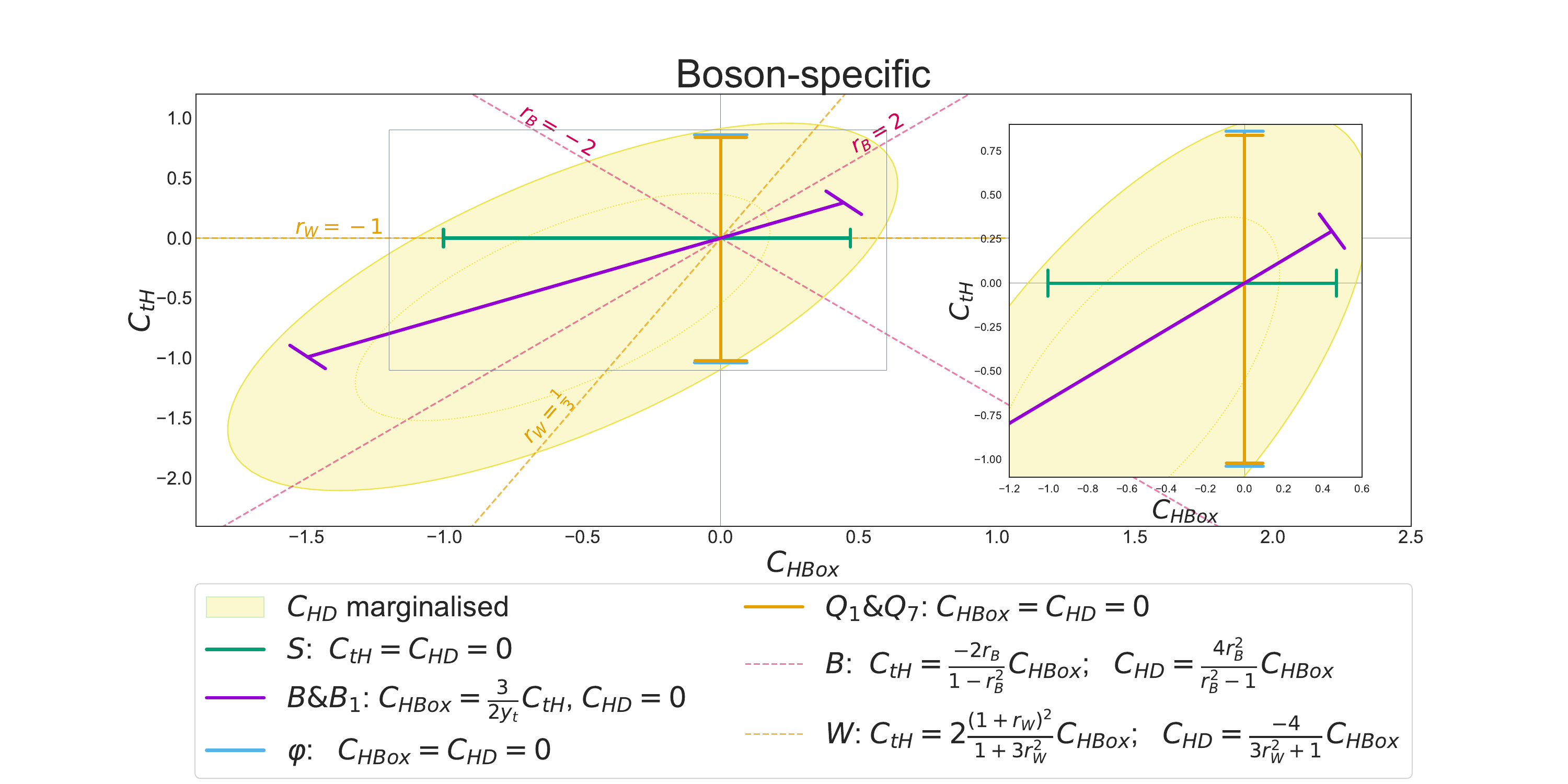}
\includegraphics[width=0.84\textwidth]{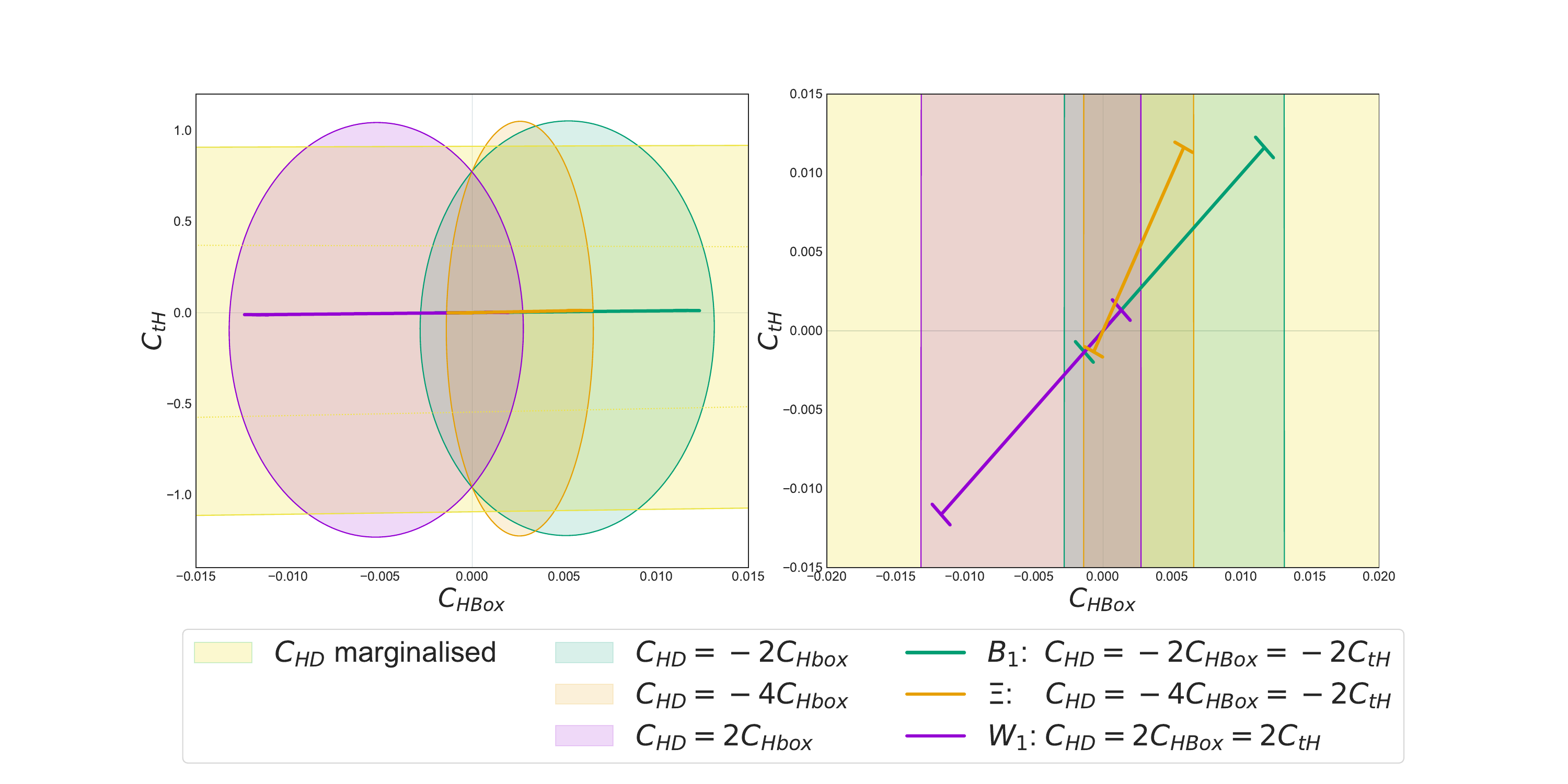}
\includegraphics[width=0.84\textwidth]{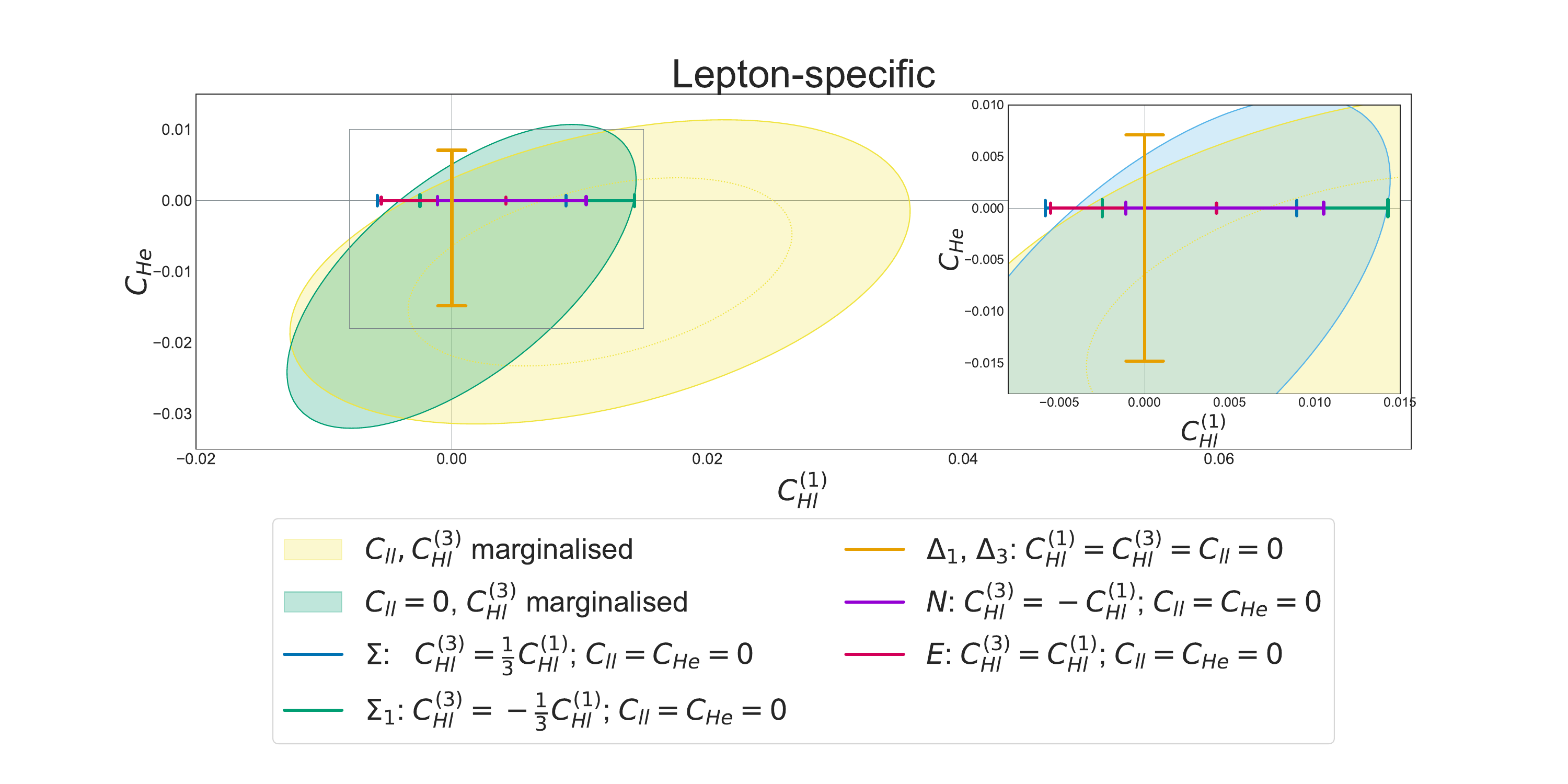}
\caption{\it Constraints at the 95\% CL on $(C_{H\Box}, C_{tH})$ in the \textit{boson-specific} scenario (upper two panels) and $(C_{Hl}^{(1)}, C_{He})$ in the \textit{lepton-specific scenario (bottom panel).}  The lines correspond to the 2-$\sigma$ limits obtained when we restrict the operators to the relations generated by integrating out the indicated single-field extensions of the SM.  }
\label{fig:2DBS}
\end{figure}

Results for the \textit{boson-specific scenario} 
are shown in the top two panels of Fig.~\ref{fig:2DBS}. We display 95\% CL contours in the $(C_{H \Box}, C_{tH})$ plane as solid contours, marginalising over $C_{HD}$, and setting all other operator coefficients to zero.  In the top panel we show how these operators are constrained in the cases of four specific UV models from Table~\ref{table:fields}: $S$, $\varphi$ [which may be derived from a 2-Higgs Doublet Model (2HDM)], $Q_{1}\&Q_{7}$ and $B\&B_{1}$, showing a detail in the inset.
As well as these 1-parameter models, 
we project two of the 2-parameter models shown in Table~\ref{table:doubleop} onto the $(C_{H \Box}, C_{tH})$
plane : $B$ and $W$.   These cases are vector bosons with couplings $\hat{g}_{ H }^{B}$ and $\hat{g}_{ H }^{W}$ to the Higgs doublet, also known as $Z'$ and $W'$ bosons, respectively. Model $B$ projects onto a line in  the $(C_{H \Box}, C_{tH})$ plane, illustrated in the top panel for $r_{B} = 2,-2$, where $r_{B} = \textrm{Re}(\hat{g}_{ H }^{B})/\textrm{Im}(\hat{g}_{ H }^{B})$. Model $W$ also
projects onto a line, illustrated here for $r_{W} = -1, \frac{1}{3}$.  While the slope of the line generated by 
model $B$ is free to take any value, the line generated by model $W$ is constrained to lie within the wedge bounded by the $r_{W} = -1, \frac{1}{3}$ lines shown.

In the middle panels of Fig.~\ref{fig:2DBS} we zoom in on the yellow ellipse
shown in the top panel, so as to study the constraints in the $(C_{H \Box}, C_{tH})$ plane when $C_{H \Box} \propto C_{HD}$.  This results in tighter constraints on $C_{H \Box}$ compared to when $C_{HD}$ is treated as an independent parameter, as shown by the small ellipses in the left panel.  The constants of proportionality $(-2,-4,2)$ are those found in the patterns generated by models $B_{1}$, $\Xi$ and $W_{1}$. Zooming in
further in the right panel, these ellipses are squashed into near-vertical parallel lines showing the constraints on $C_{H \Box}$ and $C_{tH}$ in the case of each of these 1-parameter models.

The \textit{lepton-specific scenario} is shown in the bottom panel of Fig.~\ref{fig:2DBS}, where the coefficients of the operators $C_{ll}$, $C_{Hl}^{(1)}$, $C_{Hl}^{(3)}$ and $C_{He}$ are studied.  The ellipses show how the data constrain $C_{Hl}^{(1)}$ and $C_{He}$, marginalising over $C_{Hl}^{(3)}$.  In the yellow ellipse we also marginalise over $C_{ll}$, allowing for the nonzero values of $C_{ll}$.  The constraints shrink when $C_{ll}=0$, as in models $\Sigma$, $\Sigma_{1}$, $N$, $E$, $\Delta_{1}$ and $\Delta_{3}$~\footnote{See also Ref.~\cite{Crivellin:2020ebi} for a more detailed fit to these six vector-like lepton models.}.  The inset plot demonstrates how $C_{Hl}^{(1)}$ and $C_{He}$ are constrained when we restrict these operators to the patterns associated with each of these 1-parameter models.  

\begin{figure}[t!] 
\centering
\includegraphics[width=0.85\textwidth]{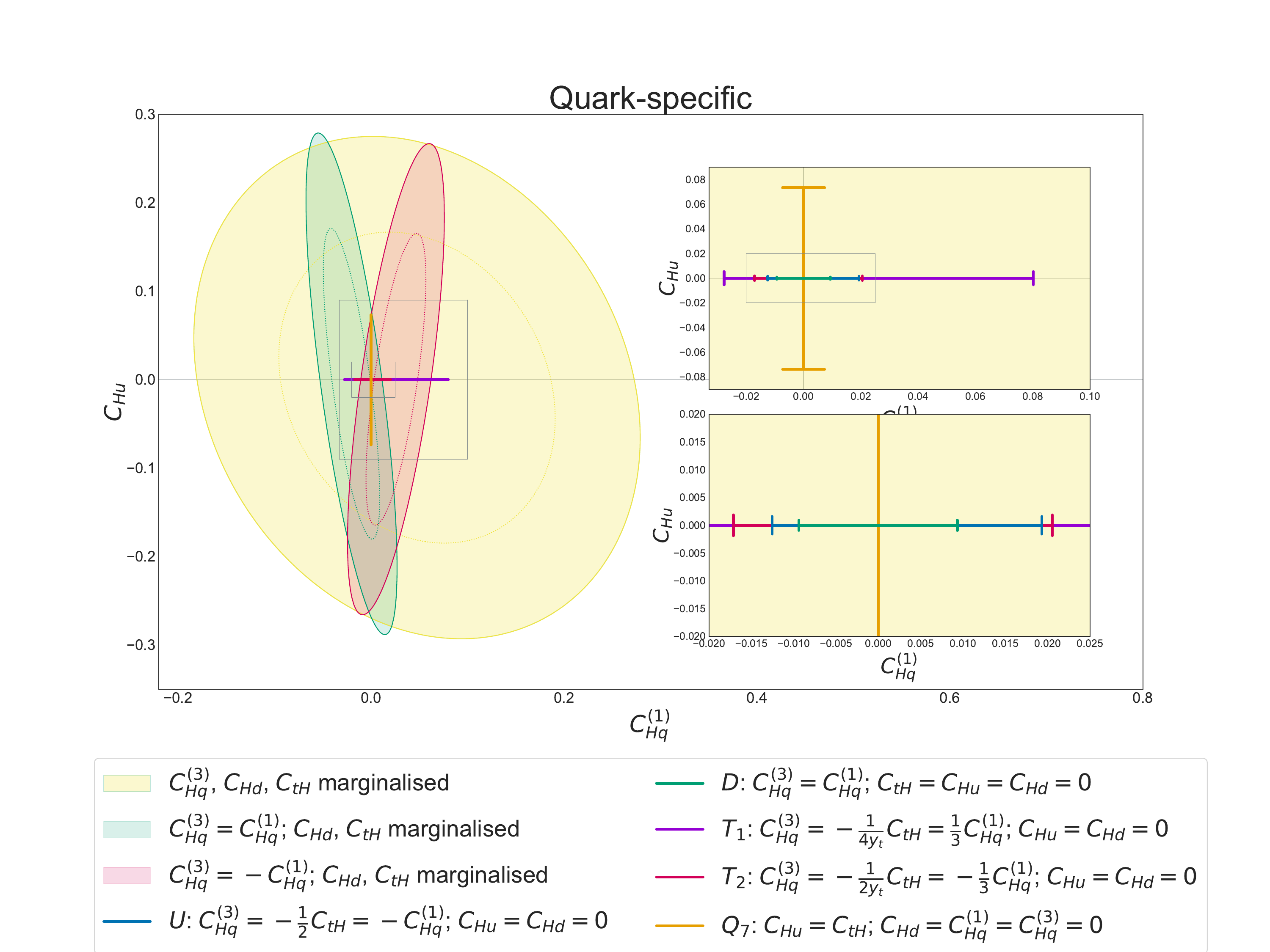}
\includegraphics[width=0.85\textwidth]{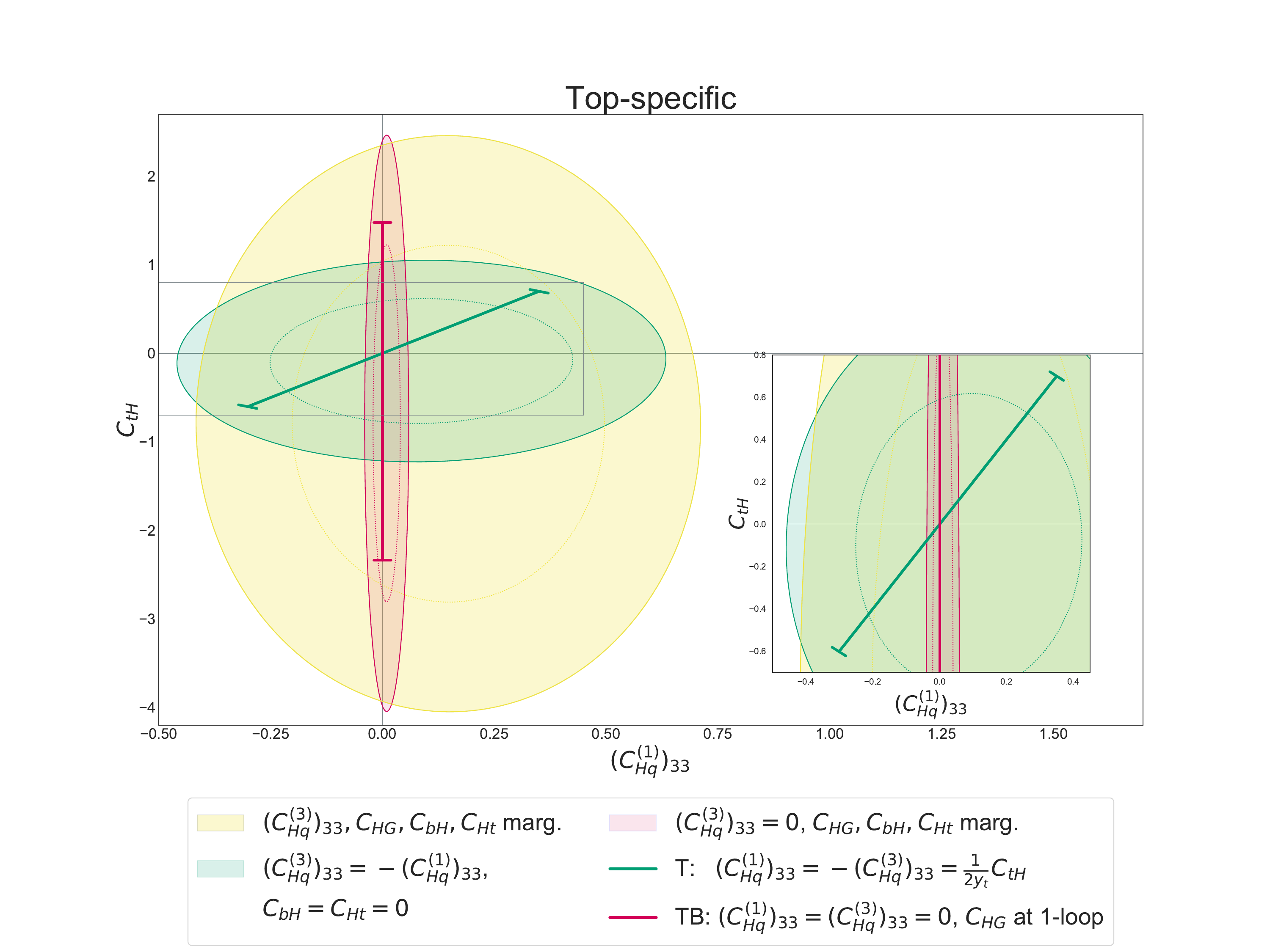}
\caption{\it Constraints at the 95\% CL on $(C_{Hq}^{(1)}, C_{Hu})$ in the \textit{quark-specific} scenario and $((C_{Hq}^{(1)})_{33}, C_{tH})$ in the \textit{top-specific scenario.}  The lines correspond to the $2 \sigma$ limits obtained when the operator coefficients are restricted to the relations generated by integrating out single-field extensions of the SM.}
\label{fig:2DQS}
\end{figure}

In Fig.~\ref{fig:2DQS} we turn to quark-Higgs interactions. The flavour-universal \textit{quark-specific scenario} is shown in the upper panel, where we constrain $(C_{Hq}^{(1)}, C_{Hu})$, marginalising over the remaining quark-Higgs operators $C_{Hq}^{(3)}$, $C_{Hd}$ and $C_{tH}$.  Models that generate these operators often lead to the pattern $C_{Hq}^{(1)} \propto C_{Hq}^{(3)}$.  The green and red ellipses show how such patterns narrow the constraints on $C_{Hu}$ and $C_{Hq}^{(1)}$ for two examples: $C_{Hq}^{(3)} =  \pm C_{Hq}^{(1)}$.  Specialising to the 1-parameter models $U$, $D$, $T_{1}$, $T_{2}$ and $Q_{7}$ further restricts the operators, leading to the 1-dimensional constraints shown in the inset plots.

Finally, the lower panel of Fig.~\ref{fig:2DQS} considers the flavour-non-universal \textit{top-specific scenario},  where we consider the operators $(C_{Hq}^{(1)})_{33}$, $(C_{Hq}^{(3)})_{33}$, $C_{HG}$, $C_{bH}$, $C_{tH}$ and $C_{Ht}$.  These operators are generated by the vector-like quark models $T$ and $TB$  with couplings to the third generation quarks only.  Integrating out the SU(2)$_{L}$ singlet $T$ generates the pattern $(C_{Hq}^{(3)})_{33} =-(C_{Hq}^{(1)})_{33}$, $C_{bH} = C_{Ht} = 0$.  The green ellipse demonstrates how this pattern tightens the constraints on $C_{tH}$.  In contrast, the SU(2)$_{L}$ doublet $TB$ does not generate the $(C_{Hq}^{(3)})_{33}$ operator.  Setting $(C_{Hq}^{(3)})_{33}=0$ results in much narrower constraints in the $(C_{Hq}^{(1)})_{33}$ direction, as shown by the red ellipse.

These patterns, and the results shown in Figs.~\ref{fig:2DBS} and \ref{fig:2DQS}, may be considered as more general explorations of the model parameter space than in the two previous Sections. Readers exploring UV completions who are searching for the indirect LHC and LEP constraints on their models can match their scenario to the allowed ellipses in these figures. For example, models linked to neutrino physics could lead to the SMEFT pattern we have denoted  as {\it lepton-specific}, whereas models with various additional scalars and gauge bosons would be contained among the  {\it boson-specific scenarios},
and models with additional coloured particles could be included among the {\it quark-specific scenarios}. 

\subsection{$R$-parity-conserving stop squarks at the 1-loop level}
\label{sec:SUSYUV}

A particularly interesting loop-induced modification of 
the SM Lagrangian is $R$-parity-conserving supersymmetry 
with a light stop sector. Whereas the discussion in the
previous Section of single-field tree-level models 
was motivated by simplicity, this scenario is motivated 
by the naturalness of the hierarchy between the electroweak scale and that of gravity or grand unification. A complete
one-loop analysis of the light-stop scenario and a comparison with the SMEFT analysis was given in~\cite{Drozd:2015kva}.

\begin{figure}[t!] 
\centering
\includegraphics[width=0.45\textwidth]{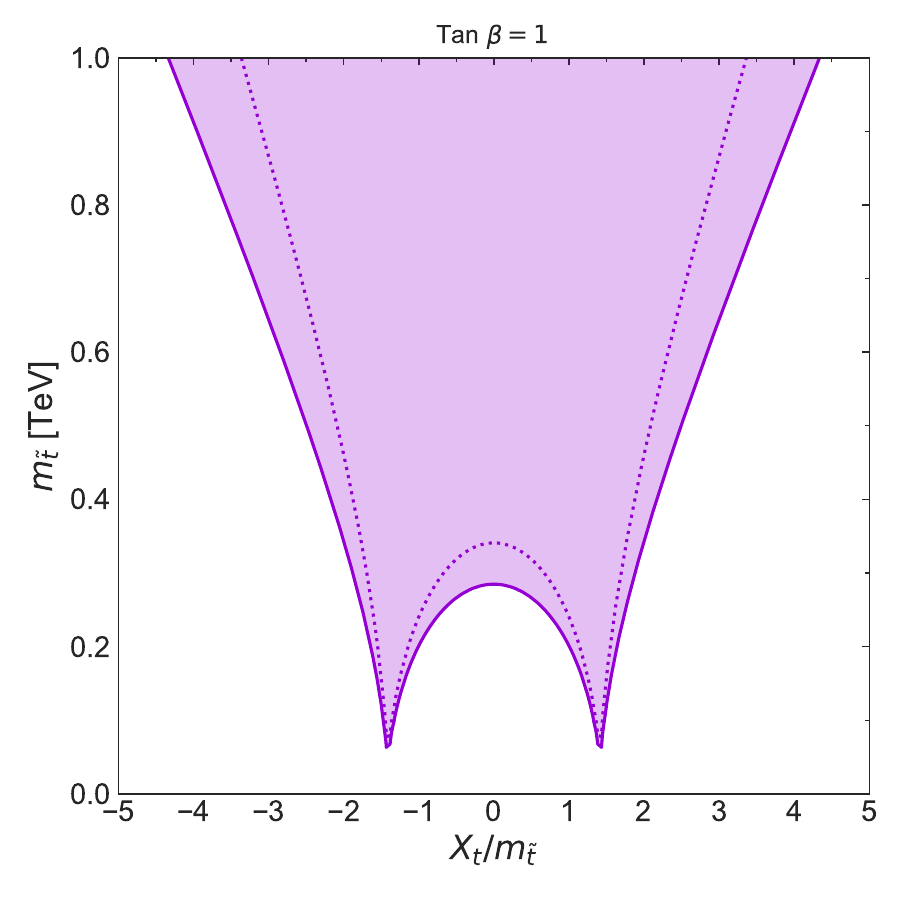}
\includegraphics[width=0.45\textwidth]{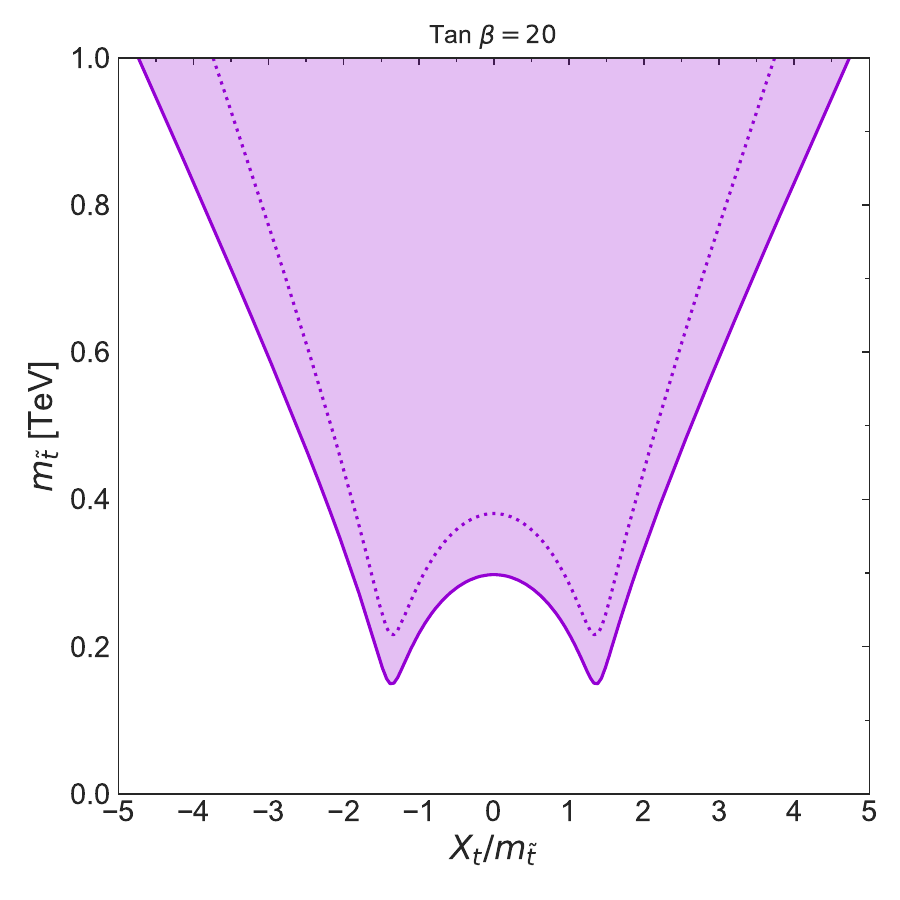}
\caption{\it Limits from the global fit in the stop parameter plane, ($\frac{X_{t}}{m_{\tilde{t}}}, m_{\tilde{t}}$,). The two panels correspond to the low and high  $\tan \beta$ choices, 1 and 20 respectively.}
\label{fig:MSSM}
\end{figure}

In presenting our results, we follow \cite{Henning:2014gca}, assuming a common  diagonal mass term $m_{\tilde{t}}$ and denoting the stop mixing parameter by $X_{t}$.  The constraints on degenerate stops are dominated by measurements of the $H \rightarrow gg$ and $H \rightarrow \gamma \gamma$ couplings, which constrain the dimension-6 operators $C_{HG}$, $C_{HB}$, $C_{HW}$ and $C_{HWB}$.  These constrain the stop parameters $m_{\tilde{t}}$ and $X_{t}$ through the following relations:
\begin{equation}
\label{stoperators}
\begin{split}
C_{HG} &= \frac{g_{s}^{2}}{12} \frac{h_{t}^{2}}{(4 \pi)^{2}}  \Big[ (1 + \frac{1}{12} \frac{c_{2 \beta} g^{'2}}{ h_{t}^{2}}) - \frac{1}{2} \frac{X_{t}^{2}}{m_{\tilde{t}}^{2}} \Big] \, ,\\
C_{HB} &=  \frac{17 g^{'2}}{144} \frac{h_{t}^{2}}{(4 \pi)^{2}}  \Big[ (1 + \frac{31}{102} \frac{c_{2 \beta} g^{' 2}}{ h_{t}^{2}}) - \frac{38}{85} \frac{X_{t}^{2}}{m_{\tilde{t}}^{2}} \Big] \, ,\\
C_{HW} &=  \frac{g^{2} }{16} \frac{h_{t}^{2}}{(4 \pi)^{2}}\Big[ (1 - \frac{1}{6} \frac{c_{2 \beta} g^{' 2}}{ h_{t}^{2}}) - \frac{2}{5} \frac{X_{t}^{2}}{m_{\tilde{t}}^{2}} \Big] \, ,\\
C_{HWB} &= - \frac{g g'}{24} \frac{h_{t}^{2}}{(4 \pi)^{2}}  \Big[ (1 + \frac{1}{2} \frac{c_{2 \beta} g^{2}}{ h_{t}^{2}}) - \frac{4}{5} \frac{X_{t}^{2}}{m_{\tilde{t}}^{2}} \Big] \, ,
\end{split}
\end{equation}
where $h_{t} \equiv \frac{m_{t}}{v}$, $m_{t}$ denotes the top mass and $\beta$ is related to the ratio of vacuum expectation values: $\tan \beta = \frac{\langle H_{u} \rangle}{\langle H_{d} \rangle}$.  We calculate the constraints in the $(\frac{X_{t}}{m_{\tilde{t}}}, m_{\tilde{t}})$ plane for the
representative values $\tan \beta = 1$ and $20$,
which are shown in the left and right panels of Fig.~\ref{fig:MSSM}, respectively.

One sees in both panels of Fig.~\ref{fig:MSSM} that current LHC data 
constrain the stop mass scale to $\gtrsim 300$~GeV, except for
$|\frac{X_{t}}{m_{\tilde{t}}}| \sim 1.5$~\cite{Espinosa:2012in}, where partial cancellations reduce the sensitivity to the stop mass scale below 200 GeV.
In these regions the SMEFT analysis gives only qualitative
results. These blind directions could be eliminated with future measurements of the $H$+jet differential distribution~\cite{Banfi:2018pki}.

\subsection{Survey of combinations of multiple operators}
\label{sec:multiple}

\begin{figure}[t!] 
\centering
\includegraphics[width=\textwidth]{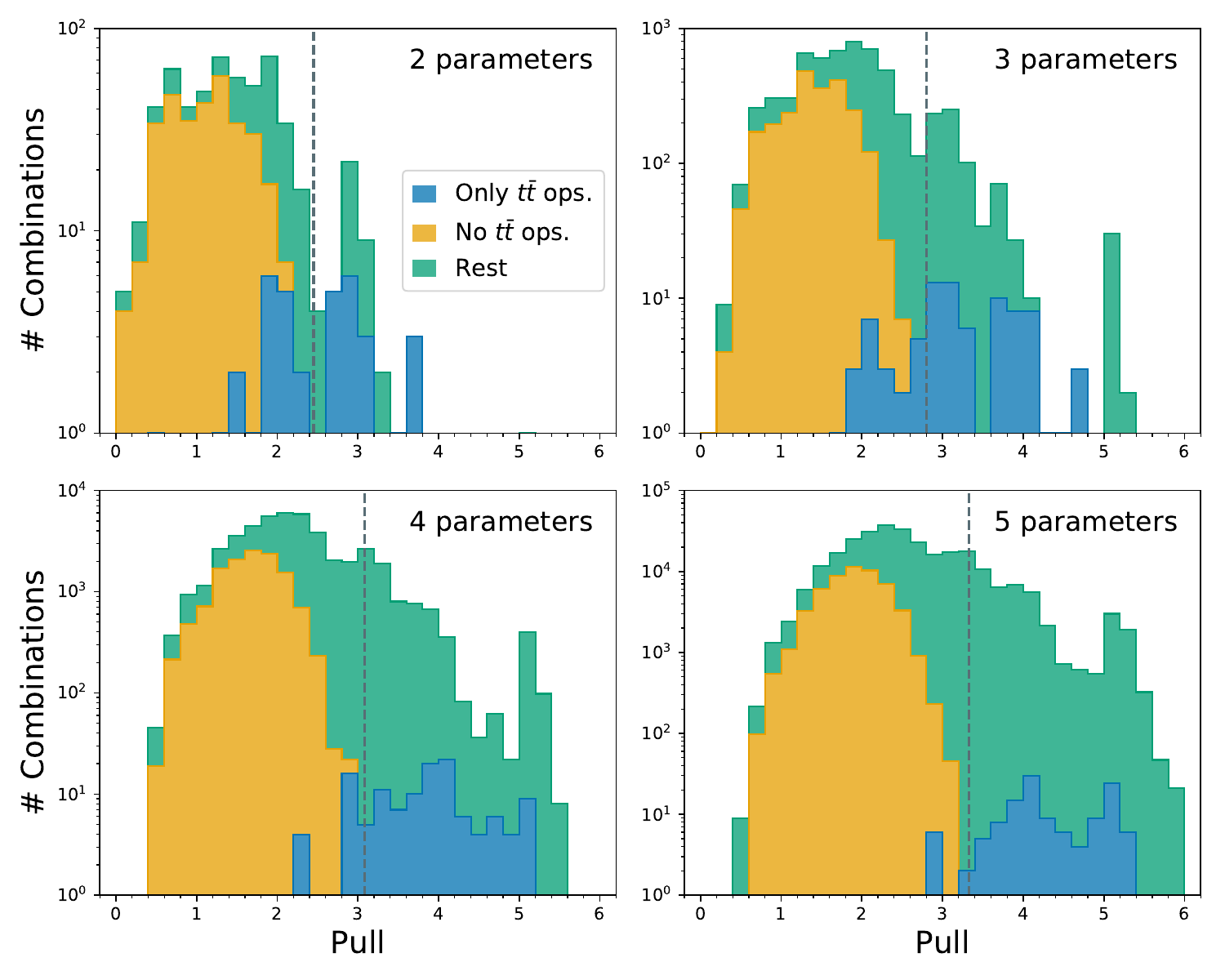}
\caption{\label{fig:pull_dist}
\it Stacked histograms of the distribution of pulls obtained in fits to 2 (upper left), 3 (upper right), 4 (lower left) and 5 (lower right) parameter subsets. The subsets have been split into three categories: those that include only operators that affect $t\bar{t}$ production (blue), those that do not include operators that affect $t\bar{t}$ production (orange), and the rest (green). The dashed vertical lines mark the expected 95\% ranges for the pull distributions.
}
\end{figure}

In general, new physics beyond the Standard Model could be expected to
contribute to the SMEFT via exchanges of more than just a single massive
particle, just as, e.g., $W$ and $Z$ exchanges both contribute to the
Fermi 4-fermion EFT of the weak interactions, and various mesons including
vectors $\rho$ and scalars $\sigma$ contribute to the low-energy pionic
EFT of QCD. Another example is provided by supersymmetry, where there 
might be a pair of relatively light stops, which contribute to four
different dimension-6 operator coefficients, as discussed in the
previous subsection. 

With this motivation, we have surveyed
all fits with contributions from any combination of two, three, four or five dimension-6 operators,
namely 561, 5984, 46736 and 278256 combinations, respectively. For
each combination $\{{\cal O}_i\}$, we calculate the 
pull that the corresponding fit exerts, given by:
\begin{equation}
\label{pulls}
    P \; \equiv \; \sqrt{\chi^2_{\rm SM} - \chi^2_{\{{\cal O}_i\}}} \, .
\end{equation}
Calculations of $P$ for all of these combinations is possible only
because, in the linear treatment that we have adopted in this paper,
the calculations of the $\chi^2_{\{{\cal O}_i\}}$ are computationally undemanding.

Fig.~\ref{fig:pull_dist} displays stacked histograms of the 
distributions of the pulls $P$ obtained in fits to combinations
of 2 (upper left), 3 (upper right), 4 (lower left) and 5 (lower right)
operators $\{{\cal O}_i\}$. In each panel, the blue histogram is for
combinations that include only operators that affect $t \bar t$ production,
see Fig.~\ref{fig:top_asym_impact},
the orange histogram is for combinations that do not include any of these
operators, and the green histogram is for the remaining combinations.

\begin{figure}[t!] 
\centering
\includegraphics[width=\textwidth]{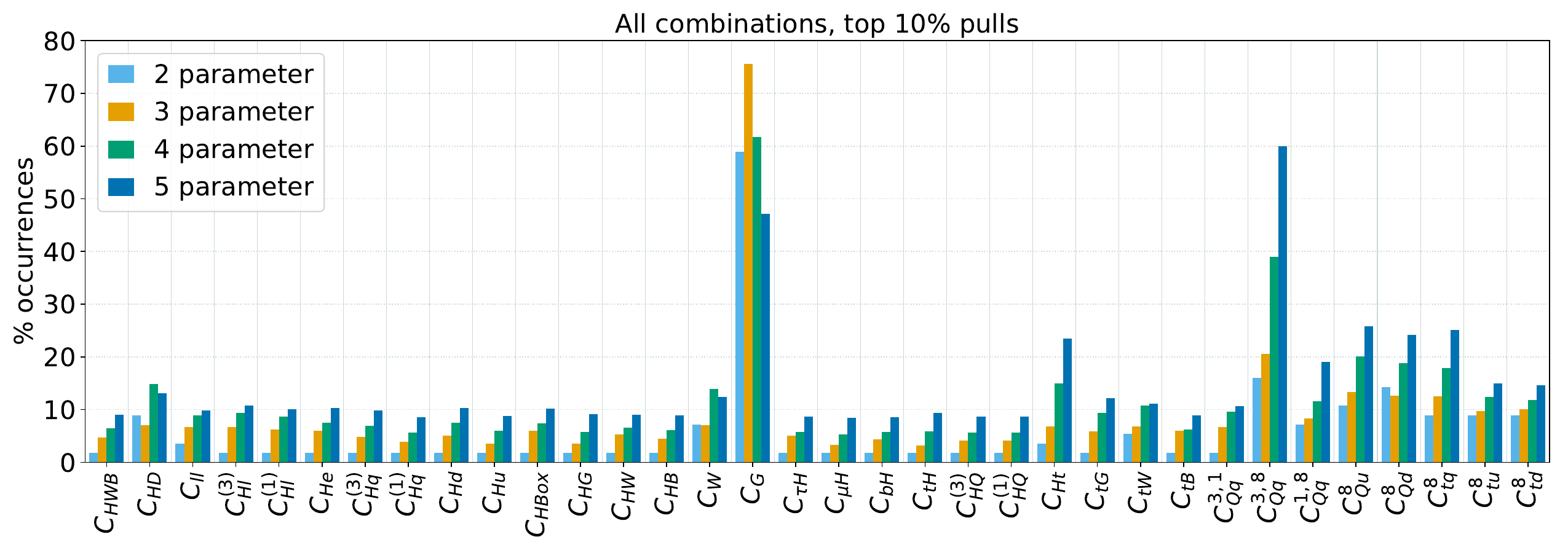}
\caption{\label{fig:pull_dist2}
\it The percentage occurrence of a given operator in the combinations ranked in the top 10\% of pulls found in fits to 2 (light blue), 3 (orange), 4 (green) and 5 (blue) parameter subsets.
}
\end{figure}

In all panels of Fig.~\ref{fig:pull_dist} we see that the blue histograms
for combinations that include operators affecting $t \bar t$ production
exhibit peaks at $P > 2$ that move to $> 4$ for combinations of 5
operators. On the other hand, the orange histograms for combinations of
operators that do not affect $t \bar t$ production are peaked at lower
values of $P \lesssim 2$ and do not have long tails extending to
large values of $P$, while the peaks of the green histograms are
intermediate. The vertical lines in the panels of Fig.~\ref{fig:pull_dist} mark
the 95\% ranges expected for Gaussian distributions of the pulls in the cases of 2, 3, 4 and 5
operators. We find that the percentage of combinations above the 95\% marks range from 9\% (2 operators) to 14\% (5 operators) of the total number of combinations: more than expected for Gaussian distributions, but not excessive. 
Neither the orange nor the green
histograms provide any indication of a significant deviation from the SM
that can be described by any combination of 2, 3, 4 or 5 operators.

Fig.~\ref{fig:pull_dist2} shows how often any given operator 
appears in the combinations whose total pulls $P$ ranked among the 
top 10\% of those obtained in fits with combinations of 2 (light blue), 
3 (orange), 4 (green) and 5 (dark blue) operators. The operators affecting
$t \bar t$ production generally appear more often among these top 10\%
combinations, particularly ${\cal O}_G$. The prevalence of
these operators in the high-pull combinations is due to the 
relatively poor quality of the global linear fit
in the top sector that we discussed in Section~\ref{sec:top}.~\footnote{As discussed there,
this issue might be mitigated by including quadratic terms in the SMEFT expansion,
but an analysis of this possibility lies beyond the scope of the present analysis.} Overall, we see that 
${\cal O}_G$ appears in more than half of the top 10\% of
combinations of $\le 4$ operators, falling to somewhat over 40\%
for combinations of 5 operators. For comparison, almost 60\% 
of the 5-operator combinations include ${\cal O}^{3, 8}_{Qq}$,
whereas this operator appears in smaller proportions of the
2-, 3- and 4-operator combinations.

This survey would suggest that the best prospects for BSM 
physics may be among the operators affecting $t \bar t$ production,
particularly ${\cal O}^{3, 8}_{Qq}$. However, we would emphasise that
the data in the top sector are currently the least precise, and
that there may be an issue with the $t \bar t$ mass distribution
near threshold, as discussed in Section~\ref{sec:top}. As the constraints on the
scales of the top operators are relatively weak, the resolution of this
issue may require including quadratic contributions of these operators,
which are not yet all available. We recall in this connection that an
analysis of $C_G$ including quadratic contributions to multijet
production found a strong constraint restricting $C_G$ to small values
below the sensitivity of our analysis.

There are many physics scenarios
that suggest the appearance of BSM physics in the top sector, such
as the light stop scenario discussed in Section~\ref{sec:SUSYUV}.
However, as we see in Eq.~(\ref{stoperators}), the operators 
$\{ {\cal O}_{HG}, {\cal O}_{HB}, {\cal O}_{HW}, {\cal O}_{HWB} \}$
that are most
constrained in this scenario contribute primarily in the EW and Higgs sectors,
rather than the top sector. We find a pull $P = 1.9$ for the combination
$\{ {\cal O}_{HG}, {\cal O}_{HB}, {\cal O}_{HW}, {\cal O}_{HWB} \}$,
which is typical for 4-operator combinations that do not include $t \bar t$
operators (the orange histogram in the lower left panel of 
Fig.~\ref{fig:pull_dist}), and the maximum pull when a fifth operator
is included is $P = 3.2$, which occurs in the combination with ${\cal O}_G$, 
and is typical of the green histogram
in the lower right panel of Fig.~\ref{fig:pull_dist}. So we find no hint
of light supersymmetry in the current data.

That said, this type of broad-brush survey of operator
combinations may be a useful way to help optimise the search for BSM
physics using the SMEFT in the future.

\section{Summary and conclusions}
\label{sec:conx}

In this paper we have introduced a new tool, {\tt Fitmaker}, to make a global analysis of
the available top, Higgs, diboson and electroweak data in the framework of the SMEFT with
dimension-6 operators included to linear order. We have presented results for fits
including each operator individually, and also when marginalising over all the other
operators. In each case, we have presented results in an SU(3)$^5$ flavour-symmetric
scenario and in an SU(2)$^2 \times$SU(3)$^3$ top-specific scenario. 
Our results are displayed in Fig.~\ref{fig:all_fit}, with numerical results for the
SU(2)$^2 \times$SU(3)$^3$ top-specific scenario presented in Table~\ref{tab:all_fit} below. We find $\chi^2/{\rm dof} = 0.94$ for our flavour-universal global fit and 
$0.81$ for our top-specific fit, to be compared with $\chi^2/{\rm dof} = 0.93$ and $0.91$, respectively, in the SM.
For $C=1$, the constraints on 
the scales $\Lambda$ of the coefficients of many operators contributing to
Higgs and electroweak measurements are individually ${\cal O}(10)$~TeV, but 
constraints in the top sector are currently less precise than the Higgs and 
electroweak data, falling to $\sim 100$~GeV in the case of the operator 
${\cal O}_{tB}$~\footnote{However, this operator is largely uncorrelated with the other
operators in our fit, as found in our principal component analysis in Section~\ref{sec:correlations}.}. This restricts the interpretation in many cases to strongly-coupled models, where we find that $\Lambda \gtrsim 1$ TeV for all operators when $C = (4\pi)^2$ at the perturbativity upper limit. 

We do not find any significant discrepancy with the SM. However, the
data in the top sector show a preference for a non-zero value for the coefficient $C_G$ 
of the triple-gluon operator ${\cal O}_G$, which is mirrored
by trends in the coefficients of other operators affecting $t \bar t$ measurements. This 
deviation can be traced back to the behaviour near threshold of the $t \bar t$ cross section,
and we await with interest future experimental measurements and developments in their
theoretical understanding. We note also that a fit to multijet data at quadratic order in
$C_G$ has constrained it to values so small that it would not contribute significantly to
the measurements we consider. However, we have not included this constraint in our linear
fit, for reasons of theoretical consistency.

We have presented the $34 \times 34$ correlation matrix
for the top-specific marginalised fit, grouping the 
operators into those affecting primarily electroweak precision observables, bosonic observables, Yukawa coupling measurements, top electroweak measurements and top-quark four-fermion operators.
We find that the most important correlations are between
operators in the electroweak, Higgs and Yukawa sectors, 
and between top electroweak and four-fermion operators. However,
there are also some notable correlations between top four-fermion
operators and bosonic operators, and between top electroweak and other electroweak operators. Overall, there are 24 instances of significant ($\ge 20\%$) correlations
between top operators and Yukawa, bosonic or electroweak operators,
confirming the relevance of making a combined analysis of all sectors.

Analyzing data to linear order in the SMEFT operator coefficients, 
the global $\chi^2$ function may be regarded as Gaussian, facilitating a principal 
component analysis, in which we diagonalise the $\chi^2$ matrix, identify 
its eigenvectors and determine their eigenvalues, which are displayed in Fig.~\ref{fig:eigenvectors}
and have the numerical values tabulated in Table~\ref{tab:eigenvectors}. The scale associated
with the best-constrained eigenvector is $> 20$~TeV, 
with Higgs and STXS measurements playing the most important roles. 
Three other eigenvectors have scales $> 10$~TeV,
with electroweak measurements also providing important constraints. The 
least-constrained eigenvector is essentially $\propto {\cal O}_{tB}$,
whose scale may be as low as $\sim 100$~GeV but is largely uncorrelated with
the other operators, and the scales of three other eigenvectors
may be $< 300$~GeV. Care must therefore be taken regarding the validity of the EFT along those poorly-constrained directions. 

We have analyzed the constraints our results provide on all the
single-field extensions of the SM that contribute to SMEFT operator
coefficients at the tree level. Normalising to unit couplings, the
lower limits on the corresponding BSM particle masses range between 
$> 1$~TeV to $> 10$~TeV. The largest pulls 
$P \; \equiv \; \sqrt{\chi^2_{\rm SM} - \chi^2_{\rm BSM}}$ are 1.6, and
hence not significant. In some instances, a particular BSM
particle may contribute to several operator coefficients, and we
have analyzed the constraints in boson-, lepton-, quark- and
top-specific subspaces of the SMEFT. We have also analyzed the 
constraints on low-mass stops, which contribute significantly 
to four operator coefficients at the one-loop level, finding 
they must weigh more than $\sim 300$~GeV when the stop mixing 
parameter $X_t = 0$.

Finally, we have surveyed the constraints on all possible 
2-, 3-, 4- and 5-operator combinations, assuming that the other
operators vanish. We find pulls that are insignificant for
combinations that do not contain operators affecting $t \bar t$
measurements, whereas the pulls for combinations of $t \bar t$
operators are larger. However, the same caveats apply to their
interpretation as to the discussion of the $t \bar t$ sector
above. The full pull distributions including all operator
combinations do not exhibit any significant features.

These examples indicate ways in which a global analysis of
all the current data may be used to obtain the broadest possible, unbiased view on 
the nature of possible BSM physics within the assumptions of the SMEFT framework. If any specific model or pattern of SMEFT
operators were to exert a significant pull, it would be a first indication of the
direction of new physics, which could be followed up with a more focused study.
However, the dataset we have included in our SMEFT analysis provides no significant indication of possible 
BSM physics: the only operator for which a non-zero coefficient is preferred is
${\cal O}_G$, and this preference is not very convincing. It is driven, in particular,
by the threshold behaviour of the $t \bar t$ production cross section, but an 
analysis of multijet data at quadratic order prefers much smaller values of $C_G$.
More $t \bar t$ data and theoretical understanding may be needed to resolve this
discrepancy.

The fact that our analysis is restricted to linear order in the dimension-6 SMEFT 
operator coefficients is clearly a limitation, but this a consequence of our 
self-imposed consistency requirement, and not a limitation of the fitter methodology.
Our {\tt Fitmaker} code could be applied equally well at quadratic order in the
dimension-6 SMEFT operator coefficients, but a consistent analysis to fourth
order in the new physics scale would require including also the contributions of
dimension-8 operators at linear order: see~\cite{Hays:2018zze} for a discussion of
their importance for Higgs measurements. There are many other extensions of our
analysis that could be tackled with {\tt Fitmaker}, including CP-violating effects,
flavour observables, RGE running, and higher-order perturbative QCD and electroweak effects with the SMEFT at NLO. 

The {\tt Fitmaker} code can be obtained from the account @kenmimasu at the following {\tt Gitlab} link: ~\href{https://gitlab.com/kenmimasu/fitrepo}{\faGithub}. Since it is built
in a modular fashion, it can readily be expanded by the user adding more data.

\acknowledgments
We thank Marcel Vos,  Maria Moreno Llacer and Victor and Marco Miralles for suggesting the inclusion of $t\bar{t}\gamma$. K.M. thanks L. Mantani, O. Mattelaer, E. Vryonidou and C. Zhang for valuable discussions and help in obtaining the predictions. M.M. thanks B. C. Allanach for useful discussions.
The work of J.E. was supported in part by the UK STFC via grants ST/P000258/1 
and ST/T000759/1, and in part by the Estonian Research Council via grant MOBTT5. M.M. is supported by the University of Cambridge Schiff Foundation studentship and in part by the UK STFC via grants ST/P000681/1 and ST/T000694/1. K.M. 
is also supported by the UK STFC via grant ST/T000759/1. V.S. acknowledges support 
from the UK STFC via Grant ST/L000504/1 and Spanish national grant FPA2017-85985-P. 
T.Y. is supported by a Branco Weiss Society in Science Fellowship and partially by 
the UK STFC via the grant ST/P000681/1.  

\newpage

\appendix

\section*{Tables of numerical fit results}
\label{app:table}

\begin{table}[h!]
\renewcommand{\arraystretch}{1.1}
\begin{adjustwidth}{-1in}{-1in}
\begin{center}
{\small
\begin{tabular}{|c||c|c|c||c|c|c|}
\hline
 & \multicolumn{3}{c||}{ Individual} & \multicolumn{3}{c|}{Marginalised} \\
 \hline
 SMEFT &  Best fit &   95\% CL &  Scale &  Best fit & 95\% CL &  Scale \\
 Coeff.   & [$\Lambda = 1$~TeV] & range & $\frac{\Lambda}{\sqrt{C}}$ [TeV] & [$\Lambda = 1$~TeV] & range & $\frac{\Lambda}{\sqrt{C}}$ [TeV] \\
\hline \hline
     $C_{HWB}$& 0.00 &[ -0.0043, +0.0026 ]& 17.0 & 0.18 &  [ -0.36, +0.73 ]  & 1.4  \tabularnewline\hline
      $C_{HD}$&-0.01 &[ -0.023, +0.0027 ] & 8.8  &-0.39 &  [ -1.6, +0.81 ]   & 0.91 \tabularnewline\hline
      $C_{ll}$& 0.01 & [ -0.005, +0.019 ] & 9.2  &-0.03 & [ -0.084, +0.02 ]  & 4.4  \tabularnewline\hline
$C_{Hl}^{(3)}$& 0.00 & [ -0.01, +0.003 ]  & 12.0 &-0.03 & [ -0.13, +0.055 ]  & 3.3  \tabularnewline\hline
$C_{Hl}^{(1)}$&0.00  &[ -0.0044, +0.013 ] & 11.0 & 0.11 &  [ -0.19, +0.41 ]  & 1.8  \tabularnewline\hline
      $C_{He}$& 0.00 &[ -0.015, +0.0071 ] & 9.6  & 0.19 &  [ -0.41, +0.79 ]  & 1.3  \tabularnewline\hline
$C_{Hq}^{(3)}$& 0.00 & [ -0.017, +0.012 ] & 8.3  &-0.05 & [ -0.11, +0.012 ]  & 4.1  \tabularnewline\hline
$C_{Hq}^{(1)}$& 0.02 &  [ -0.1, +0.14 ]   & 2.9  &-0.04 &  [ -0.27, +0.18 ]  & 2.1  \tabularnewline\hline
      $C_{Hd}$&-0.03 & [ -0.13, +0.071 ]  & 3.1  &-0.39 &  [ -0.91, +0.13 ]  & 1.4  \tabularnewline\hline
      $C_{Hu}$& 0.00 & [ -0.075, +0.073 ] & 3.7  &-0.19 &  [ -0.63, +0.25 ]  & 1.5  \tabularnewline\hline
    $C_{H\Box}$&-0.27 &   [ -1, +0.47 ]    & 1.2  & -0.9 &    [ -3, +1.2 ]    & 0.69 \tabularnewline\hline
      $C_{HG}$& 0.00 &[ -0.0034, +0.0032 ]& 17.0 & 0.00 &[ -0.014, +0.0086 ] & 9.4  \tabularnewline\hline
      $C_{HW}$& 0.00 & [ -0.012, +0.006 ] & 11.0 & 0.12 &  [ -0.38, +0.62 ]  & 1.4  \tabularnewline\hline
      $C_{HB}$& 0.00 &[ -0.0034, +0.002 ] & 19.0 & 0.07 &  [ -0.09, +0.22 ]  & 2.5  \tabularnewline\hline
       $C_{W}$& 0.18 & [ -0.071, +0.42 ]  & 2.0  & 0.15 &  [ -0.11, +0.4 ]   & 2.0  \tabularnewline\hline
       $C_{G}$&-0.46 &  [ -0.77, -0.14 ]  & 1.8  & -1.4 &  [ -2.2, -0.72 ]   & 1.2  \tabularnewline\hline
  $C_{\tau H}$& 0.01 & [ -0.015, +0.025 ] & 7.1  & 0.01 & [ -0.016, +0.028 ] & 6.7  \tabularnewline\hline
   $C_{\mu H}$& 0.00 &[ -0.0057, +0.005 ] & 14.0 & 0.00 &[ -0.0058, +0.005 ] & 14.0 \tabularnewline\hline
      $C_{bH}$&0.00  & [ -0.016, +0.024 ] & 7.1  & 0.01 & [ -0.034, +0.052 ] & 4.8  \tabularnewline\hline
      $C_{tH}$&-0.09 &   [ -1, +0.84 ]    & 1.0  & 1.5  &   [ -2.8, +5.7 ]   & 0.48 \tabularnewline\hline
$C_{HQ}^{(3)}$& 0.01 & [ -0.032, +0.048 ] & 5.0  & -0.1 &  [ -0.67, +0.46 ]  & 1.3  \tabularnewline\hline
$C_{HQ}^{(1)}$& 0.01 & [ -0.031, +0.049 ] & 5.0  &-0.01 &  [ -0.59, +0.58 ]  & 1.3  \tabularnewline\hline
      $C_{Ht}$& 0.87 &   [ -1.2, +2.9 ]   & 0.7  & 6.6  &    [ +2, +11 ]     & 0.47 \tabularnewline\hline
      $C_{tG}$&-0.01 &  [ -0.1, +0.086 ]  & 3.2  & 0.36 &  [ +0.12, +0.6 ]   & 2.0  \tabularnewline\hline
      $C_{tW}$& 0.19 &  [ -0.12, +0.51 ]  & 1.8  & 0.23 & [ -0.088, +0.55 ]  & 1.8  \tabularnewline\hline
      $C_{tB}$& -1.6 &   [ -4.5, +1.2 ]   & 0.59 & -1.4 &   [ -5.2, +2.5 ]   & 0.51 \tabularnewline\hline
$C_{Qq}^{3,1}$& 0.06 & [ -0.043, +0.16 ]  & 3.2  & 0.05 & [ -0.071, +0.17 ]  & 2.9  \tabularnewline\hline
$C_{Qq}^{3,8}$& -1.2 &  [ -2.4, +0.036 ]  & 0.91 & -6.8 &   [ -18, +4.5 ]    & 0.3  \tabularnewline\hline
$C_{Qq}^{1,8}$&-0.12 &  [ -0.56, +0.31 ]  & 1.5  &-0.65 &   [ -4.9, +3.6 ]   & 0.48 \tabularnewline\hline
  $C_{Qu}^{8}$& -0.6 &  [ -1.3, +0.06 ]   & 1.2  & 6.3  &   [ -2.5, +15 ]    & 0.34 \tabularnewline\hline
  $C_{Qd}^{8}$& -1.4 &  [ -2.9, +0.07 ]   & 0.83 & 1.8  &   [ -9.5, +13 ]    & 0.3  \tabularnewline\hline
  $C_{tq}^{8}$& -0.4 & [ -0.85, +0.059 ]  & 1.5  & -5.6 &   [ -13, +2.2 ]    & 0.36 \tabularnewline\hline
  $C_{tu}^{8}$&-0.45 &  [ -1.1, +0.23 ]   & 1.2  & 4.0  &    [ -11, +19 ]    & 0.26 \tabularnewline\hline
  $C_{td}^{8}$& -1.0 &  [ -2.5, +0.38 ]   & 0.83 &-0.42 &    [ -12, +11 ]    & 0.29 \tabularnewline\hline
\end{tabular}

}
\end{center}
\end{adjustwidth}
\caption{\it \label{tab:all_fit} Table of numerical results in Fig.~\ref{fig:all_fit} from the global fit
to the electroweak, diboson, Higgs and top data in the top-specific
SU(2)$^2\times$SU(3)$^3$ scenario.}
\end{table}

\newpage

\begin{table}[h!]
\renewcommand{\arraystretch}{1.2}
\begin{adjustwidth}{-1in}{-1in}
 \begin{center}
\begin{small}
\vspace{-5mm}

\begin{tabular}{|r||l|l|l|l|l|l|l|l|l|l|l|l|}
\hline
         $C_i$ &             EWPO & $\text{LEP}\,WW$ &         Run 1 SS &         Run 2 SS &             STXS & $\text{LHC}\,WW$ &             $WZ$ &            $Zjj$ &         $t\bar{t}$ & $W_{\text{hel.}}$ &             $tX$ &      $t\bar{t}V$ \tabularnewline\hline\hline
     $C_{HWB}$ &               51 &              $-$ &                7 &               14 &               28 &              $-$ &              $-$ &              $-$ &              $-$ &              $-$ &              $-$ &              $-$ \tabularnewline\hline
      $C_{HD}$ &              100 &              $-$ &              $-$ &              $-$ &              $-$ &              $-$ &              $-$ &              $-$ &              $-$ &              $-$ &              $-$ &              $-$ \tabularnewline\hline
      $C_{ll}$ &               99 &              $-$ &              $-$ &              $-$ &              $-$ &              $-$ &              $-$ &              $-$ &              $-$ &              $-$ &              $-$ &              $-$ \tabularnewline\hline
$C_{Hl}^{(3)}$ &               99 &              $-$ &              $-$ &              $-$ &              $-$ &              $-$ &              $-$ &              $-$ &              $-$ &              $-$ &              $-$ &              $-$ \tabularnewline\hline
$C_{Hl}^{(1)}$ &              100 &              $-$ &              $-$ &              $-$ &              $-$ &              $-$ &              $-$ &              $-$ &              $-$ &              $-$ &              $-$ &              $-$ \tabularnewline\hline
      $C_{He}$ &              100 &              $-$ &              $-$ &              $-$ &              $-$ &              $-$ &              $-$ &              $-$ &              $-$ &              $-$ &              $-$ &              $-$ \tabularnewline\hline
$C_{Hq}^{(3)}$ &               89 &                1 &              $-$ &              $-$ &                2 &              $-$ &                6 &              $-$ &              $-$ &              $-$ &              $-$ &              $-$ \tabularnewline\hline
$C_{Hq}^{(1)}$ &               99 &              $-$ &              $-$ &              $-$ &              $-$ &              $-$ &              $-$ &              $-$ &              $-$ &              $-$ &              $-$ &              $-$ \tabularnewline\hline
      $C_{Hd}$ &               99 &              $-$ &              $-$ &              $-$ &              $-$ &              $-$ &              $-$ &              $-$ &              $-$ &              $-$ &              $-$ &              $-$ \tabularnewline\hline
      $C_{Hu}$ &               98 &              $-$ &              $-$ &              $-$ &                1 &              $-$ &              $-$ &              $-$ &              $-$ &              $-$ &              $-$ &              $-$ \tabularnewline\hline
    $C_{H\Box}$ &              $-$ &              $-$ &               22 &               46 &               32 &              $-$ &              $-$ &              $-$ &              $-$ &              $-$ &              $-$ &              $-$ \tabularnewline\hline
      $C_{HG}$ &              $-$ &              $-$ &               22 &               42 &               36 &              $-$ &              $-$ &              $-$ &              $-$ &              $-$ &              $-$ &              $-$ \tabularnewline\hline
      $C_{HW}$ &              $-$ &              $-$ &               14 &               29 &               56 &              $-$ &              $-$ &              $-$ &              $-$ &              $-$ &              $-$ &              $-$ \tabularnewline\hline
      $C_{HB}$ &              $-$ &              $-$ &               14 &               29 &               57 &              $-$ &              $-$ &              $-$ &              $-$ &              $-$ &              $-$ &              $-$ \tabularnewline\hline
       $C_{W}$ &              $-$ &                3 &              $-$ &              $-$ &              $-$ &              $-$ &               13 &               84 &              $-$ &              $-$ &              $-$ &              $-$ \tabularnewline\hline
       $C_{G}$ &              $-$ &              $-$ &              $-$ &              $-$ &              $-$ &              $-$ &              $-$ &              $-$ &               43 &              $-$ &              $-$ &               56 \tabularnewline\hline
  $C_{\tau H}$ &              $-$ &              $-$ &               22 &               45 &               34 &              $-$ &              $-$ &              $-$ &              $-$ &              $-$ &              $-$ &              $-$ \tabularnewline\hline
   $C_{\mu H}$ &              $-$ &              $-$ &                5 &               95 &              $-$ &              $-$ &              $-$ &              $-$ &              $-$ &              $-$ &              $-$ &              $-$ \tabularnewline\hline
      $C_{bH}$ &              $-$ &              $-$ &               19 &               35 &               47 &              $-$ &              $-$ &              $-$ &              $-$ &              $-$ &              $-$ &              $-$ \tabularnewline\hline
      $C_{tH}$ &              $-$ &              $-$ &               21 &               45 &               34 &              $-$ &              $-$ &              $-$ &              $-$ &              $-$ &              $-$ &              $-$ \tabularnewline\hline
$C_{HQ}^{(3)}$ &               99 &              $-$ &              $-$ &              $-$ &              $-$ &              $-$ &              $-$ &              $-$ &              $-$ &              $-$ &              $-$ &              $-$ \tabularnewline\hline
$C_{HQ}^{(1)}$ &              100 &              $-$ &              $-$ &              $-$ &              $-$ &              $-$ &              $-$ &              $-$ &              $-$ &              $-$ &              $-$ &              $-$ \tabularnewline\hline
      $C_{Ht}$ &              $-$ &              $-$ &              $-$ &              $-$ &              $-$ &              $-$ &              $-$ &              $-$ &              $-$ &              $-$ &              $-$ &              100 \tabularnewline\hline
      $C_{tG}$ &              $-$ &              $-$ &               13 &               29 &               24 &              $-$ &              $-$ &              $-$ &               24 &              $-$ &              $-$ &                9 \tabularnewline\hline
      $C_{tW}$ &              $-$ &              $-$ &              $-$ &              $-$ &              $-$ &              $-$ &              $-$ &              $-$ &              $-$ &               84 &               15 &              $-$ \tabularnewline\hline
      $C_{tB}$ &              $-$ &              $-$ &              $-$ &              $-$ &              $-$ &              $-$ &              $-$ &              $-$ &              $-$ &              $-$ &              $-$ &              100 \tabularnewline\hline
$C_{Qq}^{3,1}$ &              $-$ &              $-$ &              $-$ &              $-$ &              $-$ &              $-$ &              $-$ &              $-$ &              $-$ &              $-$ &              100 &              $-$ \tabularnewline\hline
$C_{Qq}^{3,8}$ &              $-$ &              $-$ &              $-$ &              $-$ &              $-$ &              $-$ &              $-$ &              $-$ &               87 &              $-$ &              $-$ &               13 \tabularnewline\hline
$C_{Qq}^{1,8}$ &              $-$ &              $-$ &              $-$ &              $-$ &              $-$ &              $-$ &              $-$ &              $-$ &               82 &              $-$ &              $-$ &               17 \tabularnewline\hline
  $C_{Qu}^{8}$ &              $-$ &              $-$ &              $-$ &              $-$ &              $-$ &              $-$ &              $-$ &              $-$ &               91 &              $-$ &              $-$ &                7 \tabularnewline\hline
  $C_{Qd}^{8}$ &              $-$ &              $-$ &              $-$ &                2 &              $-$ &              $-$ &              $-$ &              $-$ &               92 &              $-$ &              $-$ &                6 \tabularnewline\hline
  $C_{tq}^{8}$ &              $-$ &              $-$ &              $-$ &                1 &              $-$ &              $-$ &              $-$ &              $-$ &               89 &              $-$ &              $-$ &               10 \tabularnewline\hline
  $C_{tu}^{8}$ &              $-$ &              $-$ &              $-$ &              $-$ &              $-$ &              $-$ &              $-$ &              $-$ &               96 &              $-$ &              $-$ &                3 \tabularnewline\hline
  $C_{td}^{8}$ &              $-$ &              $-$ &              $-$ &                2 &              $-$ &              $-$ &              $-$ &              $-$ &               92 &              $-$ &              $-$ &                5 \tabularnewline\hline
\end{tabular}

\end{small}
\end{center}
\end{adjustwidth}
\caption{\it \label{tab:individual} Relative constraining power in percent of different datasets on each coefficient of the global fit individually. Entries below 1\% are not displayed. `SS', $W_{\text{hel.}}$ and $tX$ refer to Higgs signal strength, $W$-helicity fraction and single top data, respectively.}
\end{table}

\newpage

\begin{table}[h!]
\renewcommand{\arraystretch}{1.2}
\begin{adjustwidth}{-1in}{-1in}
 \begin{center}
\begin{footnotesize}
\vspace{-5mm}

\begin{tabular}{|p{2cm}|p{1cm}|p{15cm}|}
\hline
$2\sigma [\Lambda=1$TeV$]$ & $\Lambda [$TeV$]$ & Eigenvector \\ \hline
\hline

0.0021 & 22 & $+0.73 C_{HB}-0.51 C_{HWB}-0.37 C_{HG}+0.22 C_{HW}-0.07 C_{Hl}^{(3)}-0.06 C_{bH}$\tabularnewline\hline
0.0036 & 17 & $-0.79 C_{HG}+0.45 C_{HWB}+0.30 C_{Hl}^{(3)}-0.17 C_{ll}+0.15 C_{HD}-0.10 C_{bH}-0.09 C_{He}+0.09 C_{\mu H}$\tabularnewline\hline
0.0042 & 15 & $+0.51 C_{HB}+0.46 C_{Hl}^{(3)}+0.46 C_{HG}+0.35 C_{HWB}-0.25 C_{ll}+0.22 C_{HD}-0.16 C_{He}+0.15 C_{HW}-0.10 C_{\mu H}-0.09 C_{Hq}^{(3)}$\tabularnewline\hline
0.0054 & 14 & $+0.99 C_{\mu H}+0.12 C_{HG}$\tabularnewline\hline
0.0066 & 12 & $-0.75 C_{Hl}^{(1)}+0.42 C_{He}+0.31 C_{Hq}^{(3)}-0.23 C_{Hl}^{(3)}+0.21 C_{HWB}+0.14 C_{HB}-0.12 C_{ll}+0.11 C_{HD}+0.08 C_{HQ}^{(1)}+0.08 C_{HQ}^{(3)}+0.06 C_{Hu}$\tabularnewline\hline
0.015 & 8.3 & $-0.56 C_{Hq}^{(3)}+0.49 C_{He}+0.37 C_{Hl}^{(3)}-0.35 C_{HWB}-0.26 C_{ll}-0.19 C_{Hl}^{(1)}-0.18 C_{HB}-0.10 C_{Hu}-0.10 C_{HQ}^{(3)}-0.10 C_{HQ}^{(1)}+0.07 C_{Hd}+0.06 C_{HD}-0.05 C_{HW}-0.05 C_{Hq}^{(1)}$\tabularnewline\hline
0.019 & 7.3 & $-0.62 C_{He}-0.51 C_{Hl}^{(1)}-0.50 C_{HD}-0.23 C_{ll}-0.14 C_{Hq}^{(3)}-0.13 C_{HWB}-0.07 C_{\tau H}-0.07 C_{HB}+0.06 C_{Hl}^{(3)}$\tabularnewline\hline
0.019 & 7.2 & $-0.96 C_{\tau H}+0.27 C_{bH}$\tabularnewline\hline
0.03 & 5.8 & $-0.52 C_{HQ}^{(3)}-0.52 C_{HQ}^{(1)}+0.48 C_{Hq}^{(3)}-0.34 C_{bH}+0.17 C_{Hl}^{(3)}-0.16 C_{ll}-0.14 C_{HWB}-0.13 C_{\tau H}-0.09 C_{HB}+0.06 C_{Hd}$\tabularnewline\hline
0.035 & 5.3 & $+0.88 C_{bH}-0.27 C_{HQ}^{(3)}-0.27 C_{HQ}^{(1)}+0.24 C_{\tau H}+0.12 C_{Hq}^{(3)}-0.10 C_{HG}$\tabularnewline\hline
0.057 & 4.2 & $-0.85 C_{ll}+0.29 C_{Hl}^{(1)}-0.26 C_{Hl}^{(3)}+0.18 C_{Hq}^{(3)}+0.17 C_{HQ}^{(3)}+0.17 C_{HQ}^{(1)}-0.11 C_{HWB}+0.07 C_{bH}-0.07 C_{HB}$\tabularnewline\hline
0.086 & 3.4 & $-0.60 C_{HW}-0.43 C_{Hl}^{(3)}-0.37 C_{Hq}^{(3)}+0.31 C_{HB}-0.24 C_{HQ}^{(3)}+0.23 C_{HWB}-0.22 C_{HQ}^{(1)}-0.13 C_{bH}-0.11 C_{ll}+0.09 C_{Hu}+0.09 C_{Hq}^{(1)}+0.07 C_{Qq}^{3,1}-0.07 C_{HD}+0.06 C_{Hl}^{(1)}$\tabularnewline\hline
0.1 & 3.2 & $-0.98 C_{Qq}^{3,1}-0.17 C_{HW}+0.08 C_{HQ}^{(3)}$\tabularnewline\hline
0.11 & 3 & $+0.66 C_{HW}-0.39 C_{Hl}^{(3)}-0.37 C_{Hq}^{(3)}-0.21 C_{HQ}^{(1)}+0.20 C_{HD}-0.20 C_{HQ}^{(3)}+0.19 C_{Hu}-0.18 C_{Qq}^{3,1}+0.15 C_{Hq}^{(1)}-0.14 C_{HB}+0.12 C_{HWB}-0.10 C_{He}-0.07 C_{bH}+0.07 C_{W}-0.07 C_{Hd}$\tabularnewline\hline
0.14 & 2.7 & $+0.93 C_{tG}+0.30 C_{G}+0.12 C_{Qq}^{1,8}+0.11 C_{tq}^{8}+0.07 C_{Qu}^{8}+0.06 C_{tu}^{8}$\tabularnewline\hline
0.2 & 2.2 & $+0.97 C_{Hq}^{(1)}+0.13 C_{Hl}^{(3)}-0.08 C_{HD}-0.08 C_{Hd}+0.08 C_{HQ}^{(3)}+0.07 C_{HQ}^{(1)}$\tabularnewline\hline
0.25 & 2 & $-0.99 C_{W}+0.07 C_{Hu}$\tabularnewline\hline
0.28 & 1.9 & $-0.92 C_{Hu}+0.24 C_{HD}-0.19 C_{Hl}^{(3)}+0.13 C_{Hd}-0.12 C_{He}-0.08 C_{HQ}^{(3)}-0.08 C_{Hl}^{(1)}-0.07 C_{HQ}^{(1)}-0.06 C_{W}+0.05 C_{Hq}^{(1)}$\tabularnewline\hline
0.31 & 1.8 & $+0.57 C_{Qq}^{1,8}-0.53 C_{tq}^{8}+0.39 C_{tu}^{8}-0.37 C_{Qu}^{8}+0.21 C_{Qq}^{3,8}+0.17 C_{td}^{8}-0.16 C_{Qd}^{8}$\tabularnewline\hline
0.32 & 1.8 & $+1.00 C_{tW}$\tabularnewline\hline
0.38 & 1.6 & $+0.82 C_{G}-0.35 C_{tG}+0.27 C_{tq}^{8}+0.24 C_{Qq}^{1,8}+0.15 C_{Qu}^{8}+0.10 C_{Ht}+0.09 C_{tu}^{8}+0.08 C_{Qd}^{8}-0.06 C_{HBox}+0.06 C_{tH}+0.05 C_{td}^{8}$\tabularnewline\hline
0.51 & 1.4 & $+0.97 C_{Hd}+0.17 C_{Hu}+0.10 C_{HQ}^{(3)}+0.09 C_{Hq}^{(1)}-0.06 C_{Hl}^{(3)}+0.05 C_{HD}$\tabularnewline\hline
0.59 & 1.3 & $-0.49 C_{tq}^{8}-0.47 C_{Qq}^{1,8}+0.43 C_{G}-0.39 C_{Qu}^{8}-0.31 C_{tu}^{8}-0.17 C_{Qd}^{8}-0.15 C_{td}^{8}-0.14 C_{Qq}^{3,8}+0.07 C_{Ht}-0.07 C_{HBox}+0.06 C_{tH}+0.06 C_{HQ}^{(3)}-0.05 C_{HQ}^{(1)}$\tabularnewline\hline
0.77 & 1.1 & $+0.70 C_{HQ}^{(1)}-0.69 C_{HQ}^{(3)}+0.11 C_{G}-0.09 C_{Qq}^{3,1}+0.06 C_{Hd}-0.06 C_{Qq}^{1,8}-0.05 C_{Ht}$\tabularnewline\hline
1.1 & 0.96 & $+0.59 C_{HBox}-0.58 C_{HD}+0.29 C_{He}+0.27 C_{HWB}+0.23 C_{HW}-0.19 C_{Hu}+0.14 C_{Hl}^{(1)}-0.10 C_{tH}+0.09 C_{Ht}+0.08 C_{Hd}+0.08 C_{HB}-0.07 C_{Qq}^{3,8}-0.06 C_{HQ}^{(1)}$\tabularnewline\hline
1.7 & 0.78 & $-0.64 C_{Qq}^{3,8}+0.51 C_{Qq}^{1,8}-0.40 C_{tu}^{8}+0.29 C_{Ht}-0.16 C_{td}^{8}-0.12 C_{Qu}^{8}-0.12 C_{HBox}-0.12 C_{G}+0.08 C_{HQ}^{(1)}-0.07 C_{HQ}^{(3)}-0.06 C_{tB}+0.05 C_{HD}$\tabularnewline\hline
2.1 & 0.7 & $+0.73 C_{HBox}+0.44 C_{HD}-0.31 C_{tH}-0.22 C_{He}-0.20 C_{HWB}-0.19 C_{HW}+0.12 C_{Hu}-0.11 C_{Hl}^{(1)}+0.10 C_{G}-0.06 C_{tB}-0.06 C_{Qu}^{8}-0.06 C_{HB}$\tabularnewline\hline
2.8 & 0.6 & $+0.85 C_{tB}-0.31 C_{Ht}-0.20 C_{Qd}^{8}+0.19 C_{Qu}^{8}+0.17 C_{tH}-0.16 C_{Qq}^{3,8}+0.13 C_{HBox}-0.11 C_{tq}^{8}-0.10 C_{tu}^{8}+0.09 C_{Qq}^{1,8}$\tabularnewline\hline
3.4 & 0.54 & $-0.71 C_{Ht}-0.45 C_{tB}+0.40 C_{tH}+0.16 C_{Qq}^{1,8}-0.14 C_{tu}^{8}+0.13 C_{HBox}+0.13 C_{Qu}^{8}-0.13 C_{Qq}^{3,8}-0.12 C_{tq}^{8}-0.10 C_{td}^{8}+0.06 C_{Qd}^{8}+0.06 C_{G}$\tabularnewline\hline
4.4 & 0.48 & $+0.82 C_{tH}+0.46 C_{Ht}+0.25 C_{HBox}+0.10 C_{HD}+0.09 C_{tu}^{8}-0.09 C_{G}+0.07 C_{Qq}^{3,8}+0.07 C_{tq}^{8}-0.06 C_{Qd}^{8}$\tabularnewline\hline
9.0 & 0.33 & $+0.55 C_{Qu}^{8}-0.46 C_{td}^{8}-0.40 C_{Qd}^{8}+0.36 C_{Qq}^{3,8}-0.26 C_{tq}^{8}+0.19 C_{Ht}-0.19 C_{tu}^{8}+0.15 C_{Qq}^{1,8}-0.13 C_{tH}-0.12 C_{tB}$\tabularnewline\hline
9.6 & 0.32 & $+0.70 C_{td}^{8}-0.40 C_{Qd}^{8}+0.39 C_{Qu}^{8}-0.31 C_{Qq}^{3,8}-0.21 C_{tB}-0.17 C_{tq}^{8}-0.13 C_{Qq}^{1,8}+0.11 C_{Ht}$\tabularnewline\hline
13.0 & 0.28 & $+0.75 C_{Qd}^{8}-0.48 C_{tq}^{8}+0.33 C_{Qu}^{8}+0.18 C_{Ht}+0.16 C_{td}^{8}-0.16 C_{tu}^{8}+0.09 C_{tB}$\tabularnewline\hline
21.0 & 0.22 & $+0.69 C_{tu}^{8}-0.50 C_{Qq}^{3,8}-0.41 C_{td}^{8}+0.20 C_{Qu}^{8}-0.18 C_{Qq}^{1,8}-0.17 C_{tq}^{8}+0.06 C_{Qd}^{8}$\tabularnewline\hline
\end{tabular}

\end{footnotesize}
\end{center}
\end{adjustwidth}
\caption{\it \label{tab:eigenvectors} Components of the eigenvectors found in the principal component analysis of the global fit displayed in Fig.~\ref{fig:eigenvectors}. Components with coefficients of magnitude less than 0.05 are omitted.}
\end{table}

\clearpage

\section{Datasets}
\setlength{\fboxsep}{0pt}
\label{app:datasets}
The following Tables summarise the observables that have been encoded into the {\tt Fitmaker} database.
Those that are not included in the final fit are greyed out. This is usually because they are not 
statistically independent from other data that we include.

{\small
\begin{center}
\begin{tabular}{|p{11cm}|x{1.8cm}|x{0.8cm}|}
\hline
 \textbf{EW precision observables} &
 \textbf{$n_{\text{obs}}$} &
 \textbf{Ref.} \\ \hline
\hline
 Precision electroweak measurements on the $Z$ resonance. \newline 
 $\Gamma_{Z}$, $\sigma_{\text{had.}}^0$, $R_\ell^0$, $A_{FB}^\ell$, $A_\ell(\text{SLD})$, $A_\ell(\text{Pt})$, $R_b^0$, $R_c^0$ $A_{FB}^b$, $A_{FB}^c$, $A_b$ \&
 $A_c$ &
$12$ &
  ~\cite{ALEPH:2005ab} \\ \hline
 Combination of CDF and D0 $W$-Boson Mass Measurements &
  $1$ &
  ~\cite{Aaltonen:2013iut} \\ \hline
 LHC run 1 W boson mass measurement by ATLAS &
 $1$ &
~\cite{Aaboud:2017svj} \\ \hline
\end{tabular}
\\[3.5ex]
\begin{tabular}{|p{11cm}|x{1.8cm}|x{0.8cm}|}
\hline
 \textbf{Diboson LEP \& LHC} &
 \textbf{$n_{\text{obs}}$} &
 \textbf{Ref.} \\ \hline
 \hline
 $W^+\,W^-$ angular distribution measurements at LEP II. &
 $8$ &
 ~\cite{Schael:2013ita} \\ \hline
 $W^+\,W^-$ total cross section measurements at L3 in the $\ell\nu\ell\nu$,  $\ell\nu qq$ \& $qqqq$ final states for 8 energies &
 $24$ &
 ~\cite{Achard:2004zw} \\ \hline
 $W^+\,W^-$ total cross section measurements at OPAL in the $\ell\nu\ell\nu$,  $\ell\nu qq$ \& $qqqq$ final states for 7 energies &
 $21$ &
 ~\cite{Abbiendi:2007rs} \\ \hline
 $W^+\,W^-$ total cross section measurements at ALEPH in the $\ell\nu\ell\nu$,  $\ell\nu qq$ \& $qqqq$ final states for 8 energies &
 $21$ &
 ~\cite{Heister:2004wr} \\ \hline
 \lgr ATLAS $W^+\,W^-$ differential cross section in the $e \nu \mu \nu$ channel, $\tfrac{d\sigma}{dp^T_{\ell_1}}$, $p_T > 120$ GeV overflow bin &
 $1$ &
 ~\cite{Aaboud:2017qkn} \\ \hline
 ATLAS $W^+\,W^-$ fiducial differential cross section in the $e\nu\mu\nu$ channel, $\tfrac{d\sigma}{dp^T_{\ell_1}}$ &
 $14$ &
 ~\cite{Aaboud:2019nkz} \\ \hline
\textcolor{purple}{ ATLAS $W^\pm\,Z$ fiducial differential cross section in the $\ell^+\ell^-\ell^\pm\nu$ channel, $\tfrac{d\sigma}{dp^T_{Z}}$} &
 $7$ &
 ~\cite{Aaboud:2019gxl} \\ \hline
 \textcolor{purple}{CMS $W^\pm\,Z$ normalised fiducial differential cross section in the $\ell^+\ell^-\ell^\pm\nu$ channel, $\tfrac{1}{\sigma}\tfrac{d\sigma}{dp^T_{Z}}$ }&
 $11$ &
 ~\cite{Sirunyan:2019bez} \\ \hline
 ATLAS $Zjj$ fiducial differential cross section in the $\ell^+\ell^-$ channel, $\tfrac{d\sigma}{d\Delta\phi_{jj}}$ &
 $12$ &
 ~\cite{Aad:2020sle} \\ \hline
\end{tabular}
\\[3.5ex]
\begin{tabular}{|p{11cm}|x{1.8cm}|x{0.8cm}|}
\hline
  \textbf{LHC Run 1 Higgs} &
 \textbf{$n_{\text{obs}}$} &
 \textbf{Ref.} \\ \hline
 \hline
 ATLAS and CMS LHC Run 1 combination of Higgs signal strengths. \newline Production: $ggF$, $VBF$, $ZH$, $WH$ \&
 $ttH$ \newline Decay: $\gamma\gamma$, $ZZ$, $W^+W^-$, $\tau^+\tau^-$ \&
 $b\bar{b}$ &
 $21$ &
 ~\cite{Khachatryan:2016vau} \\ \hline
 ATLAS inclusive $Z\gamma$ signal strength measurement &
 $1$ &
 ~\cite{Aad:2015gba} \\ \hline
\end{tabular}
\\[3.5ex]
\begin{tabular}{|p{11cm}|x{1.8cm}|x{0.8cm}|}
\hline
 \textbf{LHC Run 2 Higgs (new)} &
 \textbf{$n_{\text{obs}}$} &
 \textbf{Ref.} \\ \hline
 \hline
 ATLAS combination of signal strengths and stage 1.0 STXS in $H\to 4\ell$ including ratios of branching fractions to $\gamma\gamma$, $WW^\ast$, $\tau^+\tau^-$ \&
 $b\bar{b}$ \newline \lgbox{Signal strengths$|$coarse STXS bins$|$} fine STXS bins  &
 $\lgbox{$16|19|$}25$ &
 ~\cite{Aad:2019mbh} \\ \hline
 CMS LHC combination of Higgs signal strengths. \newline Production: $ggF$, $VBF$, $ZH$, $WH$ \&
 $ttH$ \newline Decay: $\gamma\gamma$, $ZZ$, $W^+W^-$, $\tau^+\tau^-$, $b\bar{b}$ \&
 $\mu^+\mu^-$ &
 $23$ &
 ~\cite{CMS:2020gsy} \\ \hline
 \lgr  CMS stage 1.0 STXS measurements for $H\to\gamma\gamma$. \newline 13 parameter fit $|$ 7 parameter fit &
 $13 | 7$ &
 ~\cite{CMS:1900lgv} \\ \hline
 \lgr  CMS stage 1.0 STXS measurements for $H\to\tau^+\tau^-$ &
 $9$ &
 ~\cite{CMS:2019pyn} \\ \hline
 \lgr  CMS stage 1.1 STXS measurements for $H\to 4\ell$ &
 $19$ &
 ~\cite{CMS:2019chr} \\ \hline
 \lgr  CMS differential cross section measurements of inclusive Higgs production in the $WW^\ast\to\ell\nu\ell\nu$ final state. \newline$\tfrac{d\sigma}{dn_{\text{jet}}}\quad\big|\quad\tfrac{d\sigma}{dp^T_{H}}$ &
 $5 | 6$ &
 ~\cite{CMS:2019kqw} \\ \hline
 ATLAS $H\to Z\gamma$ signal strength. &
 $1$ &
 ~\cite{Aad:2020plj} \\ \hline
 ATLAS $H\to \mu^+\mu^-$ signal strength. &
 $1$ &
 ~\cite{Aad:2020xfq} \\ \hline
\end{tabular}
\\[3.5ex]
\begin{tabular}{|p{11cm}|x{1.8cm}|x{1cm}|}
\hline
  \textbf{Tevatron \&
 Run 1 top} &
 \textbf{$n_{\text{obs}}$} &
\textbf{Ref.} \\ \hline
\hline Tevatron combination of differential $\mathrm{t}\overline{\mathrm{t}}$ forward-backward asymmetry, $A_{FB}(m_{t\bar{t}})$.  &
 $4$ &
 ~\cite{Aaltonen:2017efp} \\ \hline
 ATLAS $t\bar{t}$ differential distributions in the dilepton channel. \newline 
 $\tfrac{d\sigma}{dm_{t\bar{t}}}$ &
 $6$ &
 ~\cite{Aaboud:2016iot} \\ \hline
 ATLAS $t\bar{t}$ differential distributions in the $\ell$+jets channel.  \newline
 $\lgbox{$\tfrac{d\sigma}{dm_{t\bar{t}}}\quad\big|\quad\tfrac{d\sigma}{d|y_{t\bar{t}}|}\quad\big|$}\quad\tfrac{d\sigma}{dp^T_{t}}\quad\lgbox{$\big|\quad\tfrac{d\sigma}{d|y_{t}|}$}$. &
 $\lgbox{7$|$5$|$}8\lgbox{$|$5}$ &
 ~\cite{Aad:2015mbv} \\ \hline
 CMS $t\bar{t}$ differential distributions in the $\ell$+jets channel.  \newline 
 $\lgbox{$\tfrac{d\sigma}{dm_{t\bar{t}}}\quad\big|\quad\tfrac{d\sigma}{dy_{t\bar{t}}}\quad\big|$}\quad\tfrac{d\sigma}{dp^T_{t}}\quad\lgbox{$\big|\quad\tfrac{d\sigma}{dy_{t}}$}$. &
 \lgbox{7$|$10$|$}8\lgbox{$|$10} &
 ~\cite{Khachatryan:2015oqa,Khachatryan:2016yzq} \\ \hline
  CMS  measurement of differential $\mathrm{t}\overline{\mathrm{t}}$ charge asymmetry, $A_C(m_{t\bar{t}})$ in the dilepton channel.  &
 $3$ &
 ~\cite{Khachatryan:2016ysn} \\ \hline
 ATLAS inclusive measurement $\mathrm{t}\overline{\mathrm{t}}$ charge asymmetry, $A_C(m_{t\bar{t}})$  in the dilepton channel.  &
 $1$ &
 ~\cite{Aad:2016ove} \\ \hline
 ATLAS \&
 CMS  combination of differential $\mathrm{t}\overline{\mathrm{t}}$ charge asymmetry, $A_C(m_{t\bar{t}})$, in the $\ell$+jets channel.  &
 $6$ &
 ~\cite{Sirunyan:2017lvd} \\ \hline
 CMS $t\bar{t}$ double differential distributions in the dilepton channel.  \newline 
 $\lgbox{$\tfrac{d\sigma}{dm_{t\bar{t}}dy_{t}}\quad\big|$}\quad\tfrac{d\sigma}{dm_{t\bar{t}}dy_{t\bar{t}}}\quad\lgbox{$\big|\quad\tfrac{d\sigma}{dm_{t\bar{t}}dp^T_{t\bar{t}}}\quad\big|\quad\tfrac{d\sigma}{dy_{t}dp^T_{t}}$}$. &
 \lgbox{16$|$}16\lgbox{$|$16$|$16} &
 ~\cite{Sirunyan:2017azo,Chatrchyan:2013faa} \\ \hline
 ATLAS \&
 CMS Run 1 combination of $W$-boson helicity fractions in top decay. $f_0,\, f_L\,\& \, f_R$ &
 $3$ &
  ~\cite{Aad:2020jvx} \\ \hline
 \lgr ATLAS measurement of $W$-boson helicity fractions in top decay. $f_0,\, f_L\,\& \, f_R$ &
 $3$ &
  ~\cite{Aaboud:2016hsq} \\ \hline
 \lgr CMS measurement of $W$-boson helicity fractions in top decay. $f_0,\, f_L\,\& \, f_R$ &
 $3$ &
 ~\cite{Khachatryan:2016fky} \\ \hline
 ATLAS $t\bar{t}W$ \& $t\bar{t}Z$ cross section measurements. $\sigma_{t\bar{t}W}|\sigma_{t\bar{t}Z}$ &
 $2$ &
 ~\cite{Aad:2015eua} \\ \hline
 CMS $t\bar{t}W$ \& $t\bar{t}Z$ cross section measurements. $\sigma_{t\bar{t}W}|\sigma_{t\bar{t}Z}$ &
 $2$ &
  ~\cite{Khachatryan:2015sha} \\ \hline
   ATLAS $t\bar{t}\gamma$ cross section measurement in the $\ell+$ jets channel.&
 $1$ &
 ~\cite{Aaboud:2017era} \\ \hline
  CMS $t\bar{t}\gamma$ cross section measurement in the $\ell+$ jets channel.&
 $1$ &
  ~\cite{Sirunyan:2017iyh} \\ \hline
 ATLAS $t$-channel single-top differential distributions.  \newline 
 $\tfrac{d\sigma}{dp^T_{t}}\quad\lgbox{$\big|\quad\tfrac{d\sigma}{dp^T_{\bar{t}}}\quad\big|\quad\tfrac{d\sigma}{d|y_{t}|}\quad\big|$}\quad\tfrac{d\sigma}{d|y_{\bar{t}}}|$ &
 $4\lgbox{$|$4$|$4$|$}5$ &
 ~\cite{Aaboud:2017pdi} \\ \hline
 CMS $s$-channel single-top cross section measurement. &
 $1$ &
 ~\cite{Khachatryan:2016ewo} \\ \hline
 CMS $t$-channel single-top differential distributions.  \newline 
 $\tfrac{d\sigma}{dp^T_{t+\bar{t}}}\quad\lgbox{$\big|\quad\tfrac{d\sigma}{d|y_{t+\bar{t}}|}$}$ &
 $6$\lgbox{$|6$} &
 ~\cite{CMS:2014ika} \\ \hline
 CMS measurement of the $t$-channel single-top and anti-top cross sections. $\lgbox{$\sigma_t\,|\,\sigma_{\bar{t}}\,|\,\sigma_{t+\bar{t}}\,|$}\,
 R_t$. &
 $\lgbox{$1|1|1|$}1$ &
 ~\cite{Khachatryan:2014iya} \\ \hline
 ATLAS $s$-channel single-top cross section measurement. &
 $1$ &
 ~\cite{Aad:2015upn} \\ \hline
 CMS $tW$ cross section measurement. &
 $1$ &
 ~\cite{Chatrchyan:2014tua} \\ \hline
 ATLAS $tW$ cross section measurement in the single lepton channel. &
 $1$ &
 ~\cite{Aad:2020zhd} \\ \hline
 ATLAS $tW$ cross section measurement in the dilepton channel. &
 $1$ &
 ~\cite{Aad:2015eto} \\ \hline
\end{tabular}
\\[3.5ex]
\begin{tabular}{|p{11cm}|x{1.8cm}|x{0.8cm}|}
\hline
 \textbf{Run 2 top} &
 \textbf{$n_{\text{obs}}$} &
 \textbf{Ref.} \\ \hline
 \hline
 CMS $t\bar{t}$ differential distributions in the dilepton channel. \newline 
 $\tfrac{d\sigma}{dm_{t\bar{t}}}$ &
 $6$ &
 ~\cite{Sirunyan:2017mzl,Sirunyan:2018goh} \\ \hline
 CMS $t\bar{t}$ differential distributions in the $\ell+$jets channel. \newline 
 $\tfrac{d\sigma}{dm_{t\bar{t}}}$ &
 $10$ &
 ~\cite{Sirunyan:2018wem} \\ \hline
 ATLAS measurement of differential $\mathrm{t}\overline{\mathrm{t}}$ charge asymmetry, $A_C(m_{t\bar{t}})$. &
 $5$ &
 ~\cite{ATLAS:2019czt} \\ \hline
 ATLAS $t\bar{t}W$ \& $t\bar{t}Z$ cross section measurements. $\sigma_{t\bar{t}W}|\sigma_{t\bar{t}Z}$ &
 $2$ &
 ~\cite{Aaboud:2019njj} \\ \hline
 CMS $t\bar{t}W$ \& $t\bar{t}Z$ cross section measurements. $\sigma_{t\bar{t}W}\lgbox{$|\sigma_{t\bar{t}Z}$}$ &
 $1\lgbox{$|1$}$ &
 ~\cite{Sirunyan:2017uzs} \\ \hline
 CMS $t\bar{t}Z$ differential distributions. \newline 
 $\tfrac{d\sigma}{dp^T_{Z}}\quad\lgbox{$\big|\quad\tfrac{d\sigma}{d\cos\theta^\ast}$}$ &
 $4 \lgbox{$|4$}$ &
 ~\cite{CMS:2019too} \\ \hline
ATLAS $t\bar{t}\gamma$ differential distribution. \newline 
 $\tfrac{d\sigma}{dp^T_{\gamma}}$ &
 $11$ &
 ~\cite{Aad:2020axn} \\ \hline
 CMS measurement of differential cross sections and charge ratios for $t$-channel single-top quark production. \newline
 $\tfrac{d\sigma}{dp^T_{t+\bar{t}}}\quad\big|\quad R_t\left(p^T_{t+\bar{t}}\right)$ &
 $5 | 5$ &
 ~\cite{Sirunyan:2019hqb} \\ \hline
 \lgr CMS measurement of $t$-channel single-top and anti-top cross sections. \newline
 $\sigma_{t},\,\sigma_{\bar{t}},\,\sigma_{t+\bar{t}} \,\& \, R_t$. &
 $4$ &
 ~\cite{Sirunyan:2016cdg} \\ \hline
 CMS measurement of the $t$-channel single-top and anti-top cross sections. $\lgbox{$\sigma_t\,|\,\sigma_{\bar{t}}\,|$}\,\sigma_{t+\bar{t}}\,|\, R_t$. &
 \lgbox{1$|$1$|$}1$|$1 &
 ~\cite{Aaboud:2016ymp} \\ \hline
 \lgr CMS $t$-channel single-top differential distributions. \newline 
 $\tfrac{d\sigma}{dp^T_{t+\bar{t}}}\quad\big|\quad\tfrac{d\sigma}{d|y_{t+\bar{t}}|}$ &
 $4 | 4$ &
 ~\cite{CMS:2016xnv} \\ \hline
 ATLAS $tW$ cross section measurement. &
 $1$ &
 ~\cite{Aaboud:2016lpj} \\ \hline
 \lgr CMS $tZ$ cross section measurement. &
 $1$ &
 ~\cite{Sirunyan:2017nbr} \\ \hline
 CMS $tW$ cross section measurement. &
 $1$ &
 ~\cite{Sirunyan:2018lcp} \\ \hline
 ATLAS $tZ$ cross section measurement. &
 $1$ &
 ~\cite{Aaboud:2017ylb} \\ \hline
 CMS $tZ\,(Z\to\ell^+\ell^-)$ cross section measurement &
 $1$ &
 ~\cite{Sirunyan:2018zgs} \\ \hline
 \lgr ATLAS four-top search in the multi-lepton and same-sign dilepton channels. &
 $1$ &
 ~\cite{ATLAS:2020hrf} \\ \hline
 \lgr ATLAS four-top search in the single-lepton and opposite-sign dilepton channels. &
 $1$ &
 ~\cite{Aaboud:2018jsj} \\ \hline
 \lgr CMS four-top search in the multi-lepton and same-sign dilepton channels. &
 $1$ &
 ~\cite{Sirunyan:2019wxt} \\ \hline
 \lgr CMS four-top search in the single-lepton and opposite-sign dilepton channels. &
 $1$ &
 ~\cite{Sirunyan:2019nxl} \\ \hline
 \lgr CMS $t\bar{t}b\bar{b}$ cross section measurement in the all-jet channel. &
 $1$ &
 ~\cite{Sirunyan:2019jud} \\ \hline
 \lgr CMS $t\bar{t}b\bar{b}$ cross section measurement in the dilepton channel. &
 $1$ &
 ~\cite{Sirunyan:2020kga} \\ \hline
\end{tabular}
\end{center}
}


 \newpage
 \section{Numerical fits with nested sampling}
 \label{app:nestedsampling}
 
Throughout this paper we make use of the fitting procedure outlined in Section \ref{sec:procedure}, in which we minimise a $\chi^{2}$ function and determine the least-squares estimator for each coefficient $\hat{\vec{C}}$.  This procedure has the advantage of being analytic, as it is linearised in the coefficients $\vec{C}$, and therefore fast to implement.  However, it relies on a number of key assumptions: the linear approximation to the SMEFT predictions $\mu(\vec{C})$ must be accurate, and the experimental covariance matrix $V$ must be symmetrised.  Even if these assumptions are satisfied and we can write down a Gaussian likelihood $\mathcal{L}(\vec{C}|D)$ where $D$ denotes the dataset, the analytic fitting procedure excludes the possibility that a nontrivial prior $\pi(\vec{C})$ may lead to a non-Gaussian posterior $p(\vec{C} | D)$ through Bayes' theorem: $ p(\vec{C} | D) \propto \mathcal{L}(\vec{C}| D) \pi(\vec{C})$.  

With this motivation, we implement in \texttt{Fitmaker} an option to constrain parameters numerically using the method of nested sampling.  This is provided through {\tt MultiNest}~\cite{Feroz:2008xx}, incorporated into our code using {\tt Pymultinest}~\cite{Buchner:2014nha}.

As in traditional Markov Chain Monte Carlo (MCMC) parameter estimation, nested sampling is a method of sampling from the posterior distribution.  {\tt MultiNest} uses the ellipsoidal nested sampling algorithm, described in more detail in \cite{Feroz:2008xx}, in which the samples are drawn from ellipsoids in parameter space.  As the algorithm progresses the ellipsoids close in on the regions of high likelihood.  Overlapping and distinct ellipsoids allow for the possibility of degeneracies and multiple modes in the posterior respectively.  This is an advantage over traditional MCMC methods, in which the Markov chain may get stuck and fail to explore more than one mode of the posterior.  The importance of this feature in the context of quadratic SMEFT contributions will be seen below.

We apply nested sampling in Section \ref{sec:UV} to find constraints on the 1-parameter UV models in Table~\ref{table:fields}.  In many of these models we are constraining a positive quantity $|\lambda|^{2}$, where $\lambda$ denotes a coupling of the new field to the SM.  Nested sampling allows us to produce the constraints due to positivity bounds on $|\lambda|^{2}$ using a Heaviside prior: $\pi(|\lambda|^{2}<0)=0$.  The constraints in Figure \ref{fig:1Dlimits} are found in this way.

 \begin{figure}[t] 
\centering
\includegraphics[width=1.0\textwidth]{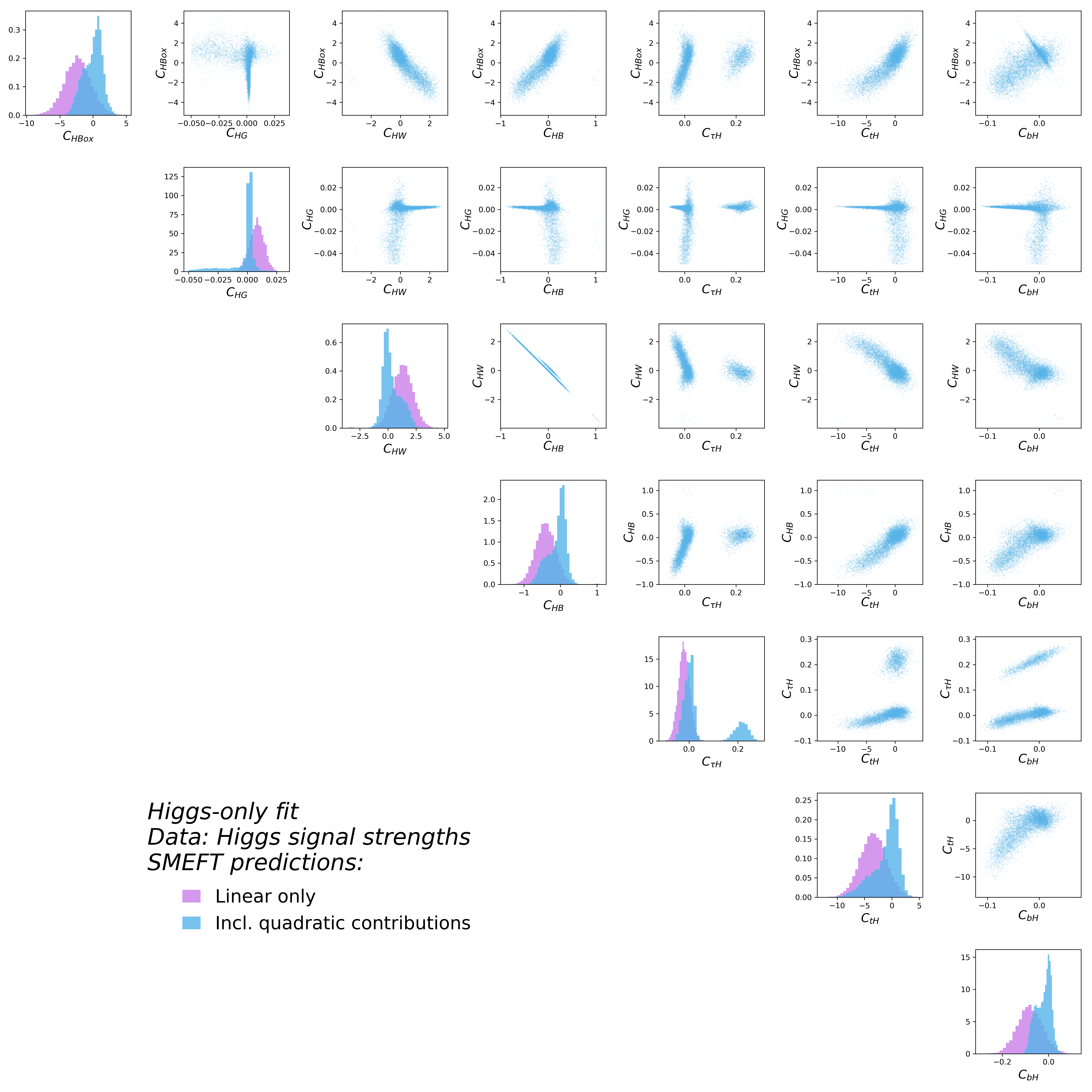}\caption{\it \label{fig:quadfit} Samples produced by nested sampling, projected onto 2- and 1-dimensional subspaces of the 7-dimensional parameter space.  Along the diagonal we compare the distributions found with and without quadratic SMEFT contributions in blue and purple, respectively.}
\end{figure}

As a proof-of-concept of the capabilities of nested sampling, we investigate in this Section the effects of including quadratic contributions from dimension-6 operators in SMEFT predictions from Higgs data.  We perform a Higgs-only fit using just the 
Run~2 signal strength measurements from ATLAS \cite{Aad:2019mbh} and CMS \cite{CMS:2020gsy}, and constrain 7 operators: $C_{H \Box}$, $C_{HG}$, $C_{HW}$, $C_{HB}$, $C_{\tau H}$, $C_{tH}$ and $C_{bH}$.  For the purpose of this proof-of-concept fit, we take our SMEFT predictions from \cite{Hays:2019cbc}, rotating the SILH basis operators into the Warsaw basis using the {\tt Rosetta} code~\cite{Falkowski:2015wza}.  This fit differs from the results in the main text: as well as the differences in the SMEFT predictions, we use just a subset of the full dataset and use only signal strengths, not STXS measurements.

\begin{figure}[t] 
\centering
\includegraphics[width=1.0\textwidth]{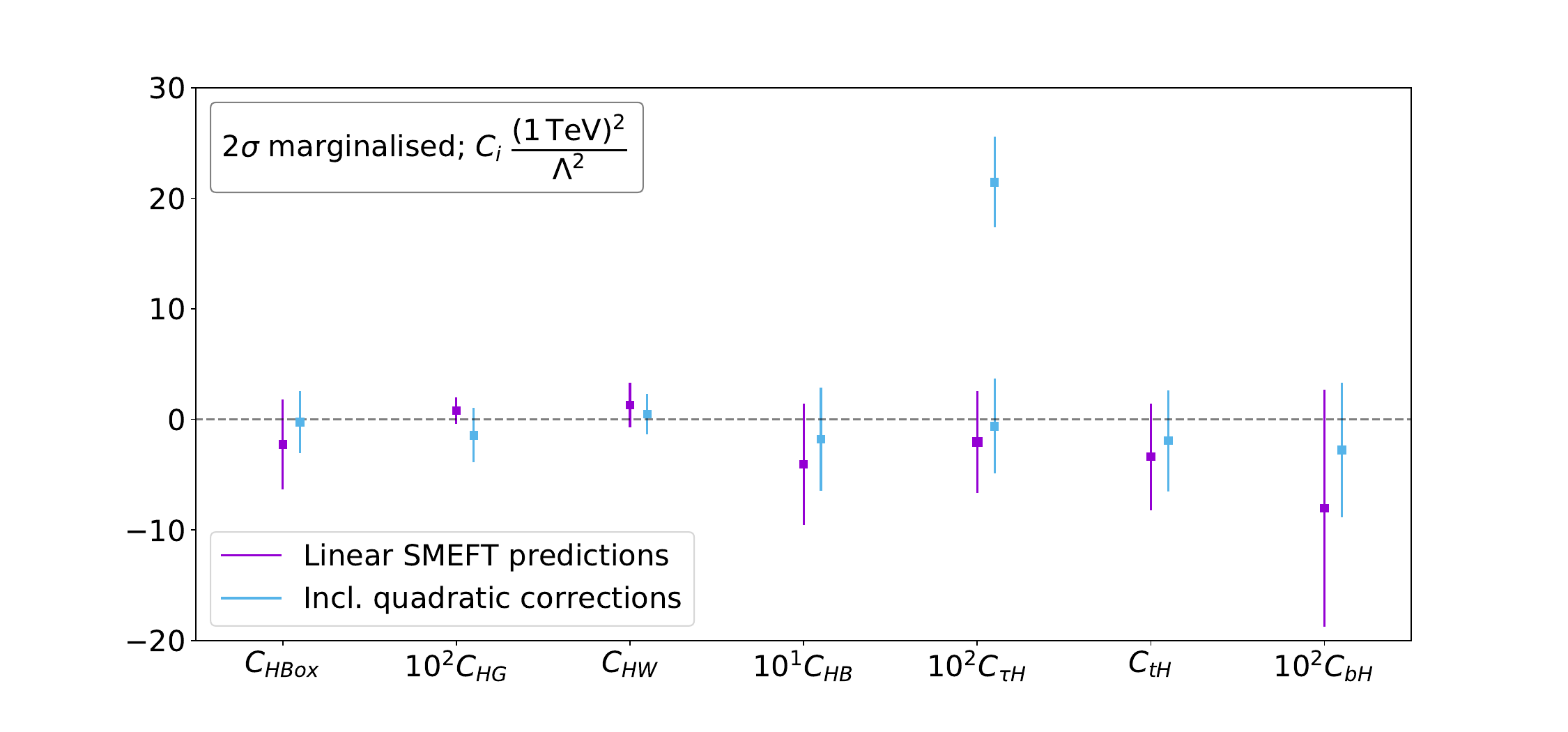}\caption{\it
\label{fig:quadcons} Marginalised 95 \% credible intervals for each of the indicated operator coefficients resulting from a fit to Higgs signal strength data using nested sampling.  We compare the effects of including quadratic contributions from dimension-6 operators (blue) to the case of linearised SMEFT predictions (purple). }
\end{figure}

Figure~\ref{fig:quadfit} shows the distribution of 30,000 samples produced by nested sampling (with a sampling efficiency of 0.8 and an evidence tolerance of 0.5, taking approximately 1 hour).    Each 2-dimensional distribution is a projection of the full 7-dimensional posterior distribution onto a 2-parameter subspace.  These plots highlight the non-Gaussianity of the posterior distributions when quadratic contributions are included.  In particular, we see multiple modes in the distribution of $C_{\tau H}$ as well as highly skewed distributions in $C_{tH}$ and $C_{bH}$. 

The histograms along the diagonal in Fig.~~\ref{fig:quadfit} show the distributions of samples in each of the 7 parameters.  Here we compare the results of nested sampling with and without quadratic contributions in blue and purple respectively.  We see that although the distributions are generally peaked close to the same value, the shapes of the distributions differ, with the quadratic contributions leading to more asymmetric and multimodal distributions.  For example, the distribution of $C_{HG}$ is more skewed towards the negative region when quadratic corrections are included, and $C_{bH}$ is much more narrowly constrained than in the linear-only case.  These differences in distributions translate into differences in the marginalised 95 \% credible intervals shown in Fig.~\ref{fig:quadcons}, computed as highest posterior density intervals.
 
Although there are visible differences between the credible regions found in the linear and quadratic SMEFT fits, they are sufficiently similar that one may consider the linear approximation to be usefully robust.  The most notable difference is in $C_{\tau H}$, in which a distinct second mode is found in the quadratic SMEFT fit, while the first mode is in good agreement with the linear SMEFT fit.  There is only one other instance, namely $C_{HG}$, where the mode in the quadratic fit lies outside the 95\% CL range found in the linear approximation, and only one instance, namely $C_{bH}$, where the size of the quadratic 95\% credible interval is much smaller than the linear 95\% CL range. Apart from these exceptions, the ranges estimated in the linear fit are encouraging approximations to the results from the quadratic fit. We note that a global quadratic fit would require calculations of many currently unknown quadratic operator contributions and, for consistency, a full treatment of the linear contributions of dimension-8 operators as discussed in the Higgs sector in~\cite{Hays:2018zze}.

\FloatBarrier

\newpage

\thispagestyle{empty}

\newpage

\bibliographystyle{JHEP}
\bibliography{biblio}

\end{document}